\documentclass[useAMS,usenatbib]{mn2e_mead}

\usepackage{times}

\usepackage{graphicx}

\usepackage{wasysym}

\usepackage{subfig}

\usepackage{color}

\usepackage{lastpage}

\newcommand{\Mpc}{h^{-1}\,\mathrm{Mpc}}
\newcommand{\iMpc}{h\,\mathrm{Mpc}^{-1}}
\newcommand{\Msun}{h^{-1}\,\mathrm{M_\odot}}

\newcommand{\eg}{e.g.,\,} 
\newcommand{\ie}{i.e.\,}

\newcommand{\sform}[2]{{#1}\times 10^{#2}}
\newcommand{\average}[1]{\langle{#1}\rangle}
\newcommand{\new}[1]{{\color{black}{#1}}}
\newcommand{\Om}{\Omega_\mathrm{m}}
\newcommand{\Ov}{\Omega_\Lambda}
\newcommand{\Ob}{\Omega_\mathrm{b}}
\newcommand{\lagr}{\mathcal{L}}

\def\m@th{\mathsurround=0pt }
\def\sqr#1#2{{\phantom{\vrule width.15em}
    \lower.05em\vbox{\hrule height.#2pt
    \hbox{\vrule width.#2pt height#1pt \kern#1pt
    \vrule width.#2pt}
    \hrule height.#2pt}
    \phantom{\vrule width.15em}}}    
\def\dalem{\mathchoice\sqr64\sqr64\sqr43\sqr33}
\def\eqalign#1{\null\,\vcenter{\openup1\jot \m@th\ialign{\strut\hfil$\displaystyle{##}$&$\displaystyle{{}##}$\hfil\crcr#1\crcr}}\,}
\def\gtsim{\mathrel{\lower0.6ex\hbox{$\buildrel {\textstyle >}\over {\scriptstyle \sim}$}}}
\def\ltsim{\mathrel{\lower0.6ex\hbox{$\buildrel {\textstyle <}\over {\scriptstyle \sim}$}}}

\title[Rapid simulation rescaling to modified gravity]{Rapid simulation rescaling from standard to modified gravity models}
\author[A. J. Mead et al.]{A. J. Mead$^{1}$\thanks{E-mail: am@roe.ac.uk}, J. A. Peacock$^{1}$, L. Lombriser$^{1}$ and B. Li$^{2}$\\
$^{1}$Institute for Astronomy, University of Edinburgh, Royal Observatory, Blackford Hill, Edinburgh EH9 3HJ\\
$^{2}$Institute for Computational Cosmology, Department of Physics, Durham University, South Road, Durham DH1 3LE\\}

\begin{document}

\date{Accepted 2015 July 02.  Received 2015 May 29; in original form 2014 December 16.}
\pagerange{\pageref{firstpage}--\pageref{lastpage}}
\pubyear{2015}

\maketitle

\label{firstpage}

\begin{abstract}
We develop and test an algorithm to rescale a simulated dark-matter particle distribution or halo catalogue from a standard gravity model to that of a modified gravity model. This method is based on that of Angulo \& White but with some additional ingredients to account for (i) scale-dependent growth of linear density perturbations and (ii) screening mechanisms that are generic features of viable modified gravity models. We attempt to keep the method as general as possible, so that it may plausibly be applied to a wide range of modified theories, although tests against simulations are restricted to a subclass of $f(R)$ models at this stage. We show that rescaling allows the power spectrum of matter to be reproduced at the $\sim 3$ per cent level in both real and redshift space up to $k=0.1\iMpc$ \new{if we change the box size and alter the particle displacement field}; this limit can be extended to $k=1\iMpc$ if \new{we additionally alter halo internal structure}. We simultaneously develop an algorithm that can be applied directly to a halo catalogue, in which case the halo mass function and clustering can be reproduced at the $\sim 5$ per cent level. Finally we investigate the clustering of halo particle distributions, generated from rescaled halo catalogues, and find that a similar accuracy can be reached.
\end{abstract}

\begin{keywords}
cosmology: theory -- dark energy -- large-scale structure of Universe
\end{keywords}

\section{Introduction}
\label{sec:introduction}


The accelerated expansion of the cosmos currently lacks a unique explanation. Either all of space is pervaded by an invisible dark energy with negative pressure, and this accelerates the expanding cosmos, or the gravitational field equations of Einstein are incorrect on cosmological scales and accelerated expansion arises naturally within the framework of the correct theory. Such `modified gravity' (MG) theories are the subject of this paper. In order to comply with contemporary observational data, these theories are designed to yield a nearly standard background expansion, but have a modified growth rate for perturbations. Gravity is then restored to the standard by some `screening' mechanism in environments such as the Solar system where modifications to gravity are limited by high-accuracy experiments. In most cases the theory can be understood as the interaction of gravity with some new scalar field that produces a new (sometimes called fifth) force in the Universe. In order to constrain such models, it is necessary to map both the gross expansion history of the Universe, for example via supernovae standard candles (\eg \citealt{Schmidt1998}; \citealt{Perlmutter1999}; \citealt{Suzuki2012}) and the evolution of density fluctuations -- either directly using gravitational lensing (\eg \citealt{Heymans2013}) or via some tracer population (\eg  \citealt{delaTorre2013a}; \citealt{Samushia2013}).

Increasingly, the ability to extract information from cosmological surveys requires the use of simulated mock data. In the case of lensing, mock mass distributions are required for large numbers of realizations because lensing necessarily mixes the linear scales of the underlying Gaussian field with non-linear information that is not as well understood (\eg \citealt{White2004}). This mixing also makes the data covariance complicated (\eg \citealt{Harnois-Deraps2015}). In the case of galaxy surveys the relation between the underlying dark matter and tracer galaxies is complicated (\eg \citealt{Peacock2000}; \citealt{Seljak2000}) and mock galaxy catalogues based on simulations are required to understand how statistics deduced from galaxy surveys (such as the power spectrum) relate to the corresponding property for the mass density field. Additionally, simulations are necessary in order to model complicated biases that arise due to the observation process, such as the effect of the geometry of the survey selection function, or close pairs of galaxies not being sampled owing to the need to avoid fibre collisions in a multiplexed spectroscopic survey.

In principle, direct simulations of any model under consideration, including MG theories, can be used to test the model against data. Recently codes have been developed to simulate MG models (\eg \citealt{Oyaizu2008a}; \citealt{Li2012a}; \citealt{Puchwein2013}; \citealt{Llinares2014}) but the simulations are complicated by the need to solve non-linear equations for the scalar field in tandem with the standard gravitational Poisson equation; this results in an increased run time.

It would therefore be useful to have a way of running simulations of MG models more rapidly. Recently \cite{Winther2015} developed a method to run approximate MG simulations by using a linear prescription for the scalar field equations, combined with a screening mechanism that is input to the simulation by hand. This reduces the run-time for an MG simulation to a similar level to that of a standard gravity simulation at the expense of some accuracy. In this paper we adopt a different approach and attempt to apply the cosmological rescaling algorithm developed by \citeauthor{Angulo2010} (\citeyear{Angulo2010}; hereafter AW10). AW10 showed that it is possible to rescale an evolved $N$-body particle distribution in order to approximate the results of a simulation with a different set of cosmological parameters, by changing the size and redshift of the evolved box so as to best match the halo mass function and then correcting the linear modes using the \cite{Zeldovich1970} approximation. This method is extremely fast, and yet it still generates a fully non-linear matter distribution. The AW10 method contains no free parameters whatsoever and simply requires the parameters of the original and target cosmologies together with fitting functions that are standard in the literature. AW10 showed that their method successfully reproduces the halo mass function and clustering statistics of the target cosmology in both real and redshift space. Subsequently AW10 has been applied by \cite{Guo2013} to look at theoretical differences in galaxy formation between WMAP1 and WMAP7 cosmologies and by \cite{Simha2013}, who looked at measuring cosmological parameters by comparing the galaxy two-point correlation function of the Sloan Digital Sky Survey with that computed from galaxy catalogues rescaled using the AW10 method. \new{Recently, \cite{Angulo2015} showed that simulation rescaling could be used to generate useful predictions for lensing correlation functions and the authors were able to carry out an analysis of the CFHTLenS data that avoided the use of standard non-linear fitting formulae.}

\new{In \cite{Ruiz2011} it was shown the AW10 method could be applied to halo catalogues, but the authors did not implement the displacement field step. In \citeauthor{Mead2014a} (\citeyear{Mead2014a}; hereafter MP14a) this was remedied and it was shown that the AW10 could be applied directly to halo catalogues in a completely self contained manner, and that the mass function and power spectrum of haloes were well matched post rescaling.} This is advantageous because halo catalogues require very much less disc space than particle data and catalogues are often all that are required to subsequently produce a mock galaxy catalogue. Applying the algorithm to haloes directly also saves the computational expense of running a halo finder on the rescaled particle data and is also much faster because run-time scales roughly in proportion to the number of particles or haloes being rescaled. Additionally MP14a showed that the original AW10 algorithm (applied to dark matter simulation particles) could be improved if the properties of individual haloes were altered to match the deeply non-linear clustering. In MP14a we showed that these methods worked well on haloes in real space, although a biased displacement field was required in order to preserve the mass-dependent clustering of haloes. In a follow up paper (\citealt{Mead2014b}; hereafter MP14b) it was shown that the method also produces good results in redshift space.

Given the current interest in MG models, and the relatively poor speed of direct MG simulations, it therefore seems interesting to ask if it is possible to approximate the results of such models using the rescaling approach. This is the main aim of the current paper, which is set out as follows: In Section~\ref{sec:mg} we review MG theories and particularly the subclass of \citeauthor{Hu2007a} (\citeyear{Hu2007a}; hereafter HS07) $f(R)$ models. In doing so we discuss perturbation theory and the chameleon mechanism which screens modifications to gravity in dense environments in HS07 models. Those familiar with MG may wish to skip straight to Section~\ref{sec:simulations}, in which we discuss details of simulations that were run in order to test the rescaling algorithm. In Section~\ref{sec:rescaling} we present our rescaling method in tandem with results for the power spectrum of particles and haloes in both real and redshift space. We show that in applying the AW10 method to MG models one must take into account both the modified perturbation growth rate and screening mechanism in the apparatus used in the original AW10/MP14 algorithms. Particularly the differences induced in halo mass function, linear fluctuation growth and halo internal structure. Finally we sum up in Section~\ref{sec:discussion}. The appendix contains the mapping from $f(R)$ to \cite{Brans1961} type theories together with the necessary machinery that would be required to generalize the method presented in this papers to these models. 

\section{Modified gravity}
\label{sec:mg}



Viable MG theories can be characterized as involving a modified growth rate of density perturbations, which may be scale dependent, combined with a screening mechanism to restore gravity to the standard in environments where gravity is well measured, such as the Solar system. In chameleon theories (\citealt{Khoury2004}) the screening is a function of halo mass and environment while in \cite{Vainshtein1972} models the screening depends primarily on the local density. 

In this paper we work in the \cite*{b:MisnerThorneWheeler} defined metric convention ($---$) and use units such that $c=1$.

Physically motivated theories typically change the Einstein-Hilbert action, from which the gravitational field equations are derived, thus retaining all the principal apparatus of general relativity. One may, for instance, consider non-linear functions of the Ricci Scalar ($R$), rather than just a linear $R$ term, to appear in the action. These are so-called $f(R)$ theories (\citealt{Buchdal1970}; \citealt{Capozziello2003}; \citealt{Nojiri2003}; \citealt{Carroll2005}). In this paper we specialise to $f(R)$ theories because the simulations available to us were cast in this form, but we emphasize that we expect our approach to be easily generalized to other theories. $f(R)$ models are derived from an action of the form
\begin{equation}
S=\int\,\mathrm{d}^4x\,\sqrt{|g|}\left[\frac{R+f(R)}{16\pi G}+\lagr_\mathrm{m}(\psi_i,g_{ab})\right]\ ,
\label{eq:fr_action}
\end{equation}
where $\psi_i$ indicates the matter fields, which follow geodesics of the metric $g_{ab}$.  Standard gravity is restored in the limit $f\rightarrow 0$ (or $-2\Lambda$). Minimizing the action with respect to the metric results in a modified field equation:
\begin{equation}
\eqalign{
R_{ab}&-\frac{1}{2}g_{ab}\left[R+f(R)\right]\cr&+(g_{ab}\dalem-\nabla_a\,\nabla_b+ R_{ab})f_R=-8\pi G T_{ab}\ ,
}
\label{eq:fr_field_equation}
\end{equation}
where 
\begin{equation}
f_R\equiv\frac{\mathrm{d}f}{\mathrm{d}R}\ ,
\end{equation}
$T_{ab}$ is the stress-energy tensor and $\dalem\equiv\nabla_a\nabla^a$. 



In this work we use the high curvature limit of the HS07 $f(R)$ function that is widely employed throughout the literature:
\begin{equation}
f(R)=-2\Lambda-\bar{R}_0\frac{f_{R0}}{n}\left(\frac{\bar{R}_0}{R}\right)^n\ ,
\label{eq:hu_sawicki_approx}
\end{equation}
where $f_{R0}$ and $n$ are the model parameters and $\bar{R}_0$ is the background value of $R$ measured today. Here $f(R)$ has the form of a (cosmological) constant plus a correction term. One should note that the mechanism for accelerated expansion ($-2\Lambda$) is entirely divorced from that which directly modifies gravitational forces (the inverse $R$ term). We work in the limit where $|f_{R0}|\ll 1$ (which covers values that are interesting observationally) so that the inverse $R$ term is negligible when considering the evolution of the background. 



It should be noted that any $f(R)$ theory can be mapped to a scalar-tensor theory (see Appendix~\ref{sec:bd_gravity}) and it is convenient to consider $f_R$ as an additional scalar degree of freedom whose value is locked to $R$ by the derivative condition. The equation of motion for $f_R$ can be derived by taking the trace of equation~(\ref{eq:fr_field_equation}):
\begin{equation}
\dalem f_R=\frac{1}{3}[R+2f(R)-Rf_R-8\pi G T]\ .
\label{eq:fr_trace_equation}
\end{equation}
In this way one can consider $f_R$ to evolve as a separate field, that is sourced by curvature.


For HS07, at the level of the homogeneous background, equation~(\ref{eq:fr_trace_equation}) simplifies to 
\begin{equation}
\bar{R}+4\Lambda=8\pi G \bar{\rho}_\mathrm{m}\ ,
\label{eq:background_R}
\end{equation}
where $\bar{R}$ and $\bar{\rho}_\mathrm{m}$ indicate background values of the curvature and matter density respectively. This can also be written as
\begin{equation}
\bar{R}(a)=3 H_0^2\left(\Om a^{-3}+4\Ov\right)\ ,
\label{eq:bar_R}
\end{equation}
where $a$ is the cosmic scale factor, $H_0$ is the current Hubble parameter, and $\Om$ and $\Ov$ are the dimensionless cosmological densities in matter and vacuum.

In the HS07 model $f_R$ is related to $R$ via:
\begin{equation}
f_R=f_{R0}\left(\frac{\bar{R}_0}{R}\right)^{n+1}\ ,
\label{eq:f_R_HS07}
\end{equation}
which at the background level implies
\begin{equation}
\bar{f}_R(a)=f_{R0}\left(\frac{1+4\Ov/\Om}{a^{-3}+4\Ov/\Om}\right)^{n+1}\ .
\label{eq:background_fr}
\end{equation}
If explicit time dependence is neglected (the quasi-static limit) in equation~(\ref{eq:fr_trace_equation}), and the homogeneous background subtracted, we arrive at the equation that governs the evolution of departures of $f_R$ from the background value:
\begin{equation}
\frac{1}{a^2}\nabla^2\delta f_R=\frac{1}{3}\delta R-\frac{8\pi G}{3}\bar{\rho}_\mathrm{m}\delta\ ,
\label{eq:quasistatic_fr}
\end{equation}
where, $\delta f_R\equiv f_R-\bar{f}_R$, $\delta R\equiv R-\bar{R}$, $\delta$ is the matter perturbation and the Laplacian is comoving. The right hand side of this equation can be considered an effective potential in which the $f_R$ field evolves. Equation~(\ref{eq:quasistatic_fr}) is only valid below the size of the current horizon but does \emph{not} assume that $|\delta f_R|$ is small in comparison with $|f_{R0}|$. \cite{Noller2014} has shown that the quasi-static approximation is valid for viable $f(R)$ models, even on some super-horizon scales, due to the slow-roll nature of the fields in viable screened models.


Non-relativistic particles in an HS07 model feel a modified acceleration compared to standard gravity counterparts, as extra forces arise due to gradients in the $f_R$ field. This can be seen most easily via the perturbed metric in flat space:
\begin{equation}
\mathrm{d}s^2=(1+2\Psi)\,\mathrm{d}t^2-a^2(t)(1-2\Phi)\,\mathrm{d}\mathbf{x}^2\ ,
\label{eq:weak_field_metric}
\end{equation}
from which equations for the time-gravitational potential $\Psi$ and space-gravitational potential $\Phi$ can be derived:
\begin{equation}
\frac{1}{a^2}\nabla^2\Psi=\frac{16\pi G}{3}\bar{\rho}_\mathrm{m}\delta-\frac{1}{6}\delta R\ ,
\label{eq:psi}
\end{equation}
\begin{equation}
\frac{1}{a^2}\nabla^2\Phi=\frac{8\pi G}{3}\bar{\rho}_\mathrm{m}\delta+\frac{1}{6}\delta R\ .
\label{eq:phi}
\end{equation}
Non-relativistic particles are accelerated by the time potential, $\mathbf{\ddot{x}}=-\nabla\Psi$, and are thus affected by the $\delta f_R$ field via its relation to $\delta R$ (equations~\ref{eq:f_R_HS07} and~\ref{eq:quasistatic_fr}). Since the value of $\delta f_R$ can change depending on environment, modifications to gravity that depend on environment are possible via the Poisson equation~(\ref{eq:psi}).

Photon trajectories (and thus lensing) are governed by the sum of space and time potentials:
\begin{equation}
\nabla^2(\Psi+\Phi)=8\pi G \bar{\rho}_\mathrm{m}a^2\delta\ ,
\end{equation}
a result that is unchanged compared to standard gravity for all $f(R)$ models. Lensing is therefore not \emph{directly} sensitive to the modification (as long as $|f_{R0}|\ll 1$), which means that dynamical mass and lensing mass estimates will be different for $f(R)$ models (\citealt{Schmidt2010}). Obviously lensing is still able to probe the enhanced clustering of matter in an $f(R)$ model relative to $\Lambda$CDM in the standard way.

\subsection{Linear perturbation theory}
\label{sec:perturbations}

\begin{table}
\centering
\caption{Simulations of standard gravity (GR) and HS07 (F4, F5, F6) models analysed in this paper. The cosmological parameters are $h = 0.697$, $\Om =0.281$, $\Ob = 0.046$, $\Ov = 0.719$, $n_\mathrm{s} = 0.971$ and $\sigma_8 = 0.82$. All $f(R)$ models have $n=1$ but differing values of $f_{R0}$ (see equation~\ref{eq:hu_sawicki_approx}). Simulations begin at $z_\mathrm{i}=49$ in a cube of side $512\Mpc$ from exactly the same initial conditions, which themselves are generated on a perfect initial Cartesian grid. It follows that $\sigma_8$ at $z=0$ will be different in each case due to the enhanced linear growth in the HS07 models (equation~\ref{eq:perturbations}), the true $\sigma_8$ is shown in the table for each model. Note that the F4 model has a very different $\sigma_8(z=0)$ from GR, despite having the same initial conditions, whereas the F6 model is very similar to GR. The Compton scale (equation~\ref{eq:compton}) at $z=0$ is also shown for each model, and indicates the approximate scale at which the modification is active.}
\begin{tabular}{c c c c c}
\hline
Simulation & $n$ & $f_{R0}$ & True $\sigma_8$ & $1/\lambda$ \\ [0.5ex] 
\hline
GR & -- & -- & 0.820 & --            \\
F6 & 1 & $-10^{-6}$ & 0.834 & $0.419\iMpc$ \\
F5 & 1 & $-10^{-5}$ & 0.875 & $0.133\iMpc$ \\
F4 & 1 & $-10^{-4}$ & 0.940 & $0.042\iMpc$ \\
\hline
\end{tabular}
\label{tab:simulations}
\end{table}

If $\delta f_R$ is small compared to the average background $\bar{f}_R$ at a particular epoch, it can be approximated as 
\begin{equation}
\delta f_R\simeq \left.\frac{\mathrm{d}f_R}{\mathrm{d}R}\right|_{\bar R}\delta R\equiv\frac{1}{3}\lambda^2\delta R\ ,
\end{equation}
where $\lambda$ is known as the (physical) Compton wavelength. In general this is defined as
\begin{equation}
\lambda^2=3\left.\frac{\mathrm{d^2}f(R)}{\mathrm{d}R^2}\right|_{\bar R}\ .
\end{equation}
In the specific case of the HS07 model,
\begin{equation}
\lambda^2=-3(n+1)\frac{f_{R0}}{\bar{R}_0}\left(\frac{\bar{R}_0}{\bar{R}}\right)^{n+2}\ .
\label{eq:compton}
\end{equation}
For the models studied as part of this work, the value of the Compton wavelength at $z=0$ is given in Table \ref{tab:simulations}. Note that larger $|f_{R0}|$ values mean the modification is felt to larger scales.

The resulting linear equation for $\Psi_k$ (in comoving Fourier Space) is
\begin{equation}
-\frac{k^2}{a^2}\Psi_k=4\pi G\bar{\rho}_\mathrm{m}\left[1+\frac{1}{3}\left(\frac{\lambda^2 k^2/a^2}{1+\lambda^2 k^2/a^2}\right)\right]\delta_k\ .
\label{eq:phi_perturbative}
\end{equation}
This is the potential that accelerates non-relativistic particles and so the growth of matter perturbations is scale dependent:
\begin{equation}
\ddot\delta_k+2H\dot\delta_k=\frac{3}{2}H^2\Omega_\mathrm{m}(a)\left[1+\frac{1}{3}\left(\frac{\lambda^2 k^2/a^2}{1+\lambda^2 k^2/a^2}\right)\right]\delta_k\ ,
\label{eq:perturbations}
\end{equation}
where the over-dots denote time derivatives. On large scales, $\lambda k/a \ll 1$, the term in square brackets is approximately equal to 1, and the perturbation equation is identical to that in standard gravity. But on scales smaller than the comoving Compton wavelength, $\lambda k/a \gg 1$, gravity is enhanced by a factor $4/3$. The only part of the linear theory calculation that depends on the specific form of $f(R)$ is how $\lambda$ relates to parameters in the specific choice of $f(R)$ function and therefore the maximum linear gravitational enhancement possible in \emph{any} $f(R)$ theory is a factor $4/3$ in the quasi-static limit. More general scalar-tensor theories can be made to give different linear enhancements to gravity (see Appendix~\ref{sec:bd_gravity}).



\subsection{The Chameleon Mechanism}
\label{sec:chameleon}

A remarkable feature of HS07 models is that they have the potential to screen the effect of the modification in some regions and this `chameleon screening' exhibits itself naturally, without it having to be introduced by hand. Screening was first discussed for scalar-tensor models by \cite{Khoury2004} and allows stringent tests of gravity within the Solar system to remain satisfied (see \eg \citealt{Will2006}), while modifying gravity on larger scales. It was shown that $f(R)$ models can exhibit the chameleon mechanism in HS07 and \cite{Brax2008}.

The Solar system is far removed from the perturbative regime, so one needs to explore exactly how gravity in an $f(R)$ model behaves in dense environments in order to say what deviations from standard gravity are predicted within the Solar system. Behaviour is governed by the quasi-static Poisson equations for $\delta f_R$ and $\Psi$, given in equations~(\ref{eq:quasistatic_fr}) and~(\ref{eq:psi}). If a region of space exists where $\nabla^2 \delta f_R=0$, (\ie minima of the effective potential), then
\begin{equation}
\frac{1}{3}\delta R=\frac{8\pi G}{3}\bar{\rho}_\mathrm{m}\delta\ ,
\end{equation}
and gravitational forces ($\nabla\Psi$) are restored to the standard. This is the regime of screening and it then remains to discover for a given model in which environments the screening condition is satisfied. The combined equations~(\ref{eq:quasistatic_fr}) and~(\ref{eq:psi}) must be solved for a given density field from the external field value all the way into the internal structure of the density distribution in question. This can either be solved in a cosmological context by simulations (\eg \citealt{Oyaizu2008a}; \citealt{Li2012a}; \citealt{Puchwein2013}; \citealt{Llinares2014}) or by direct calculations in idealized situations with symmetry properties (\eg HS07; \citealt{Schmidt2010}; \citealt{Lombriser2012c}). The result of calculations and simulations is that the modification to gravity is able to be screened in \emph{some} environments, depending on model parameter values. For $n=1$ models the transition of the field from the cosmological regime into the Solar system can be used to place limits of $|f_{R0}|\ltsim 10^{-6}$ (HS07). Alternatively, limits can be placed by looking at samples of similar objects in screened and unscreened environments (\eg dwarf galaxies -- \citealt{Jain2013}; \citealt{Vikram2013}) and constraints of $|f_{R0}|\ltsim 10^{-7}$ are obtained. Independent constraints can be placed from large-scale structure measurements -- particularly from the abundance of clusters, which increases in HS07 models due to the enhanced gravity for set initial conditions (\ie the same primordial CMB). Constraints from clusters yield $|f_{R0}|\ltsim 10^{-4}$ (\citealt{Schmidt2009}; \citealt{Ferraro2011}; \citealt{Lombriser2012a}; \citealt{Lombriser2012b}). As this paper was nearing completion \cite{Cataneo2014} reported constraints of $|f_{R0}|\ltsim 10^{-5}$ from cluster abundance. \cite{Terukina2014} use the difference between hydrostatic and lensing masses in HS07 models to infer constraints of $|f_{R0}|< 6\times 10^{-5}$. Cosmologically \cite{Dossett2014} used redshift space distortions in the WiggleZ survey to place limits of $|f_{R0}|\ltsim 10^{-5}$.

We note that it is theoretically feasible that the $f_R$ field couples only to dark matter (if the HS07 model is thought of in terms of a scalar field with non-universal couplings in the Einstein frame), and that this would invalidate Solar system and Galactic constraints on HS07 parameters, meaning that the model may \emph{only} be constrained on cluster or cosmological scales. If Solar system and baryonic constrains are excluded then a conservative bound on current limits is $|f_{R0}|\ltsim 10^{-5}$, whereas if they are not excluded this limit is more like $|f_{R0}|\ltsim 10^{-7}$. Note that all constraints quoted are $2\sigma$ for $n=1$ HS07 models; constraints placed on $|f_{R0}|$ that use data over a redshift range are degraded slightly for models with $n>1$ because these models transition to mimic $\Lambda$CDM more quickly in the recent past.

Although one would expect the enhanced gravity to change the halo density profile it has been shown (\eg \citealt{Lombriser2012c}) that HS07 haloes can be well described by the halo profile of \citeauthor*{Navarro1997} (\citeyear{Navarro1997}; NFW) as well as in standard gravity. The NFW profile is 
\begin{equation}
\rho(r)=\frac{\rho_\mathrm{N}}{(r/r_\mathrm{s})(1+r/r_\mathrm{s})^2}\ ,
\label{eq:nfw}
\end{equation}
which is truncated at the halo virial radius $r_\mathrm{v}$, $\rho_\mathrm{N}$ is a normalization which is set by the halo mass and $r_\mathrm{s}$ is a scale radius that is related to the virial radius via the concentration parameter $r_\mathrm{v}=c r_\mathrm{s}$. The fraction of halo mass enclosed at radius $r$ for an NFW profile is
\begin{equation}
M(r)=M\frac{F(r/r_\mathrm{s})}{F(c)}\ ,
\label{eq:nfw_enclosed_mass}
\end{equation}
where $F(x)=\ln(1+x)-x/(1+x)$. While $r<r_\mathrm{v}$, the Newtonian potential felt by a test particle as a function of radius from the centre of the potential is
\begin{equation}
\Psi_\mathrm{N}=-\frac{GM}{r}\frac{1}{F(c)}\left[\ln(1+r/r_\mathrm{s})-\frac{r/r_\mathrm{s}}{1+c}\right]\ .
\label{eq:nfw_potential}
\end{equation}

\begin{figure}
\begin{center}
\includegraphics[width=60mm,trim=0cm 1.cm .5cm 1cm,angle=270]{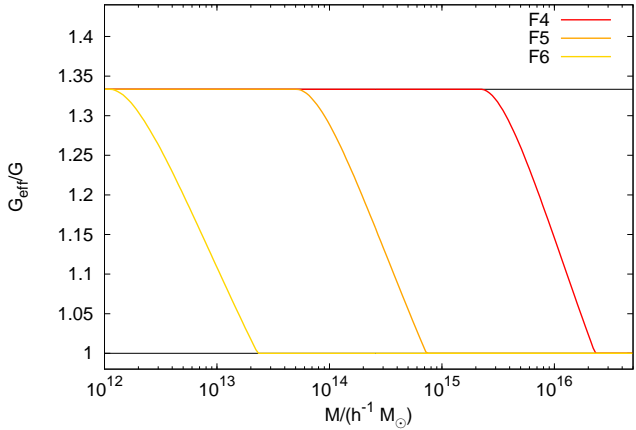}
\end{center}
\caption{The effective gravitational constant felt in haloes as a function of halo mass for cosmologies that we simulate (see Section~\ref{sec:simulations}) according to the simple model in equations~(\ref{eq:full_screening}) and ~(\ref{eq:screening}). Only the highest mass haloes (above $\sim\sform{8}{15}\Msun$) are screened in the F4 model, which corresponds to a tiny fraction of haloes at $z=0$. Whereas the screening mass is $\sim\sform{3}{14}\Msun$ in the F5 case and $\sim\sform{5}{12}\Msun$ in the F6 case. There is a broad transition between screened and unscreened haloes that takes place over approximately a decade in halo mass for each model. This toy calculation agrees well with measurements of screening in simulations.}
\label{fig:geff}
\end{figure}

A simple model for screening is that a region of the Universe can be considered to be screened when 
\begin{equation}
\bar{f}_{R}(a)\ltsim \frac{2}{3} \Psi_\mathrm{N}\ ,
\end{equation}
essentially the $f_R$ field is forced into the minimum of the effective potential when the local gravitational potential is of the order of the background $f_R$ value (\citealt{Schmidt2010}). Using equation~(\ref{eq:background_fr}) and the NFW potential in equation~(\ref{eq:nfw_potential}) results in a chameleon screening radius $r_\mathrm{c}$ as a function of $M$, $f_{R0}$ and $n$:
\begin{equation}
\eqalign{
f_{R0}\left(\frac{1+4\Ov/\Om}{a^{-3}+4\Ov/\Om}\right)^{n+1}&=\cr
-\frac{2GM}{3r_\mathrm{c}}\frac{1}{F(c)}&\left[\ln(1+r_\mathrm{c}/r_\mathrm{s})-\frac{r_\mathrm{c}/r_\mathrm{s}}{1+c}\right]\ ,
}
\label{eq:full_screening}
\end{equation}
which can be solved numerically to find $r_\mathrm{c}$. The effective gravitational `constant', felt by particles, in a halo of a given mass can then be estimated via the fraction of the mass of the halo that is screened
\begin{equation}
\frac{G_\mathrm{eff}}{G}=1+\frac{1}{3}\frac{M-M(r_\mathrm{c})}{M}\ .
\label{eq:screening}
\end{equation}
Note that this simple model ignores any environmental dependence of the screening mechanism, although this could be included if required. A result of this calculation of $G_\mathrm{eff}$ for models that we later simulate (see Section~\ref{sec:simulations}) is shown in Fig.~\ref{fig:geff}, where it can be seen that there is quite a broad transition that takes place over approximately a decade in halo mass between low mass haloes, that feel enhanced gravity, to screened high mass haloes. This toy calculation agrees well with results of full numerical calculations of screening in idealized symmetric haloes and $N$-body simulations (see fig. 3 of \citealt{Schmidt2010}, although note that our result differs from the theoretical model shown in that work because we truncate our NFW profiles and those of Schmidt are untruncated; truncating the potential seems to improve the match to data). Note that the Milky Way lies in the transition region for screening in the F6 model, and this is what drives the Solar system based $f_{R0}$ constraints.

\section{Simulations}
\label{sec:simulations}

An $N$-body simulation must calculate the gravitational forces on all particles and evolve their positions over time. This is complicated in MG models, even those with a standard background expansion, because it is additionally necessary to solve a scalar field equation. Recently $N$-body codes have been developed to carry out this calculation: initially particle-mesh methods (\citealt{Oyaizu2008a}; \citealt{Oyaizu2008b}) and more recently adaptive mesh techniques (\textsc{mlapm} -- \citealt{Zhao2011}; \textsc{ecosmog} -- \citealt{Li2012a}; \textsc{isis} -- \citealt{Llinares2014}) and tree codes (\textsc{mg-gadget} -- \citealt{Puchwein2013}).

In this paper we use simulations run using the \textsc{ecosmog} code of \cite{Li2012a}, which is based on the $N$-body code \textsc{ramses} (\citealt{Teyssier2002}). This uses adaptive meshes to solve the coupled $\Psi$ and $f_R$ Poisson equations. For our simulations \textsc{ecosmog} was run in the limit that the background expansion is \emph{exactly} $\Lambda$CDM -- so the modification due to gravity is \emph{only} present via the $\delta f_R$ field in equation~(\ref{eq:quasistatic_fr}) which impacts on particle accelerations via equation~(\ref{eq:psi}). This approximation is useful so as to be able to disentangle effects due to modified gravitational forces from those caused by a non-standard background expansion. This $\Lambda$CDM background approximation covers $f_{R0}$ values that are interesting observationally but would be incorrect if the limit $|f_{R0}|\ll 1$ ceased to be true and the second term in equation~(\ref{eq:hu_sawicki_approx}) became important for the background. 

This paper analyses data from simulations of standard gravity (GR) and HS07 models (F4, F5, F6) that all start from exactly the same initial conditions (including seed) with $512^3$ particles in a box with $L=512\Mpc$; these are summarized in Table \ref{tab:simulations}\footnote{Additionally CMB temperature $T_\mathrm{CMB} = 2.7255$ K, effective number of neutrinos $n_\mathrm{eff}=3$, neutrino mass $m_\nu=0$ and Helium mass fraction $Y_\mathrm{He} = 0.24$.}. An initial power spectrum for the simulations was generated using \textsc{mpgrafic} (\citealt{MPgrafic}). The particle mass in each case is $\simeq\sform{7.80}{10}\Msun$. Each simulation has exactly the same power spectrum at $z_\mathrm{i}=49$ and the same background cosmological parameters and therefore identical background expansion rates. Observers in each case would see exactly the same CMB sky with the exception of foreground contributions such as the integrated Sachs-Wolfe effect. Differences between models are confined to different strengths of enhanced perturbation growth at late times and different strengths of screening.

\subsection{Haloes}

In this paper we analyse halo catalogues that are generated from the simulated particle data. These were generated with the public Friends-Of-Friends (FOF) code, available at http://www-hpcc.astro.washington.edu/tools/fof.html, using a linking length of $b=0.2$ times the mean inter-particle separation. No attempt was made to reject unbound particles. To create halo mass functions we simply bin the haloes in logarithmically spaced bins in mass and assign $M$ to each bin as the logarithmic mid-point. We then convert this to the mass fraction in the simulation contained in that mass bin, normalized by the bin width:
\begin{equation}
\mathrm{d}F=\frac{M}{\bar\rho_\mathrm{m}} n(M)\,\mathrm{d}M\ .
\label{eq:mass_fraction} 
\end{equation}
We plot multiplicity functions, which are given by $M\,\mathrm{d}F$ and correspond to the mass fraction in haloes per $\ln M$.

\subsection{Variance and power}

In this paper we investigate structure formation in terms of matter and halo clustering and statistics. Halo formation and the halo mass function are intimately related to the variance in the linear power spectrum as a function of scale (\citealt{Press1974}; \citealt{Sheth1999}; \citealt*{Sheth2001}). This is defined for comoving scale $R$ (not to be confused with the Ricci scalar) as 
\begin{equation}
\sigma^2(R,z) =\int g^2(k,z)\Delta_\mathrm{lin}^2(k,0)W^2(kR)\,\mathrm{d}\ln{k}\ ,
\label{eq:sigma}
\end{equation}
where $W$ is the normalized Fourier transform of the spherical top hat filter function:
\begin{equation}
W(x)=\frac{3}{x^3}\left(\sin{x}-x\cos{x}\right)\ ;
\label{eq:tophat}
\end{equation}
$\Delta^2_\mathrm{lin}$ is the dimensionless linear matter power spectrum
\begin{equation}
\Delta^2(k)=\frac{4\pi V k^3}{(2\pi)^3}P(k)\ ;
\label{eq:delta_def}
\end{equation}
$P(k)=\average{|\delta_\mathbf{k}|^2}$ and $g(k,z)$ is the growth function, which are the growing solutions to equation~(\ref{eq:perturbations}) normalized such that $g(k,z=0)=1$. This means that $\sigma_8$ ($z=0$, $R=8\Mpc$) will be larger in the MG models due to the enhanced linear growth at small scales, despite the identical initial conditions. The true $\sigma_8(z=0)$ for the modified models is given in Table \ref{tab:simulations}. 

\subsection{Measuring power spectra}

In this paper we measure power in simulated particle distributions in both real and redshift space. To do this we assign particles to a Cartesian mesh via Cloud-In-Cell (CIC: \citealt{Hockney1988}) interpolation to create the density field, and compute the Fourier transform. We then deconvolve the density field in Fourier Space to account for the CIC mesh assignment (\eg \citealt{Jing2005}) and then bin modes in equally logarithmically spaced $|\mathbf{k}|$ bins between the fundamental box mode and half the mesh Nyquist frequency. $P(k)$ is created in bins by averaging $|\delta_\mathbf{k}|^2$ over all modes that fall into each bin. The $k$ assigned to the bin is simply the logarithmic mid point between the upper and lower boundary of the bin. Finally we multiply by the suitable factors to create $\Delta^2(k)$ (equation~\ref{eq:delta_def}). 

To analyse redshift-space effects we use the methods discussed in detail in MP14b. We first move particles to their redshift space positions under the distant-observer approximation along an arbitrarily chosen line-of-sight, and then compute anisotropic power as a function of $|\mathbf{k}|$ and $\mu=\cos\theta$ where $\theta$ is the angle of the mode to the line-of-sight. To compute monopole and quadrupole moments of this distribution we fit a model `monopole $+$ quadrupole' to $\Delta^2(k,\mu)$. This procedure is necessary in order to avoid biases induced at large scales by the Cartesian density field mesh, where there are only a few values of $\mu$ per $k$ mode. For particles we subtract shot noise after computing the power but we do not do this for haloes, as the discrete haloes \emph{are} the density field in that case, rather than tracers of it.

\begin{figure}
\begin{center}
\includegraphics[width=60mm,trim=0cm 1cm .5cm .5cm,angle=270]{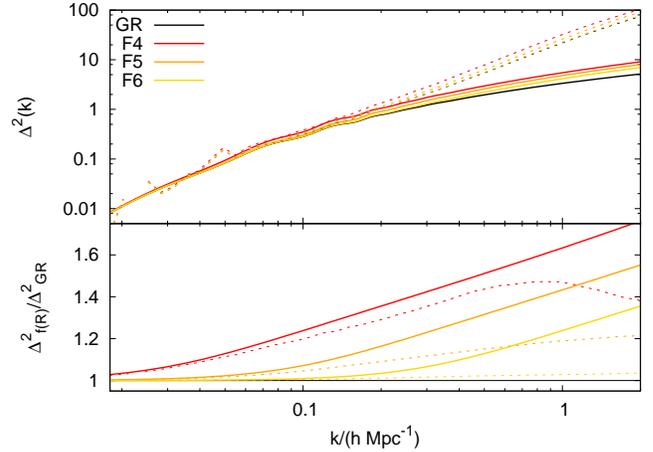}
\end{center}
\caption{The $z=0$ linear theory matter power spectrum (solid lines) together with the measured non-linear power (dashed lines) for each of the models in Table \ref{tab:simulations}. The ratio of power for each MG model compared to the GR model are shown in the lower panel for both the linear and measured non-linear power. In all cases an enhancement in power at small scales can be seen, with this being most pronounced in the F4 case. At large scales all the models agree (almost) exactly because the growth factor is equal in all models at large scales (equation~\ref{eq:perturbations}). One can see that the relative linear enhancement of power in each HS07 model is diminished in the full non-linear simulation. This is seen at its most extreme in the F6 case where the full non-linear spectrum only deviates from GR at the $\simeq 2$ per cent level at $k=1\iMpc$ compared to the $\simeq 20$ per cent deviations seen in the linear spectrum.}
\label{fig:power}
\end{figure}

The theoretical linear matter power spectrum for each model is shown in Fig.~\ref{fig:power} together with the measured non-linear spectrum measured in each simulation at $z=0$. Note that the non-linear enhancement in power is \emph{less} strong than the linear enhancement. \new{This is partly due to the chameleon effect, but also partly due to the different non-linear velocity fields, as the same suppression of power relative to the linear prediction can be found in simulations with \emph{no} screening mechanism (\eg linearised HS07 models -- \citealt{Li2013})}. The results in Fig.~\ref{fig:power} agree well with similar results for simulated matter power in \cite{Li2012a} amongst others.

\section{Rescaling}
\label{sec:rescaling}

In this section the rescaling algorithm, developed in AW10, MP14a and MP14b, is modified so that it may be applied to MG theories. In doing so an attempt is made to keep the algorithm as general as possible, however, tests are restricted to HS07 models at this stage. Results are presented along with each stage of the algorithm so as to provide a worked example. We recapitulate the essentials of rescaling here for convenience but direct the reader to AW10 and MP14a,b for a more in depth discussion. 

The AW10 algorithm works by mapping an `original' simulation at redshift $z$ in a box of size $L$ to a `target' cosmology at redshift $z'$ in a box of size $L'=sL$. In this work quantities in the original simulation are unprimed whereas quantities in the target simulation are primed. In order to conserve mass the scaling in length units simultaneously implies a scaling in mass:
\begin{equation}
M'= s^3\frac{\Om'}{\Om} M\equiv s_\mathrm{m} M\ .
\label{eq:mass_scaling}
\end{equation}
Note that we use units of $\Mpc$ for length and $\Msun$ for mass and the necessary factors of $h$ are included in the scalings.

\begin{figure}
\begin{center}
\includegraphics[width=60mm,trim=0cm 1.cm .5cm 1cm,angle=270]{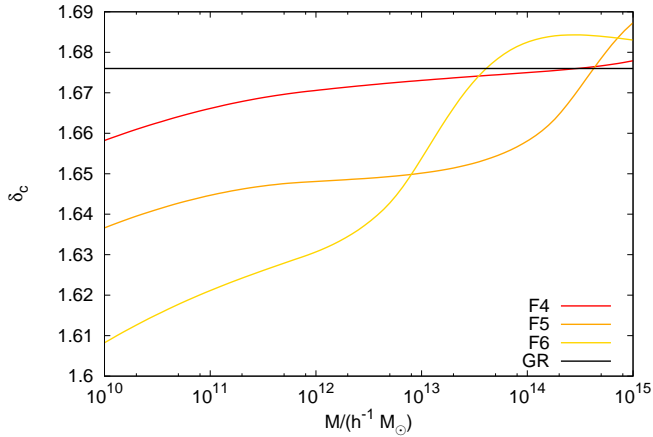}
\end{center}
\caption{The function $\delta_\mathrm{c}(M)$ at $z=0$ for the models discussed in this work, the calculation uses a mean halo environment of $\delta_\mathrm{env}\simeq 0.43$. $\delta_\mathrm{c}$ is defined such that when divided by $\sigma(R)$ -- calculated using the enhanced scale-dependent growth function -- it gives $\nu$ that enters the mass function. The flat black line is the GR prediction of $\delta_\mathrm{c}\simeq 1.676$ for the $\Lambda$CDM cosmology in question. Haloes are screened in all models with this screening being most pronounced in the F6 case where the largest deviations are observed at low masses. That the curves rise above the GR line at high masses is due to them being scaled with the linear version of $\sigma$, which does not contain any information about screening, they asymptote to $\delta_\mathrm{c}\simeq 1.692$.}
\label{fig:mg_dc}
\end{figure}

The original AW10 procedure obtains $s$ and $z$ by minimizing the difference in $\sigma(R)$ between the two models across a range of scales. This is done because, for standard gravity, the halo mass function is approximately universal when expressed in terms of the variable $\nu=\delta_\mathrm{c}/\sigma(R)$ where $\delta_\mathrm{c}\simeq 1.686$ is the value of the linear density field at which collapse occurs (\eg \citealt{Press1974}; \citealt{Sheth1999}; \citealt{Sheth2001}). Given the aim of matching the mass function one might expect that better results could be achieved by minimizing the difference in $\nu=\delta_\mathrm{c}(M)/\sigma(M)$, where $\delta_\mathrm{c}(M)$ can be calculated taking the gravitational modifications into account (see Fig.~\ref{fig:mg_dc}). However, \cite{Schmidt2010} showed that the \cite{Sheth2001} mass function works well in HS07 models if one computes $\sigma(R)$ using the linear power spectrum with the correct scale-dependent growth (equation~\ref{eq:sigma}), even though this ignores the chameleon mechanism. The potential environmental dependence of the mass function is not addressed here, and we turn the attention of the reader to \cite{Li2012b} and \cite{Lombriser2013a} for a more in-depth discussion of this.


Spherical collapse models in $f(R)$ predict that the collapse threshold for halo formation should vary as a function of halo mass. Fig.~\ref{fig:mg_dc} shows a result of a full spherical model calculation that includes screening.  We take the haloes to reside in an \emph{average} environment with $\delta_\mathrm{env}\simeq 0.43$ (defined with a filtering scale of $5\Mpc$) which is calculated using the extended excursion set methods of \cite{Lam2012}. The $\delta_\mathrm{c}$ calculation is similar to that in \cite{Lombriser2013a} where $\delta_\mathrm{c}$ is extrapolated to $z=0$ using the $\Lambda$CDM growth function. To be consistent with this, the collapse threshold $\nu$ must be calculated with $\sigma(R)$ calculated in $\Lambda$CDM. Therefore in Fig. \ref{fig:mg_dc} we show $\delta_\mathrm{c}$ multiplied by the ratio of $\sigma(R)$ in the HS07 model to that in an equivalent $\Lambda$CDM model. Thus the $\delta_\mathrm{c}$ shown is exactly $\nu$ when divided by $\sigma(R)$ calculated using the enhanced scale-dependent growth function.


\begin{figure*}
\begin{center}
\makebox[\textwidth][c]{
\subfloat{\includegraphics[width=60mm,trim=1.5cm 1cm 0cm 1cm,angle=270]{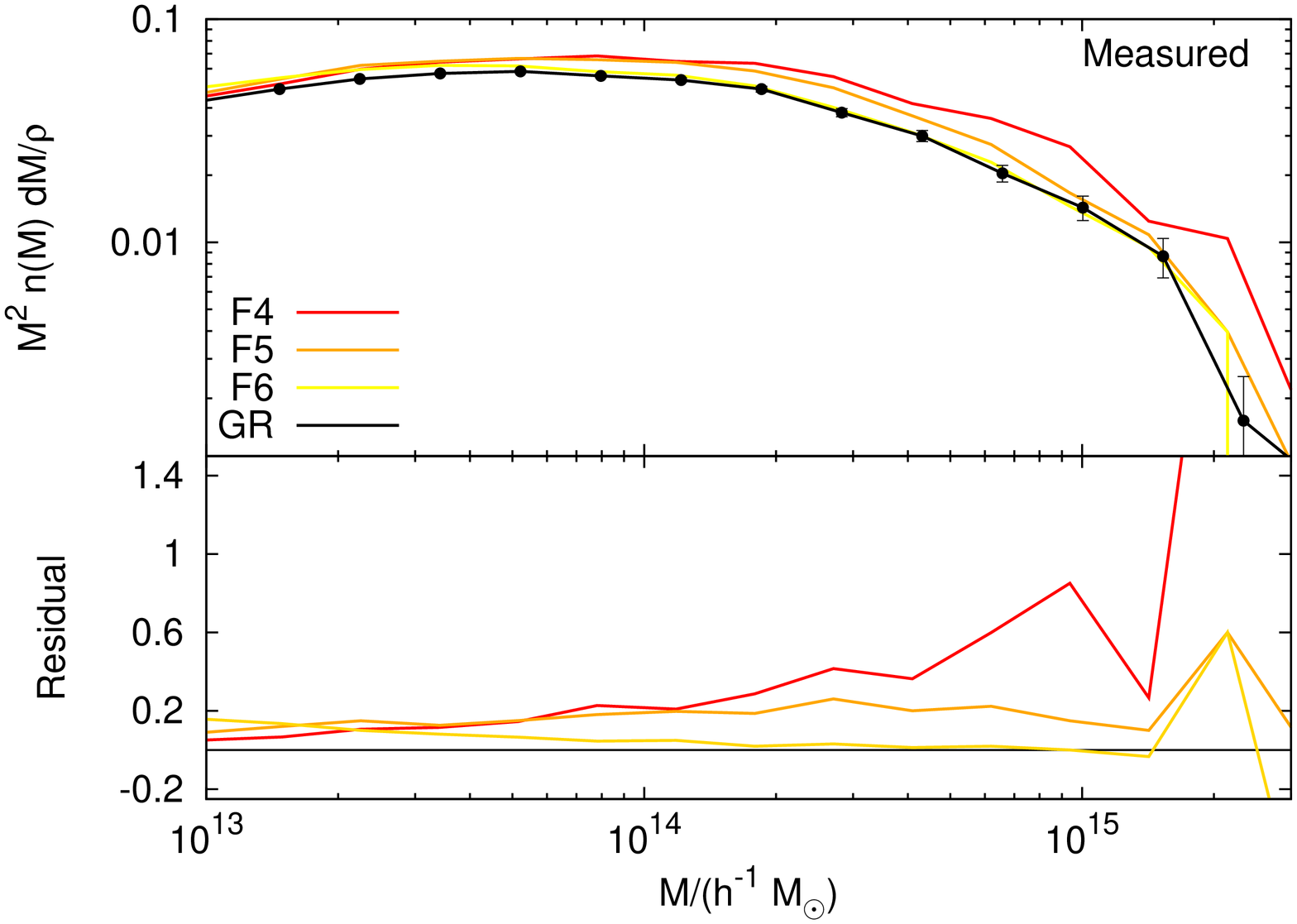}}
\subfloat{\includegraphics[width=60mm,trim=1.5cm 1cm 0cm 1cm,angle=270]{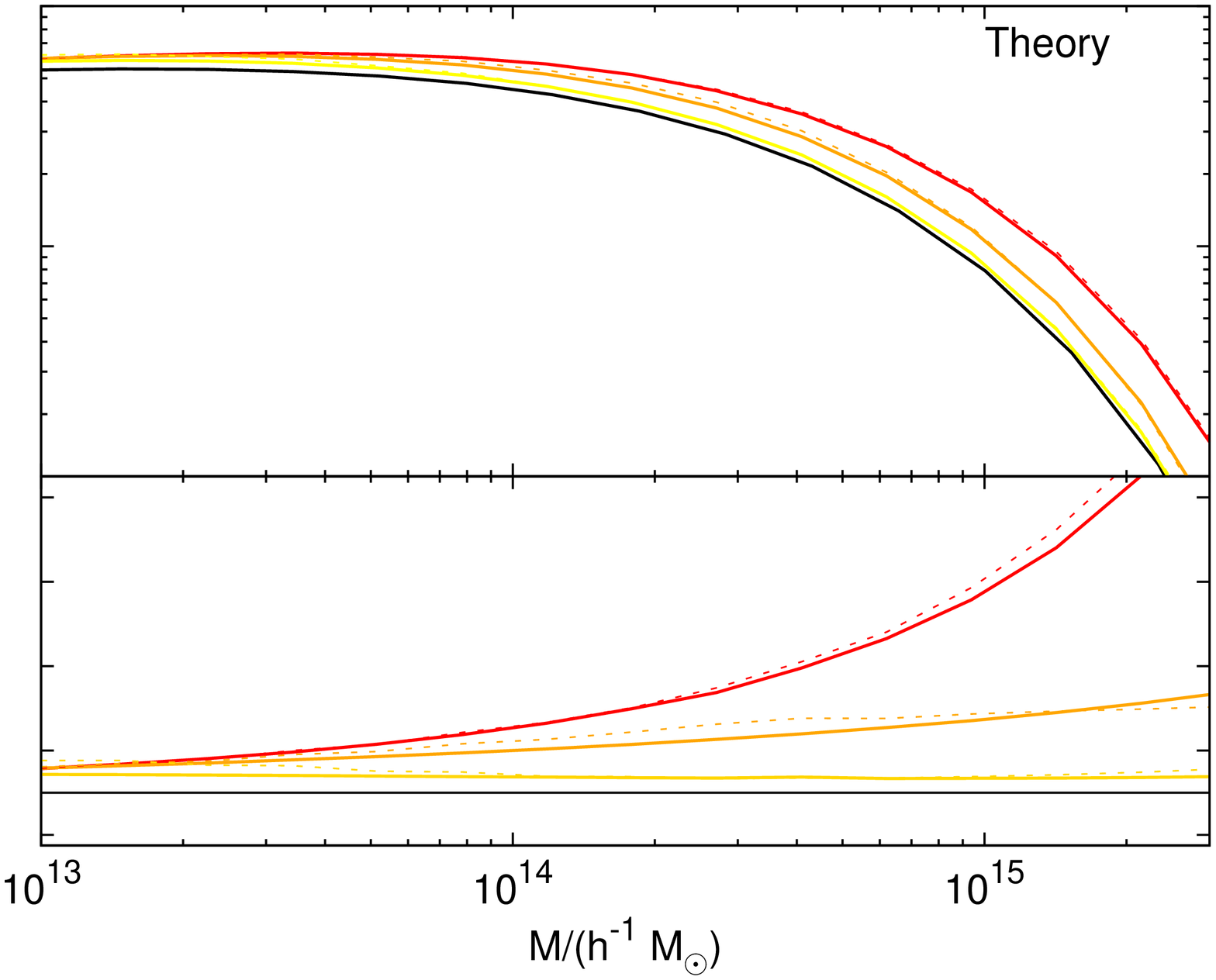}}}
\end{center}
\caption{The halo mass function measured at $z=0$ in the simulations listed in Table \ref{tab:simulations} and fractional residuals for the MG models compared to GR. The left panel shows the measured multiplicity functions, with Poisson errors due to finite halo number shown only on the GR measurement for reference. The right panel shows theoretical predictions; the solid lines being ST using $\delta_\mathrm{c}(M)$ from Fig.~\ref{fig:mg_dc} while the dashed lines shows the same mass function with fixed $\delta_\mathrm{c}=1.686$. In both cases $\sigma(R)$ has been calculated using the modified growth functions for the HS07 models. Because the mass function predictions are similar we use the simpler approach in this paper and ignore possible $\delta_\mathrm{c}$ variations.}
\label{fig:mfs}
\end{figure*}

To test theoretical predictions for the mass function, Fig.~\ref{fig:mfs} shows measurements from simulations together with predictions from the \citeauthor{Sheth1999} (\citeyear{Sheth1999}; ST) fitting formula; 
\begin{equation}
f(\nu)=A\left[1+\frac{1}{(q\nu^2)^p}\right]\mathrm{e}^{-q\nu^2/2}\ ,
\label{eq:STmf}
\end{equation}
where $A=0.216$, $q=0.707$ and $p=0.3$. $f(\nu)$ is defined such that it gives the fraction of the mass in the Universe in haloes in a range $\nu$ to $\nu+\mathrm{d}\nu$ such that $\mathrm{d}F=f(\nu)\,\mathrm{d}\nu$ with $\mathrm{d}F$ defined in equation~(\ref{eq:mass_fraction}). Fig.~\ref{fig:mfs} shows the cases of $\delta_\mathrm{c}=1.686$ fixed and $\delta_\mathrm{c}$ varied as a function of mass as per Fig.~\ref{fig:mg_dc}. In both cases $\sigma(R)$ is calculated using the scale-dependent growth of perturbations. Across the range of mass shown, which corresponds to the masses probed by our simulations, there is very little difference in using either prescription for the mass function. This relates to the fact that $\delta_\mathrm{c}$ only differs from the $\Lambda$CDM result by $\sim 1$ per cent for the range of masses shown (Fig.~\ref{fig:mg_dc}). In \cite{Schmidt2009} it was shown that for the maximum gravity enhancement of a factor $4/3$ $\delta_\mathrm{c}, \simeq 1.692$ in HS07 models and this is the value to which the curves asymptote in Fig.~\ref{fig:mg_dc}.

\new{In \cite{Angulo2015} a slightly updated version of the AW10 method was presented, in which $s$ and $z$ were chosen not only to provide a match to the mass function, but also such that the original and target cosmologies had closely matched growth histories (\ie ensuring $g(z)$ is matched across a range of $z$). The logic being that growth history is what determines halo concentrations (\eg \citealt{Bullock2001}) and so that halo structure should be in better agreement before and after rescaling if this additional constraint is imposed. \cite{Angulo2015} showed that power-spectrum matches at small scales were improved if this update was applied. However, in this paper we do not attempt to apply this because it is not obvious how to implement it given the scale-dependent growth of perturbations in HS07 models. \ie at which scale of $g(k,z)$ should we attempt to match to the growth history in the original cosmology?}


In light of the above discussion, we choose rescaling parameters $s$ and $z$ exactly as in AW10, by minimizing the difference in $\sigma(R)$ rather than $\nu(R)$. Partly this is because $\delta_\mathrm{c}$ variations are small, and certainly smaller than the error in the ST theoretical mass function in any case. A sensible way to choose scaling parameters $s$ and $z$ is to minimize the cost function
\begin{equation}
\delta_\mathrm{rms}^2(s,z\mid z')=\frac{1}{\ln(R'_2/R'_1)}\int_{R'_1}^{R'_2}
\frac{\mathrm{d}R}{R}\left[1-\frac{\sigma(R/s,z)}{\sigma'(R,z')}\right]^2\ ,
\label{eq:minimise}
\end{equation}
over $s$ and $z$, with $z'$ fixed by the desired target redshift ($0$ in our case). $R'_1$ and $R'_2$ are chosen so as to relate to the physical scale of the least and most massive haloes in the \emph{target} simulation via
\begin{equation}
M=\frac{4}{3}\pi R^3 \bar{\rho}\ ,
\label{eq:mass_to_R}
\end{equation}
with $R'=sR$.

\begin{table}
\centering
\caption{Best fitting AW10 scaling parameters between the high $\sigma_8$ $\Lambda$CDM cosmology with parameters: $h = 0.7$, $\Om =0.3$, $\Ob = 0.045$, $\Ov = 0.7$, $n_\mathrm{s} = 0.97$, $\sigma_8 = 1.2$ and the target GR, F4, F5 and F6 models at $z'=0$. $L$ gives the box size of the parent $\Lambda$CDM simulation required for rescaling to each model so as to be able to analyse the scaling without the complication of cosmic variance. $k_\mathrm{nl}$ is the non-linear scale for each model, defined by $\sigma(R=1/k_\mathrm{nl})=1$.}  
\begin{tabular}{c c c c c}
\hline 
Target & $s$ & $L$ & $z$ & $k_\mathrm{nl}$ \\ [0.5ex] 
\hline
GR & 1.063 & $482\Mpc$ & 0.844 & $0.170\iMpc$ \\
F6 & 0.956 & $536\Mpc$ & 0.644 & $0.164\iMpc$ \\
F5 & 0.850 & $602\Mpc$ & 0.381 & $0.151\iMpc$ \\
F4 & 0.838 & $611\Mpc$ & 0.225 & $0.136\iMpc$ \\
\hline
\end{tabular}
\label{tab:rescaling_parameters}
\end{table}


In order to test the rescaling method, we then ran a tailored `parent' $\Lambda$CDM simulation from which the existing GR and $f(R)$ models could be obtained via scaling. Note that from now on $\Lambda$CDM is the `original' model whereas GR is one of the target models. For each model listed in Table \ref{tab:simulations} we found best fitting $s$ and $z$ values by minimizing equation~(\ref{eq:minimise}) and then ran the parent simulation for each model with a box side given by $L'/s$ where $L'=512\Mpc$. Each parent simulation was run with exactly the same random seed for the initial conditions as its child and this enables comparisons to be made without the added complications of cosmic variance, but necessitates the running of the parent simulations in tailored box sizes. The $\Lambda$CDM cosmology and associated scaling parameters are given in Table \ref{tab:rescaling_parameters}. The cosmology was chosen to have a high value of $\sigma_8=1.2$ in order that it explored a large range of fluctuations during its evolution. This is necessary in order to permit scaling to models with higher values of $\sigma(R)$ (\eg \citealt{Harker2007}; AW10; \citealt{Ruiz2011}; MP14a), which is particularly true of the $f(R)$ models in this work.

\begin{figure}
\begin{center}
\includegraphics[width=60mm,trim=0cm 1cm .5cm .5cm,angle=270]{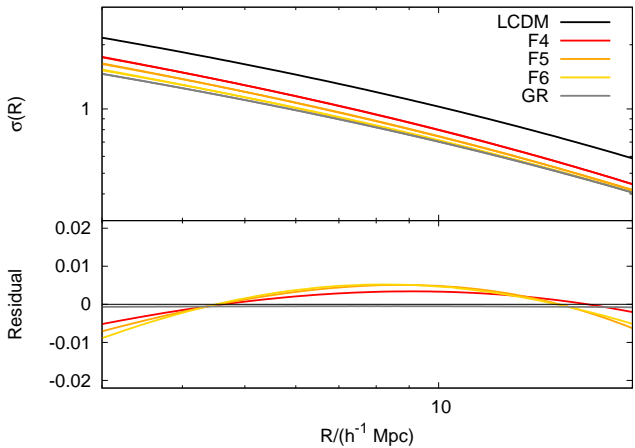}
\end{center}
\caption{The theoretical $\sigma(R)$ functions for the $\Lambda$CDM cosmology after scaling in size and redshift by values given in Table \ref{tab:rescaling_parameters} compared to the target MG and GR cosmologies. In the upper panel $\sigma(R)$ is shown for the unscaled $\Lambda$CDM model (black curve) along with that for each target and rescaled model, but differences cannot be distinguished and so fractional residual differences are shown in the lower panel. The match is good to within $1$ per cent for all models across the range of scales shown, which correspond to the mass range probed by the simulations. The match to the GR simulation is at the level of $0.1$ per cent across the range and is hard to see even in the lower panel.}
\label{fig:sigma_residuals}
\end{figure}T

The rescaled theoretical form of $\sigma(R,z)$ for all cosmologies discussed is shown in Fig.~\ref{fig:sigma_residuals} where it can be seen that the match is good to $1$ per cent across the full range of scales relevant to halo masses in the target cosmologies. The error in this is far smaller than the known non-universality in the mass function (\eg \citealt{Warren2006}, \citealt{Tinker2008}). The match in the GR case is hard to see but is within $0.1$ per cent across all $R$ shown.

\begin{figure}
\includegraphics[width=105mm,trim=0.cm 0cm 1.cm 0cm,angle=270]{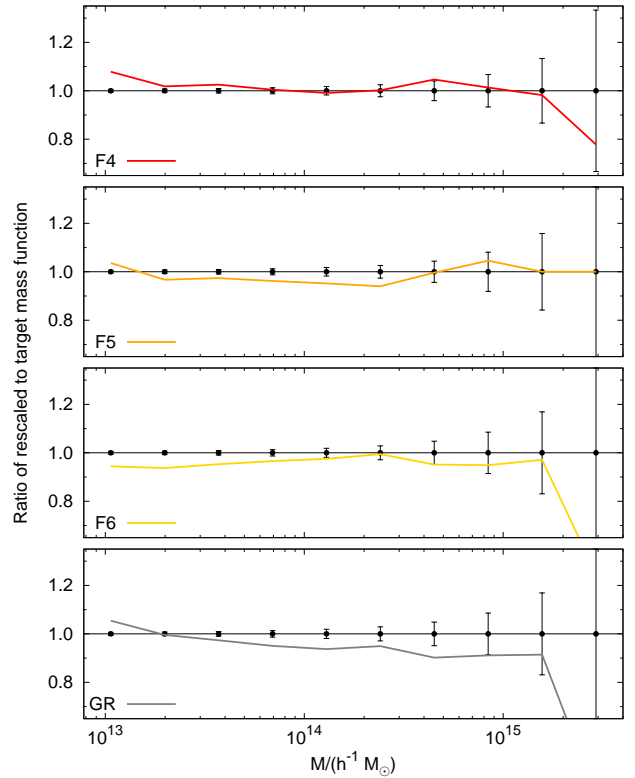}
\caption{The residual error in mass function, in 10 log-spaced bins, of the rescaled $\Lambda$CDM simulations to target F4 (top), F5, F6 and GR (bottom) models. Error bars shown are Poissonian and due to the finite numbers of haloes in each bin. The mass functions are matched well (mainly at the $5$ per cent level) across the entire range for each model. Deviations at high $M$ are of low significance owing to higher mass bins containing few haloes ($\sim 5$ for the highest mass bin). Surprisingly the match to the mass function shown here is \emph{better} than seen in previous tests of the AW10 method when scaling between more standard cosmologies, the worst match is to the GR simulation.}
\label{fig:scaled_mf}
\end{figure}

The ratio of rescaled to target halo mass functions are shown in Fig.~\ref{fig:scaled_mf} after both the size and redshift scaling have been applied. It can be seen that the mass function is matched at around the $5$ per cent level across the range of halo mass probed by the simulation. In fact the match shown here is actually better than in AW10 (WMAP1 to WMAP3 scaling) or in the scaling from $\Om=1$ case analysed in MP14a and MP14b; this supports the conclusion that the overall HS07 mass function can be well modelled using the \cite{Sheth2001} argument. 


\begin{figure}
\begin{center}
\includegraphics[width=60mm,trim=0cm 1cm .5cm .5cm,angle=270]{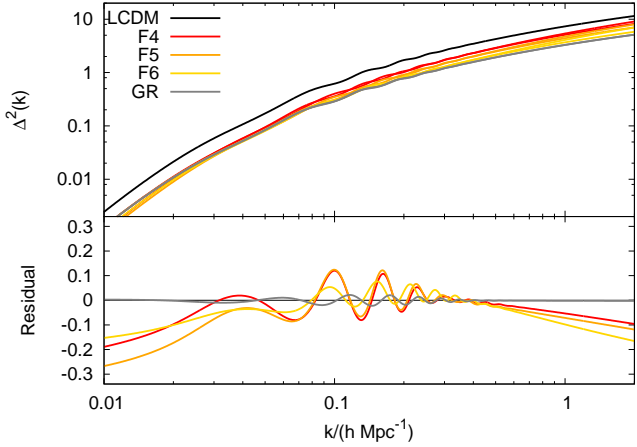}
\end{center}
\caption{Theoretical linear power spectra for the target models compared to the initial $\Lambda$CDM model after scaling in size and redshift by values given in Table \ref{tab:rescaling_parameters}. In the upper panel the black curve shows the unscaled $\Lambda$CDM power and two coloured curves are shown for each model, one being the target and the other the rescaled $\Lambda$CDM power. The fractional residuals are shown in the lower panel where it can be seen that the match is good to the $10$ per cent level around $k=0.1\iMpc$ but a residual BAO can clearly be seen. The ZA step of the AW10 algorithm aims to correct exactly these post-scaling differences in linear clustering. Note that at very large scales the power is different by as much as $30$ per cent in the case of the F5 model and that the match to GR is essentially perfect across the range, but for some small residual BAO.}
\label{fig:linear_power_residuals}
\end{figure}

Additionally when rescaling, the dimensionless velocity units of the simulation must be conserved before and after scaling (see AW10; MP14a; MP14b), which implies a scaling in peculiar velocity ($\mathbf{v}\equiv a\dot{\mathbf{x}}$ where $\mathbf{x}$ is the comoving position) of particles or haloes such that
\begin{equation}
\mathbf{v}'=s\frac{H'f'_\mathrm{g}a'}{Hf_\mathrm{g}a}\mathbf{v}\ ,
\label{eq:velocity_scaling}
\end{equation}
where $H$ is the Hubble parameter and $f_\mathrm{g}\equiv \mathrm{d}\ln g/\mathrm{d}\ln a$ is the logarithmic growth rate. Since the growth rate in HS07 models is scale dependent this approach cannot be followed exactly. Instead one can use the growth rate at the scale of the simulation box $f_\mathrm{g}(k_\mathrm{box},z)$, where $k_\mathrm{box}=2\pi/L$. For the type of cosmological volumes usually simulated the modification to gravity at the scale of the box is negligible, so the growth rate used here will be almost exactly that of a standard gravity model. As discussed in the next section, we are able to later modify the peculiar velocities of particles as a function of scale, and in doing so we can properly account for the scale-dependent growth of linear velocity perturbations.

\subsection{Particles}
\label{sec:partciles}

\begin{figure*}
\begin{center}
\makebox[\textwidth][c]{
\subfloat{\includegraphics[width=41.5mm,trim=1.cm 3.4cm 1.5cm 2.4cm,angle=270]{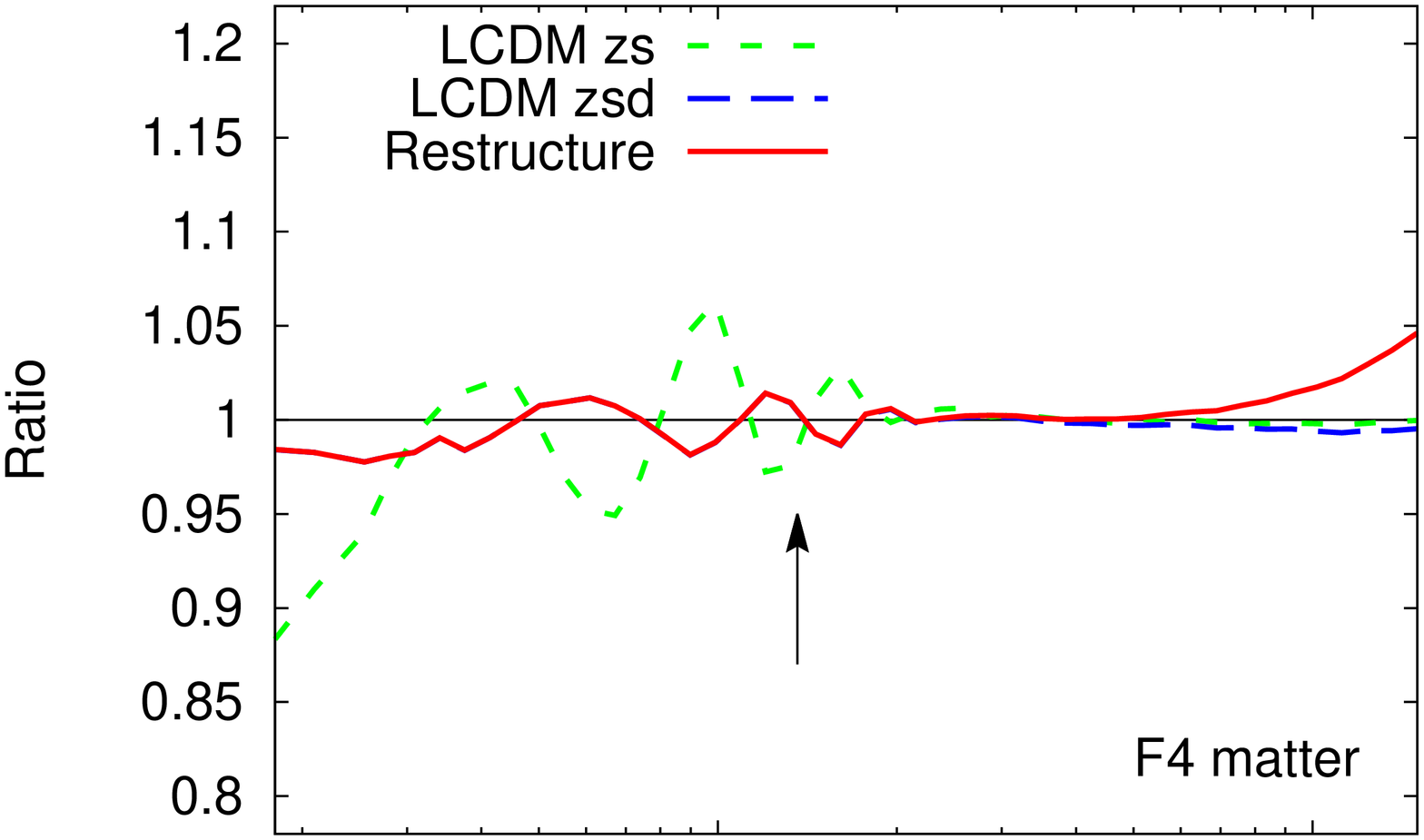}}\hspace{.1cm}
\subfloat{\includegraphics[width=41.5mm,trim=1.cm 3.4cm 1.5cm 2.4cm,angle=270]{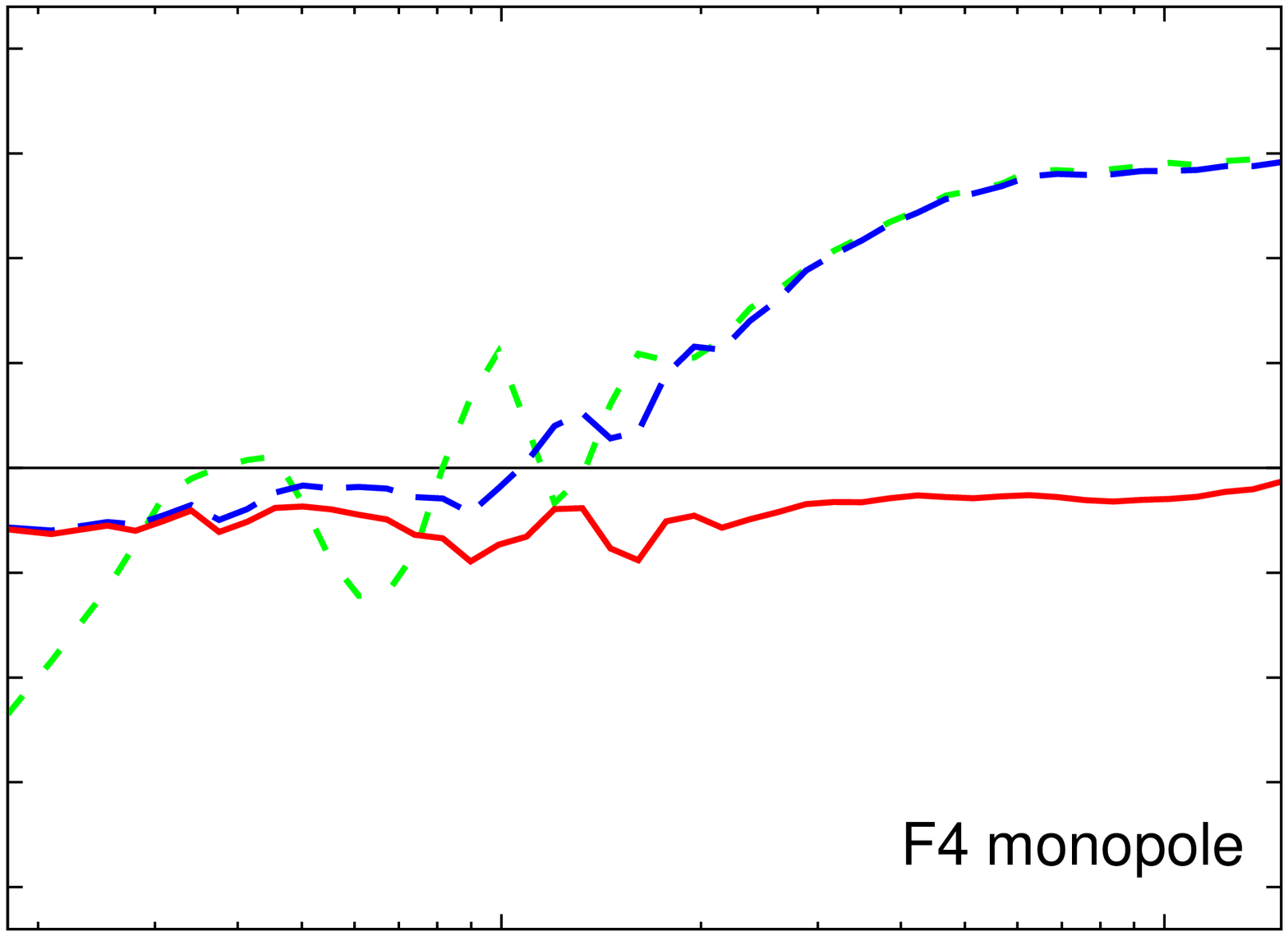}}\hspace{.1cm}
\subfloat{\includegraphics[width=41.5mm,trim=1.cm 3.4cm 1.5cm 2.4cm,angle=270]{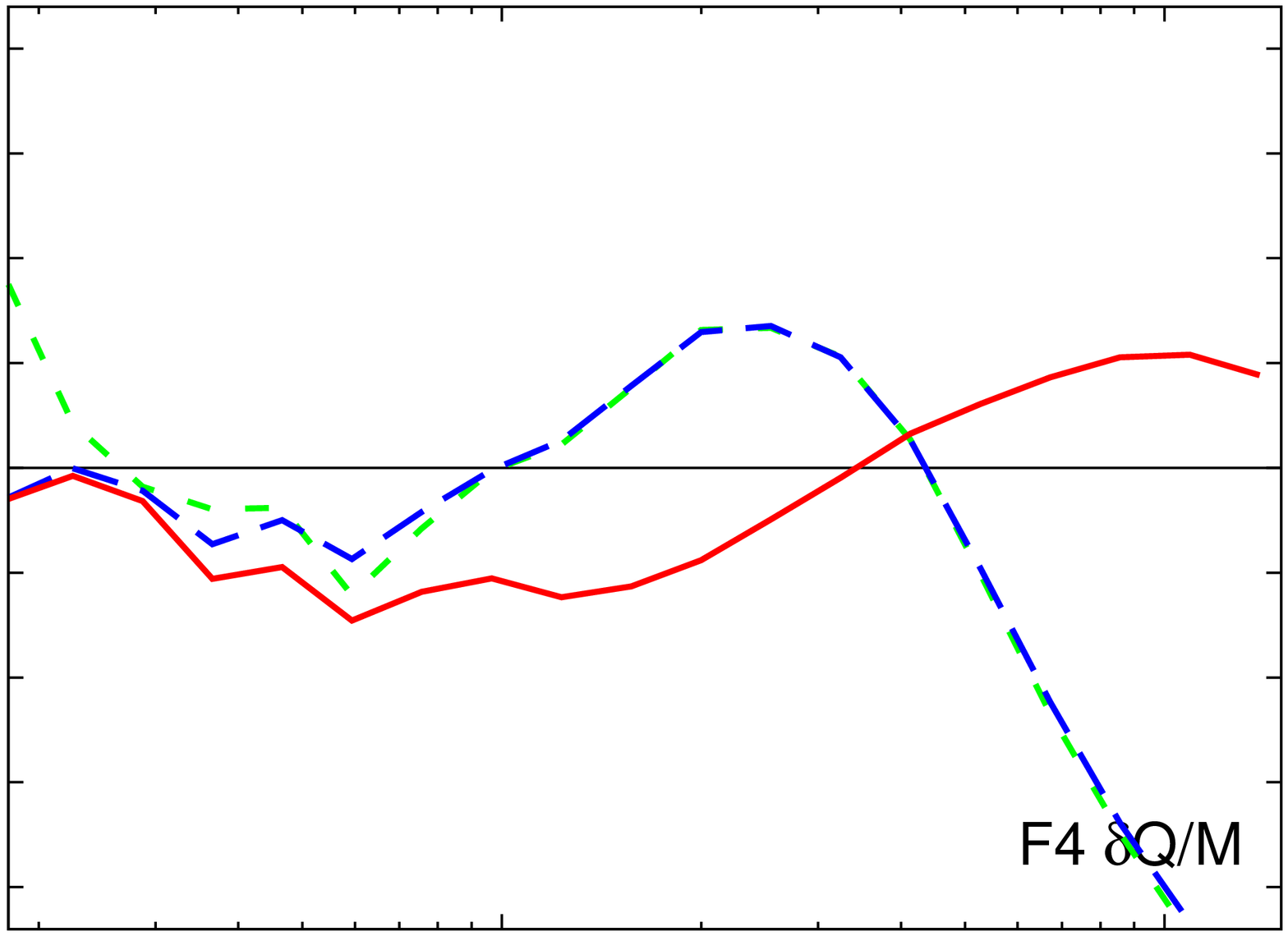}}}\\
\makebox[\textwidth][c]{
\subfloat{\includegraphics[width=41.5mm,trim=1.cm 3.4cm 1.5cm 2.4cm,angle=270]{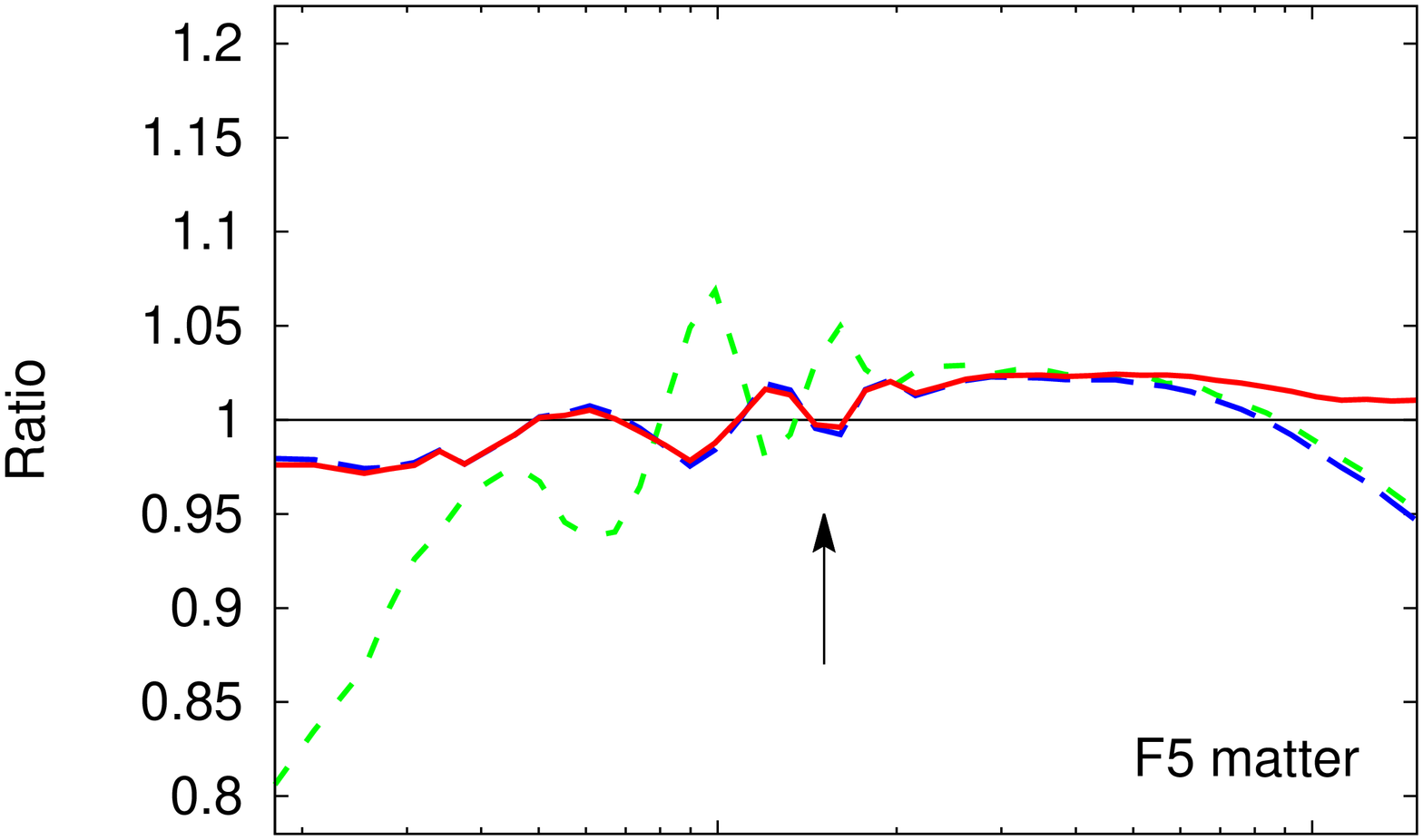}}\hspace{.1cm}
\subfloat{\includegraphics[width=41.5mm,trim=1.cm 3.4cm 1.5cm 2.4cm,angle=270]{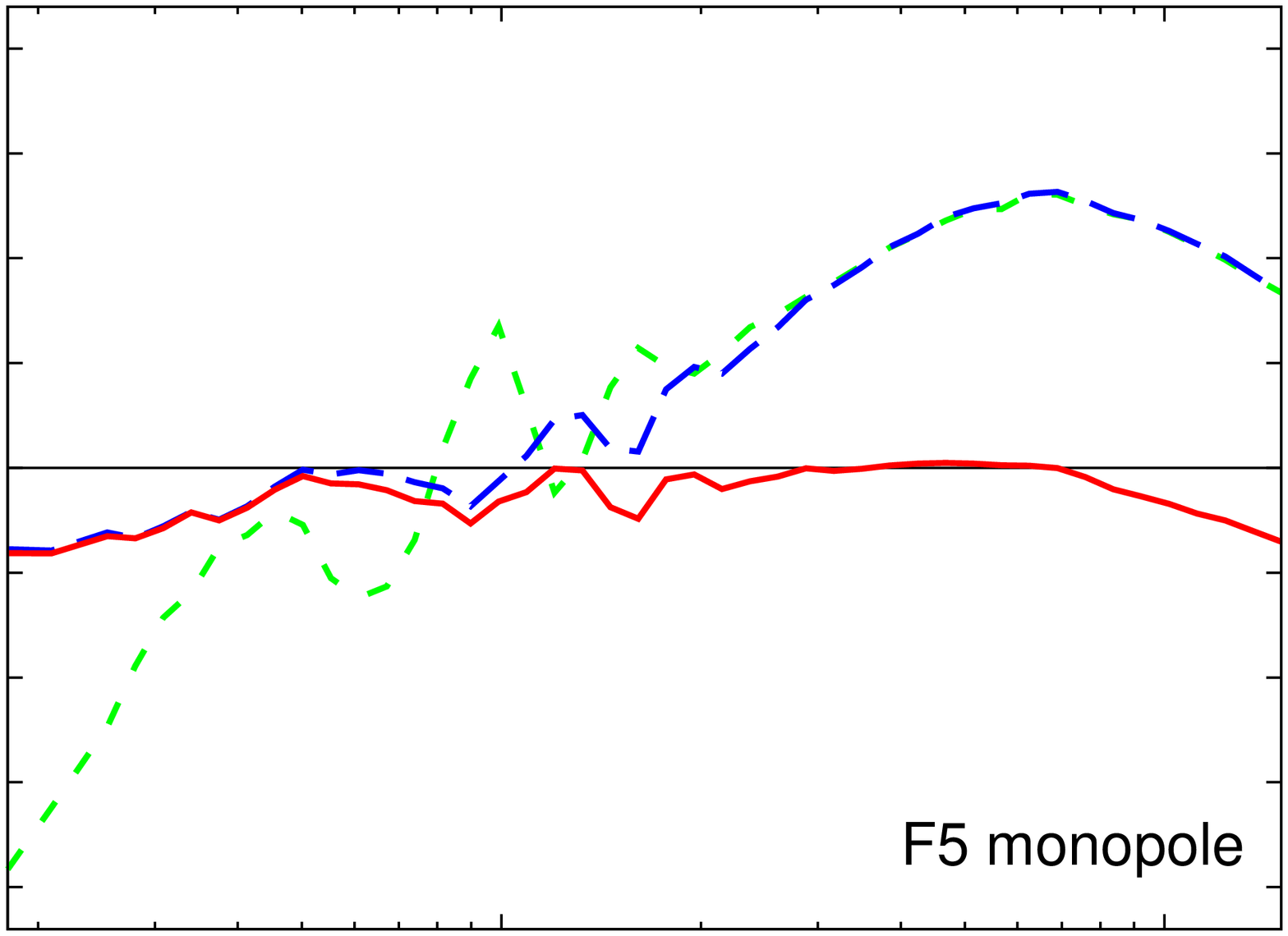}}\hspace{.1cm}
\subfloat{\includegraphics[width=41.5mm,trim=1.cm 3.4cm 1.5cm 2.4cm,angle=270]{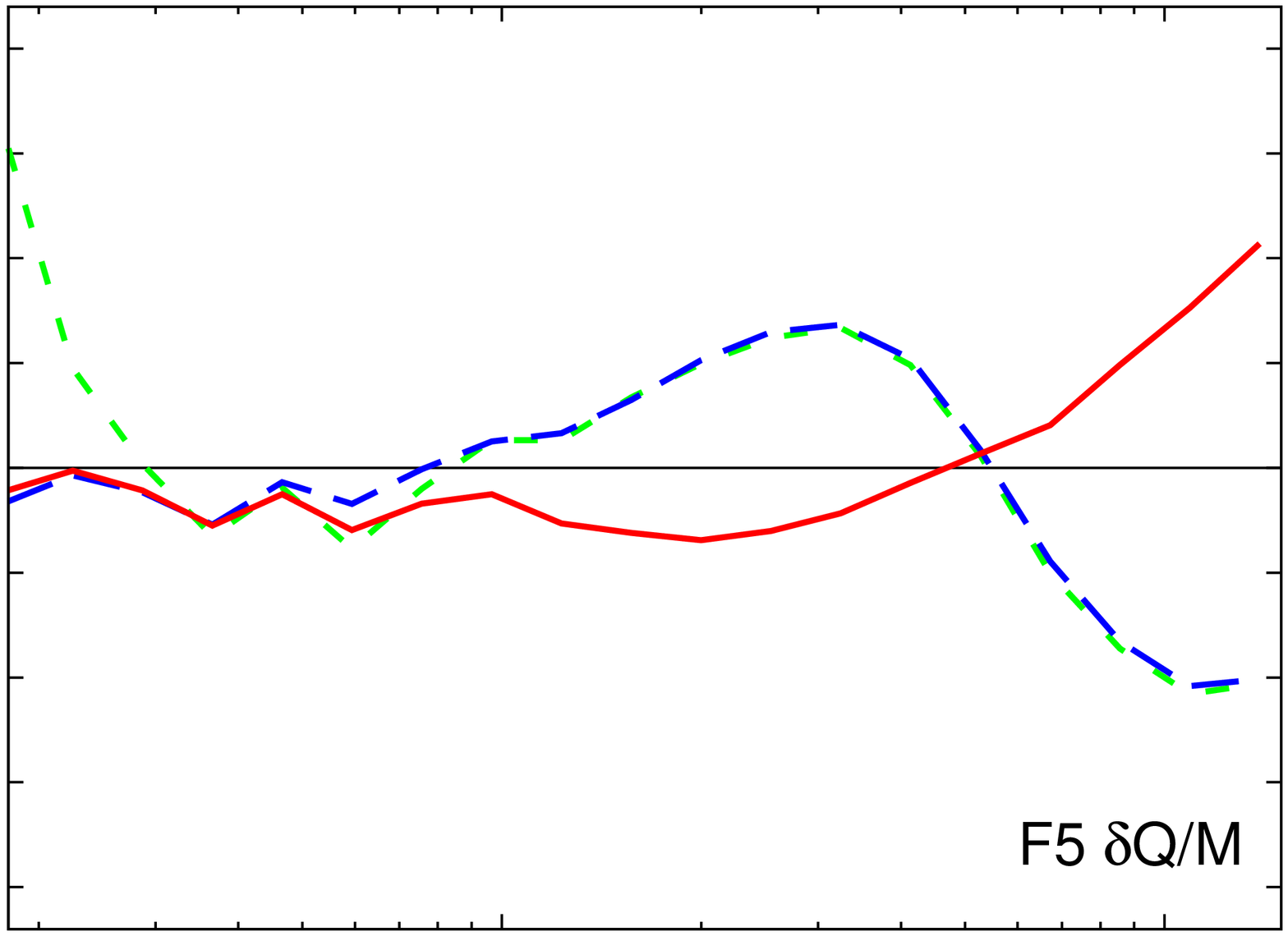}}}\\
\makebox[\textwidth][c]{
\subfloat{\includegraphics[width=41.5mm,trim=1.cm 3.4cm 1.5cm 2.4cm,angle=270]{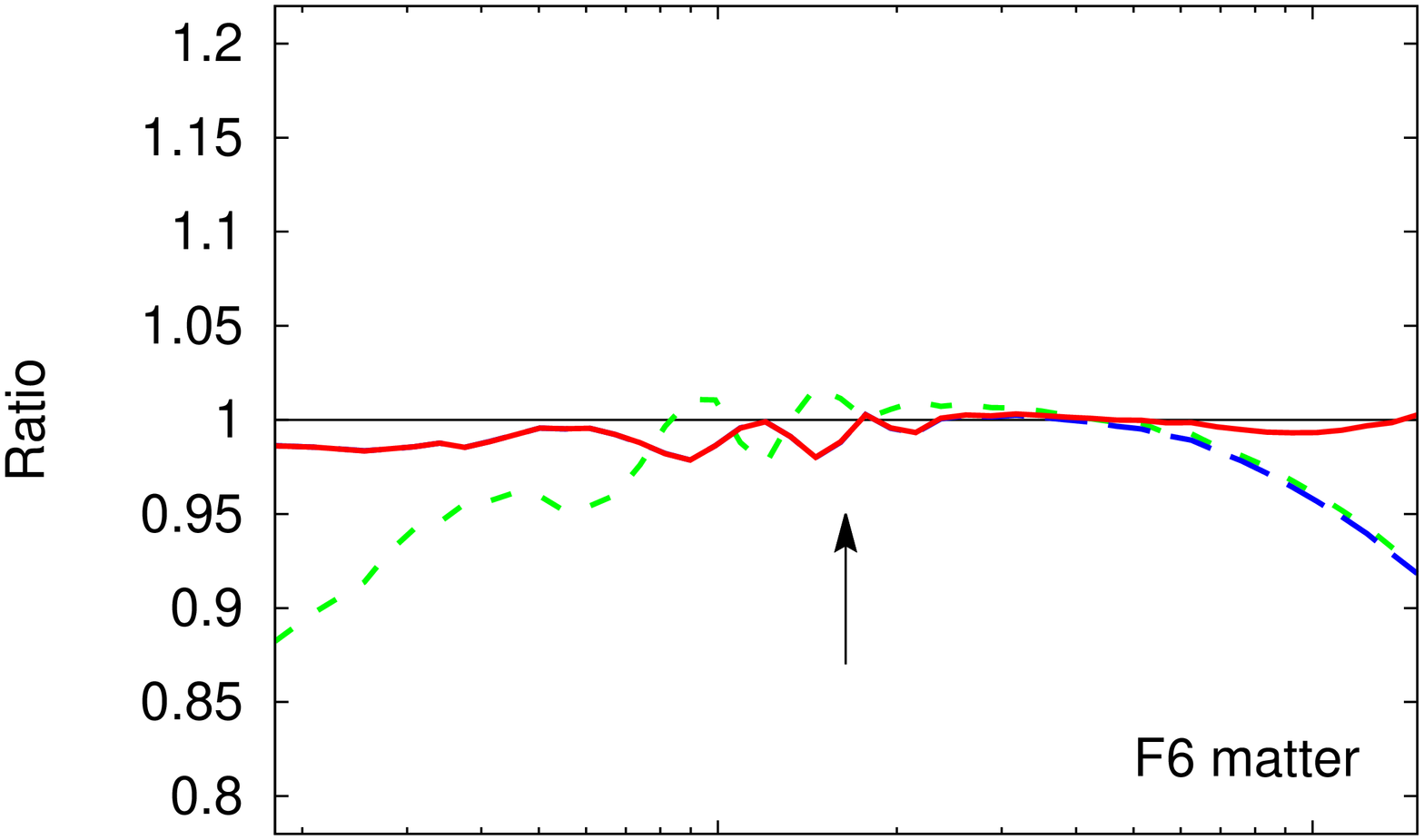}}\hspace{.1cm}
\subfloat{\includegraphics[width=41.5mm,trim=1.cm 3.4cm 1.5cm 2.4cm,angle=270]{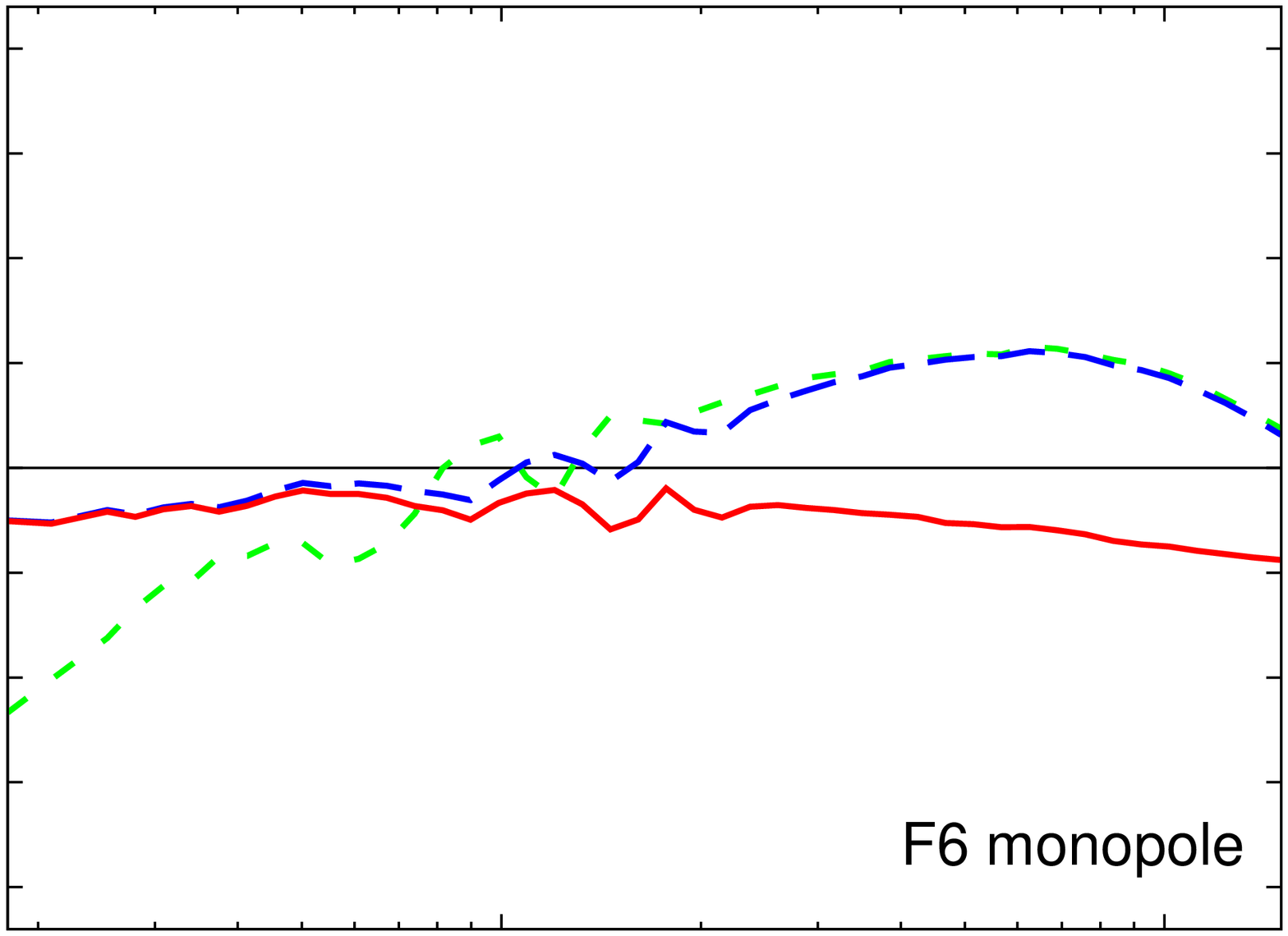}}\hspace{.1cm}
\subfloat{\includegraphics[width=41.5mm,trim=1.cm 3.4cm 1.5cm 2.4cm,angle=270]{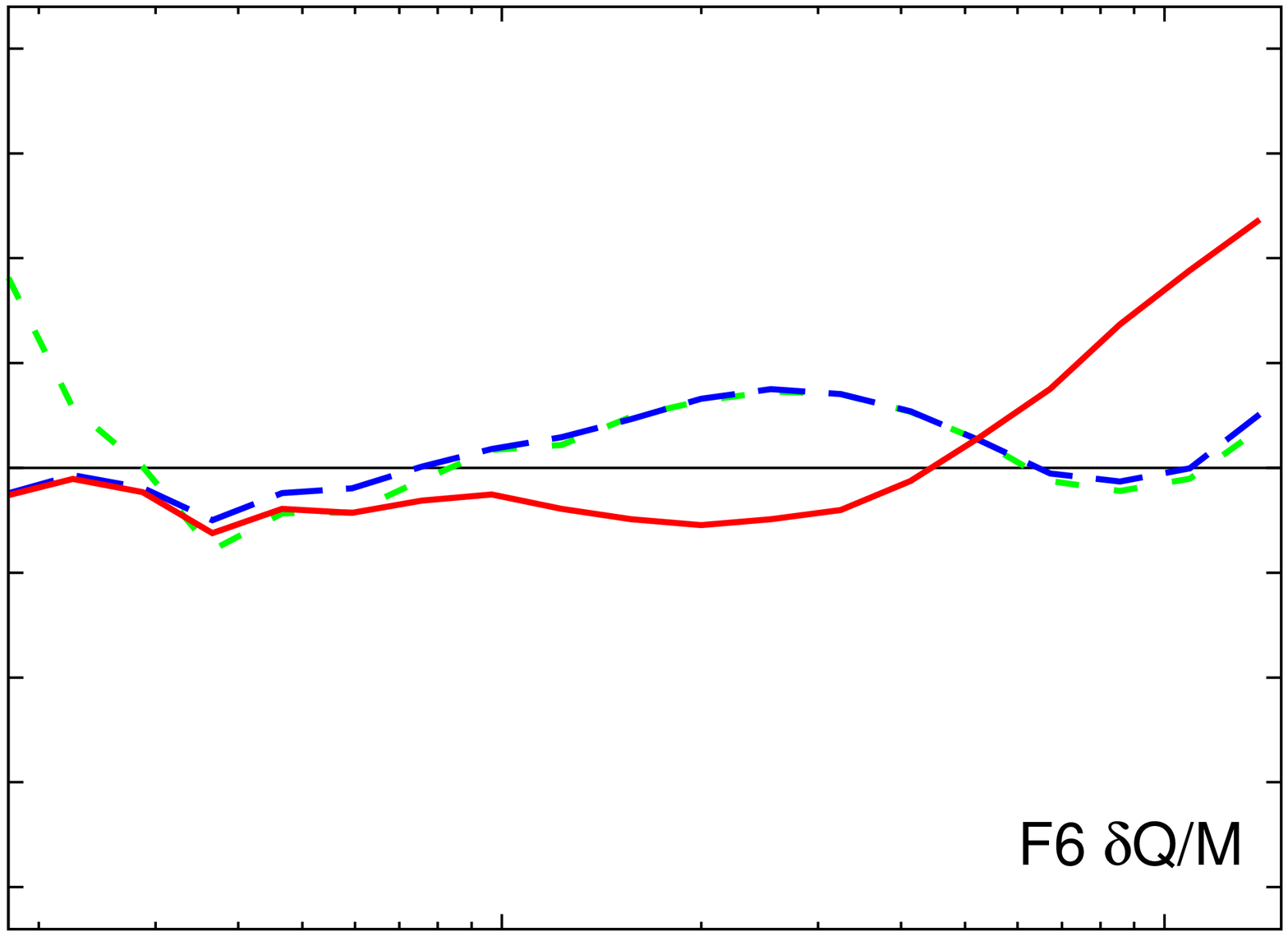}}}\\
\makebox[\textwidth][c]{
\subfloat{\includegraphics[width=41.5mm,trim=1.cm 3.4cm 1.5cm 2.4cm,angle=270]{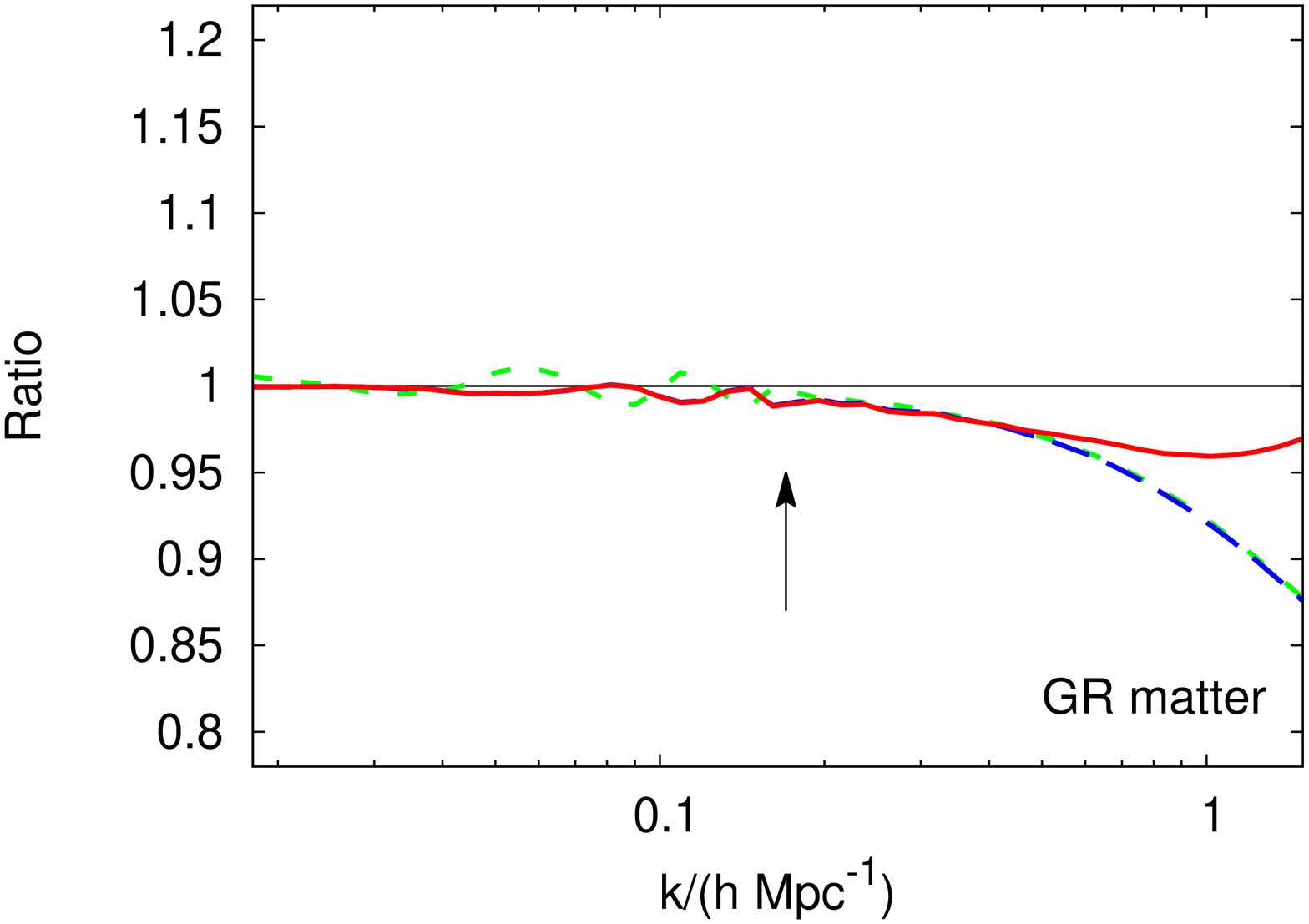}}\hspace{.1cm}
\subfloat{\includegraphics[width=41.5mm,trim=1.cm 3.4cm 1.5cm 2.4cm,angle=270]{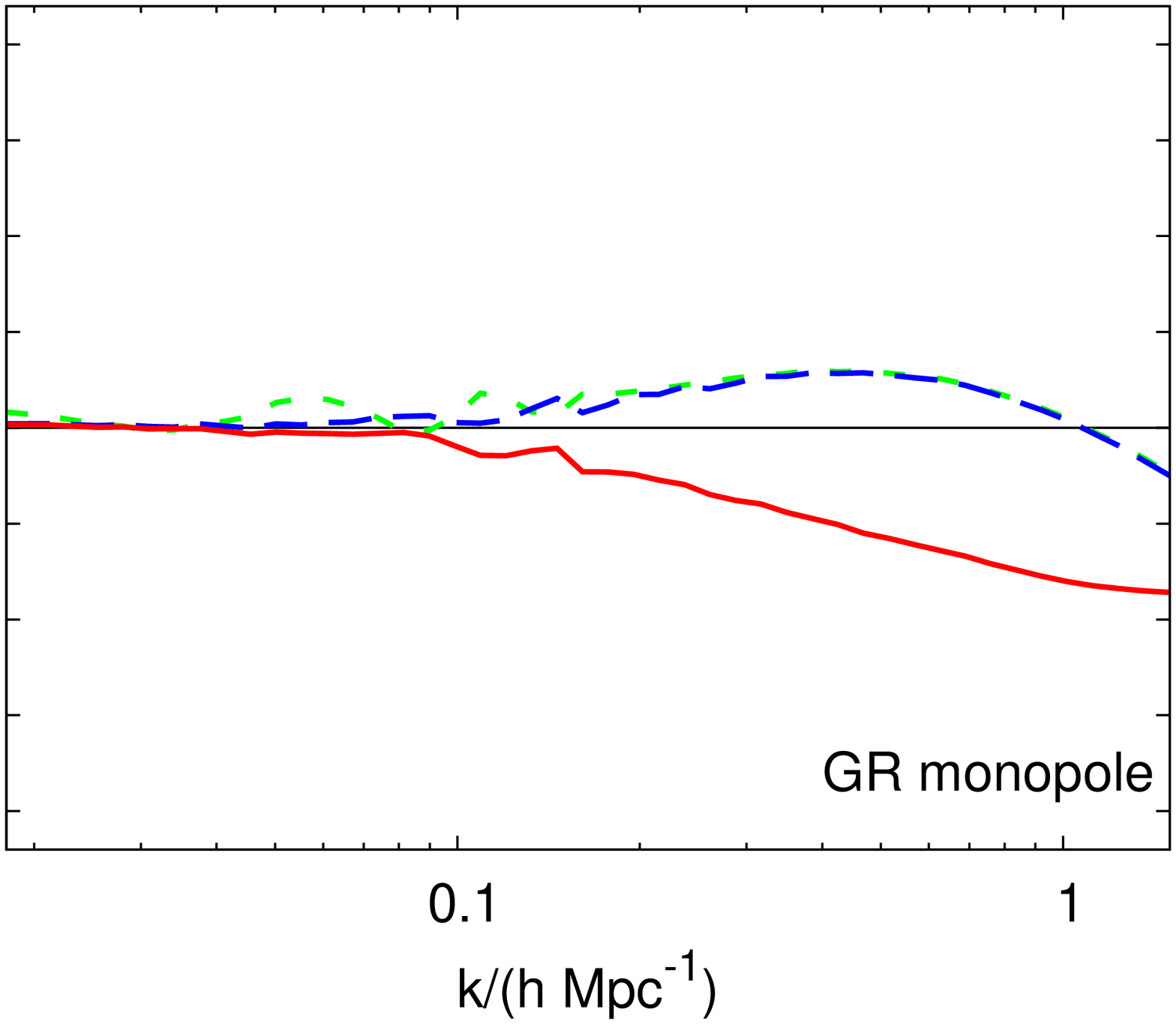}}\hspace{.1cm}
\subfloat{\includegraphics[width=41.5mm,trim=1.cm 3.4cm 1.5cm 2.4cm,angle=270]{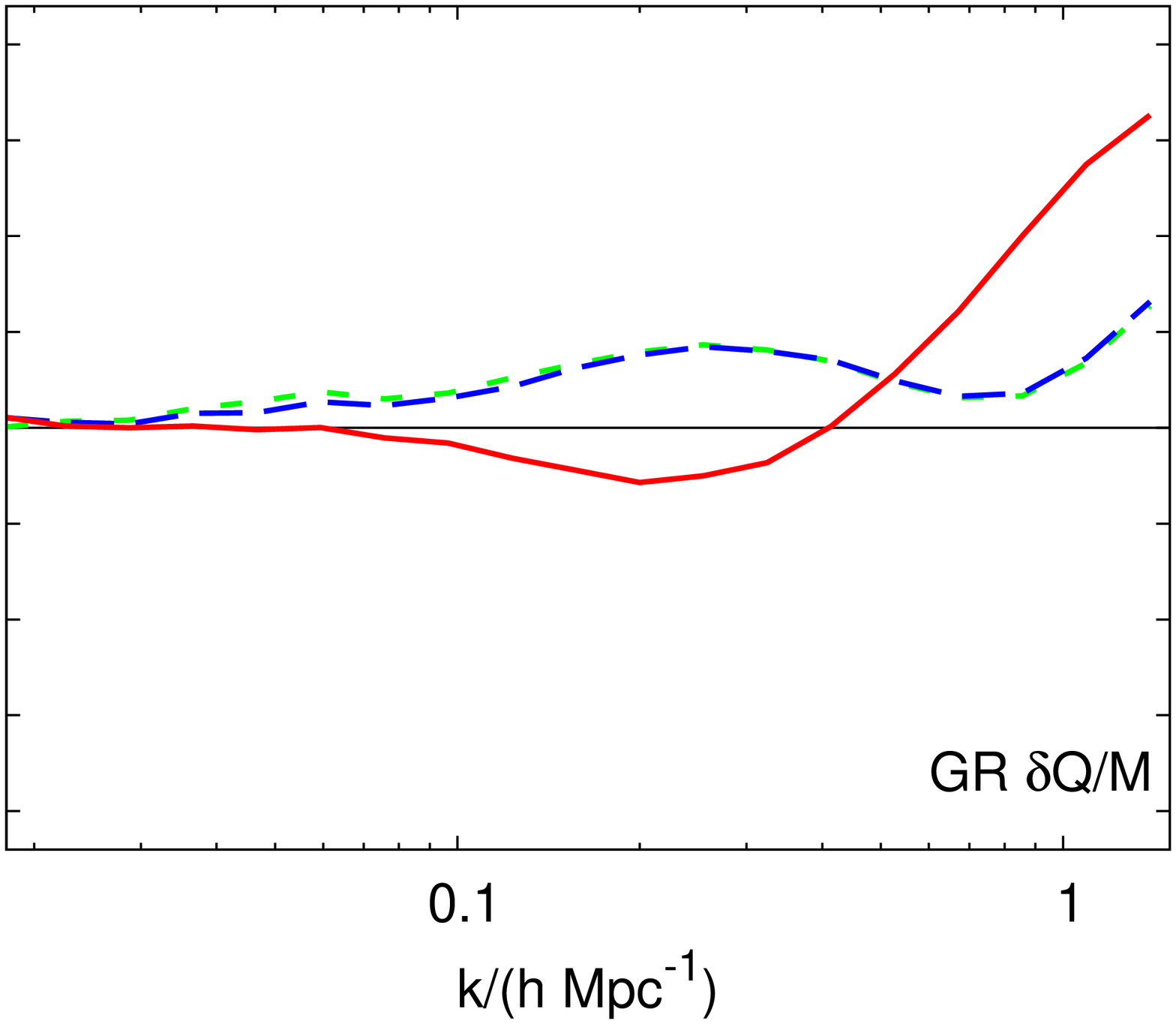}}}\\
\end{center}
\caption{The ratio of rescaled to target matter power spectra (left column), redshift-space monopole (central column) and quadrupole to monopole ratio (right column; see the text for definition) from scaling the full $\Lambda$CDM particle distribution to F4 (top), F5, F6 and GR (bottom) models. In each case the green (short-dashed; $zs$) curve shows the initial scaling in size and redshift while the blue curve shows the result of applying the additional extra displacements (long-dashed; $zsd$), the red (solid) curve then shows the result when additionally restructuring halo interiors. The black arrow in the matter plots shows the non-linear scale (equation~\ref{eq:nl_scale}), which is slightly different for each model. For all models the matter spectrum is matched to better than $5$ per cent across all scales with only modest improvements gained by restructuring -- this reflects the similarity in internal structure between haloes in $f(R)$ and those in $\Lambda$CDM. Conversely errors in the monopole are large at small scales (FOG are underestimated) unless the halo internal velocities are restructured, this reflects the lack of enhanced gravity in the $\Lambda$CDM model and this needs to be introduced by hand. Surprisingly restructuring worsens the match to the monopole in the GR case. The quadrupole to monopole ratio is improved at quasi-linear scales by the restructuring with a maximum error at the $5$ per cent level up to the scale where the quadrupole changes sign.}
\label{fig:scaling_power}
\end{figure*}

\begin{figure*}
\begin{center}
\makebox[\textwidth][c]{
\subfloat{\includegraphics[width=52mm,trim=3cm 8.5cm 2cm 6.8cm,angle=270]{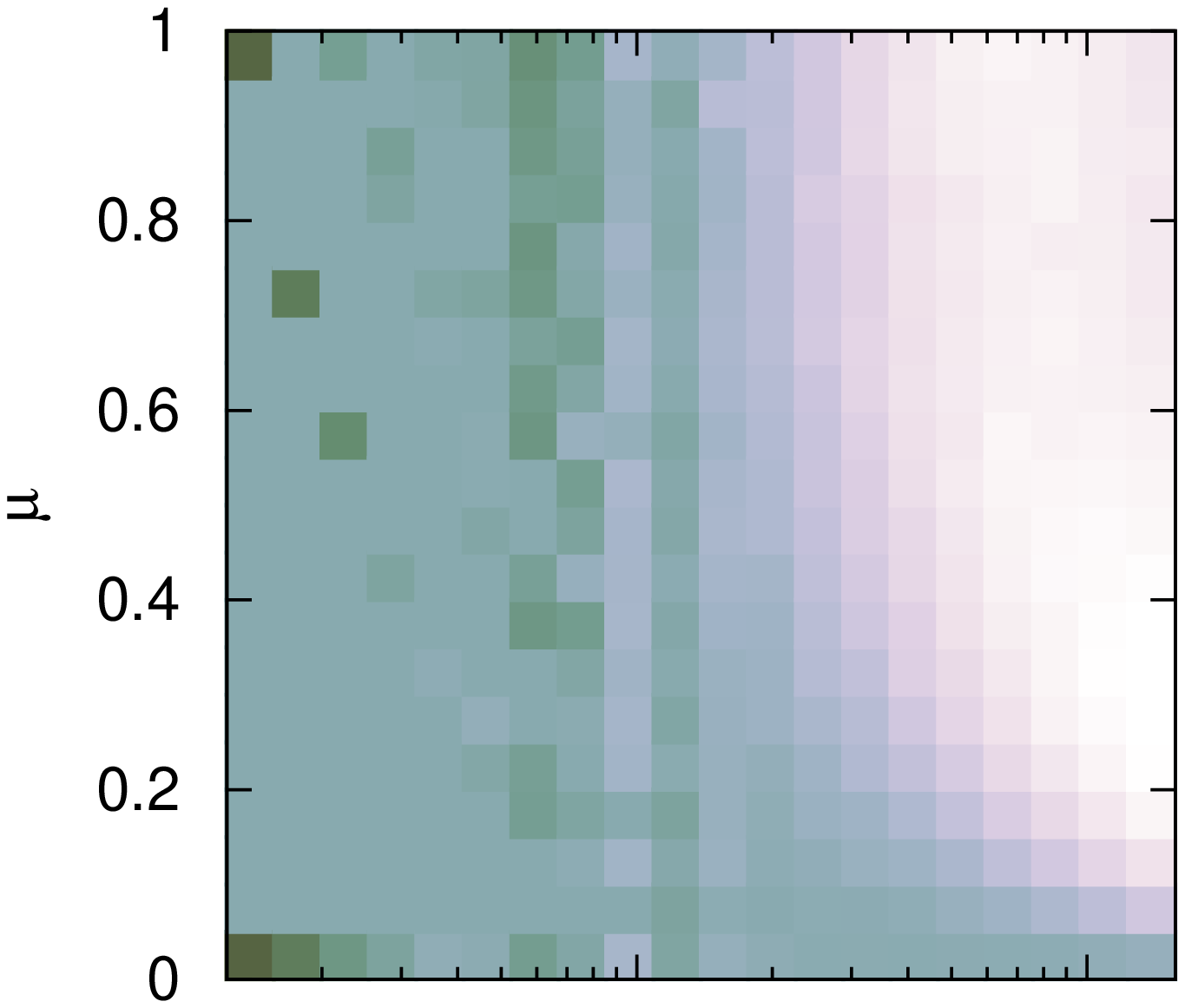}}\hspace{1cm}
\subfloat{\includegraphics[width=52mm,trim=3cm 8.5cm 2cm 6.8cm,angle=270]{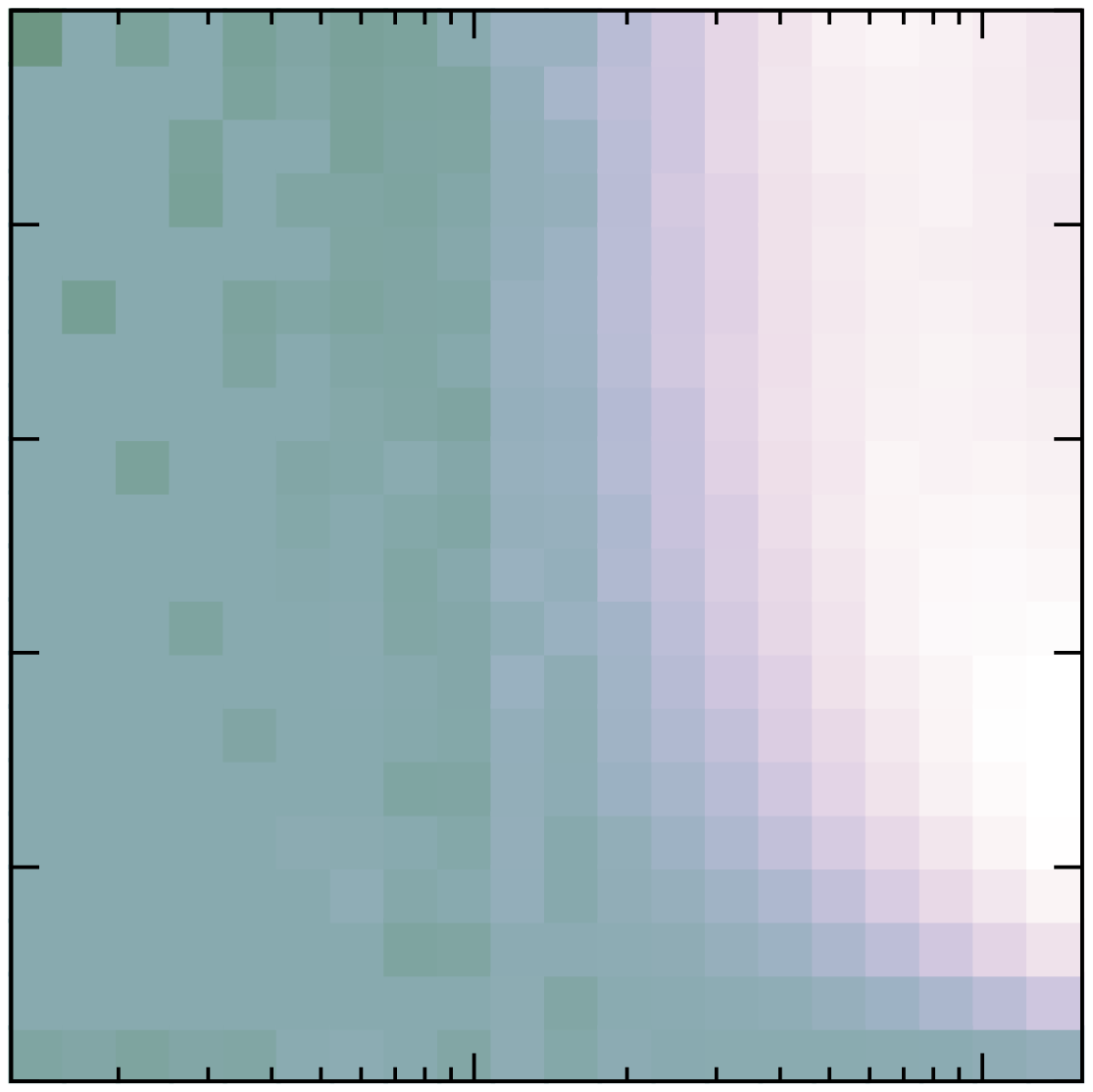}}\hspace{1cm}
\subfloat{\includegraphics[width=52mm,trim=3cm 8.5cm 2cm 6.8cm,angle=270]{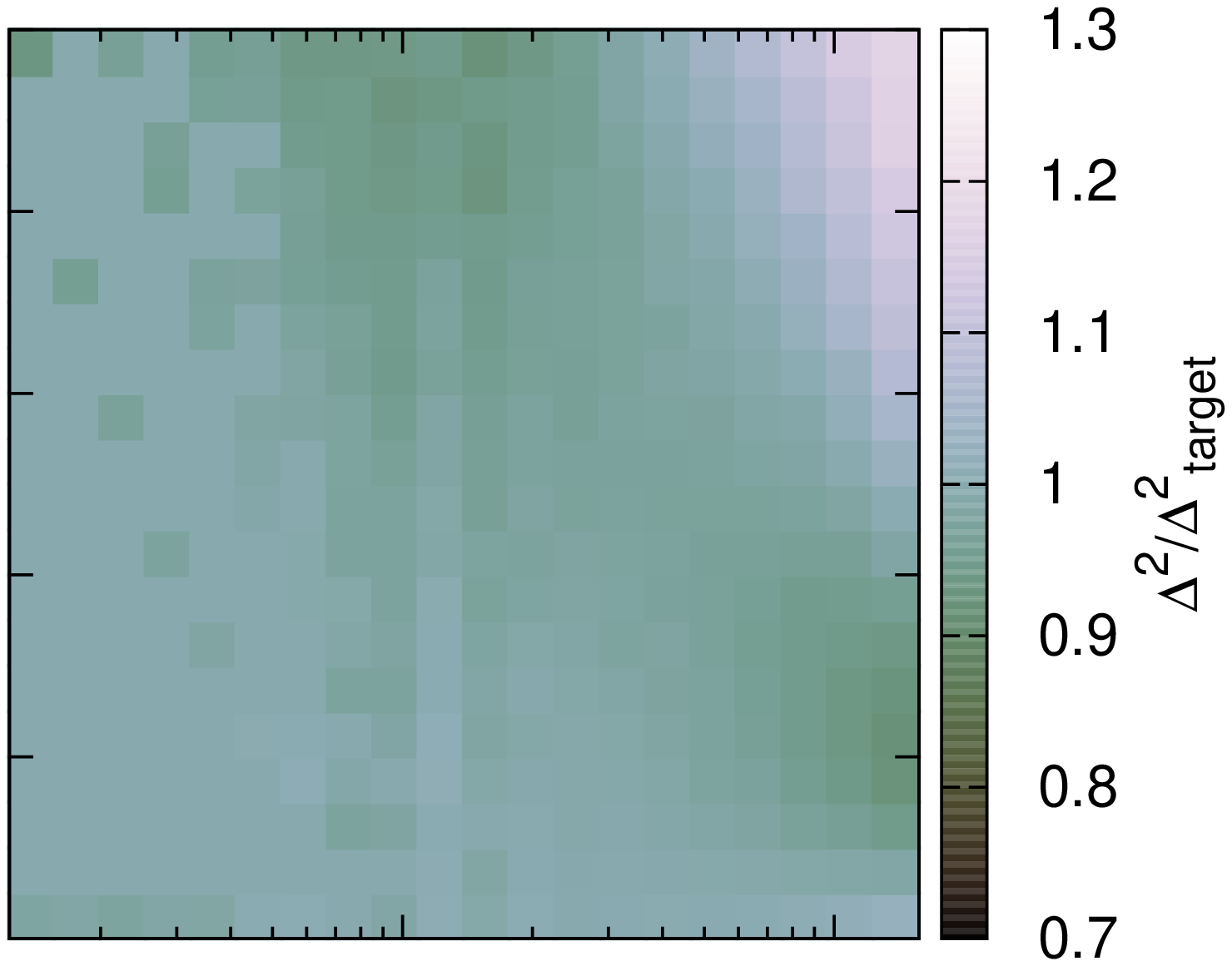}}}
\makebox[\textwidth][c]{
\subfloat{\includegraphics[width=52mm,trim=3cm 8.5cm 2cm 6.8cm,angle=270]{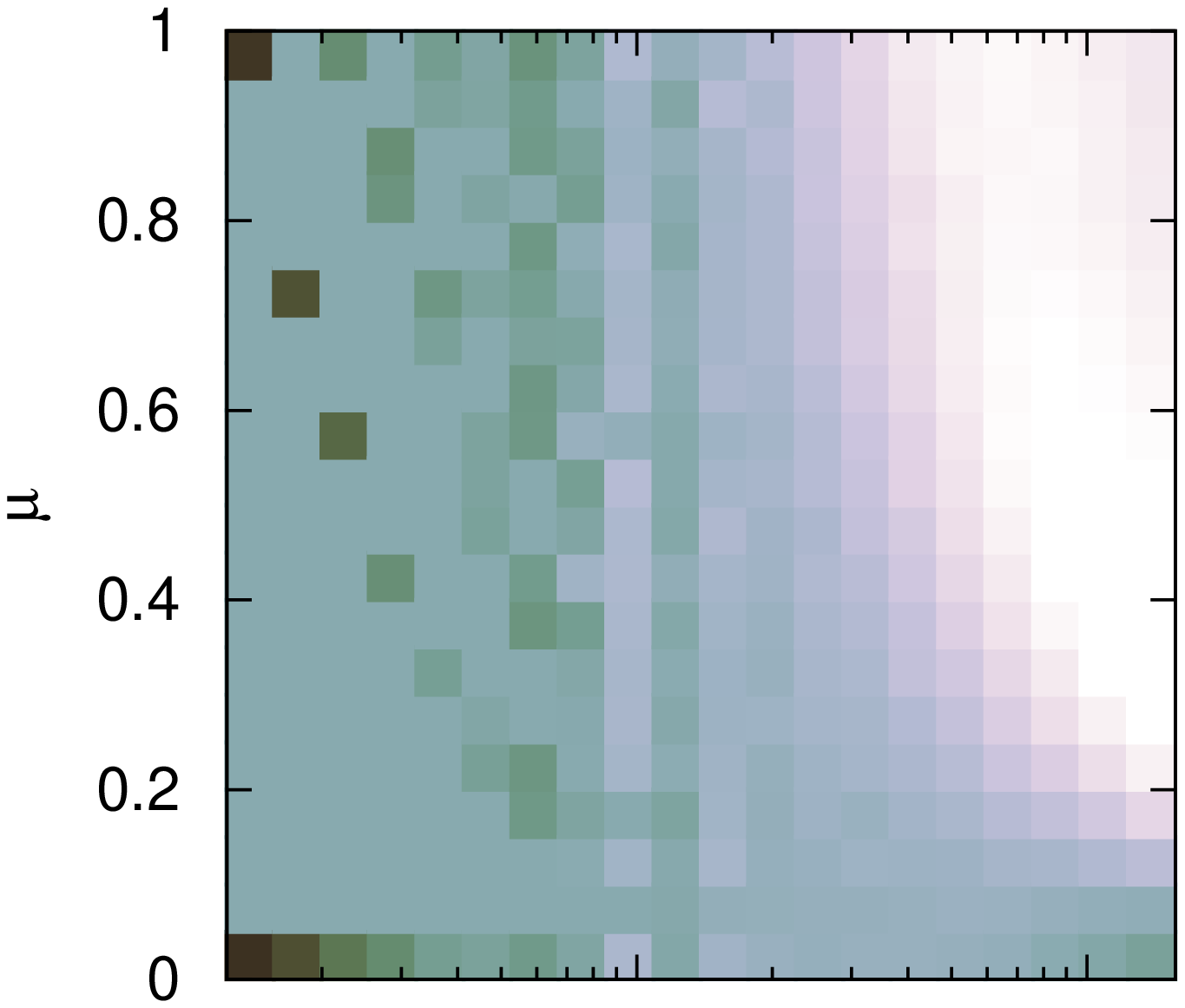}}\hspace{1cm}
\subfloat{\includegraphics[width=52mm,trim=3cm 8.5cm 2cm 6.8cm,angle=270]{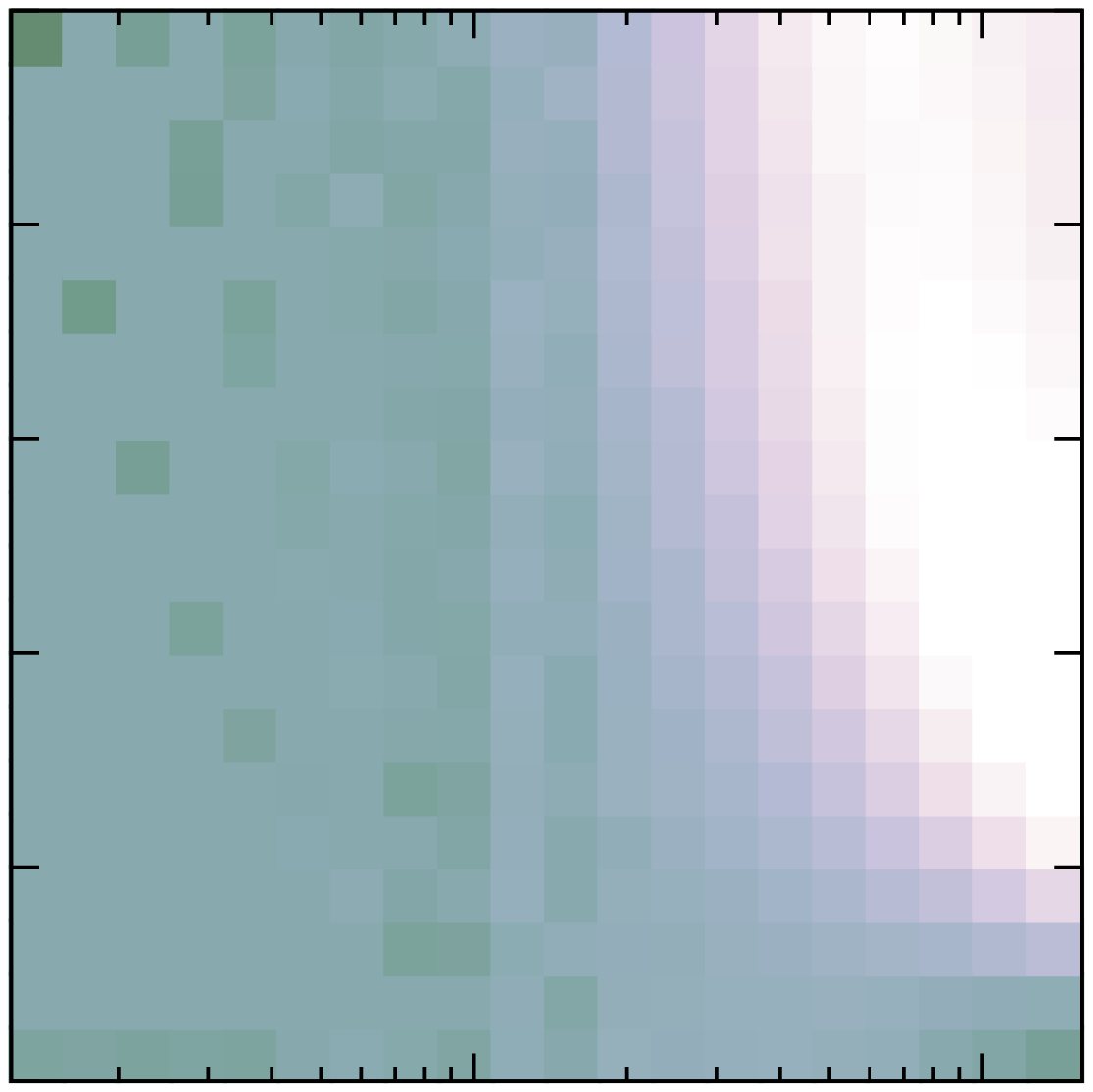}}\hspace{1cm}
\subfloat{\includegraphics[width=52mm,trim=3cm 8.5cm 2cm 6.8cm,angle=270]{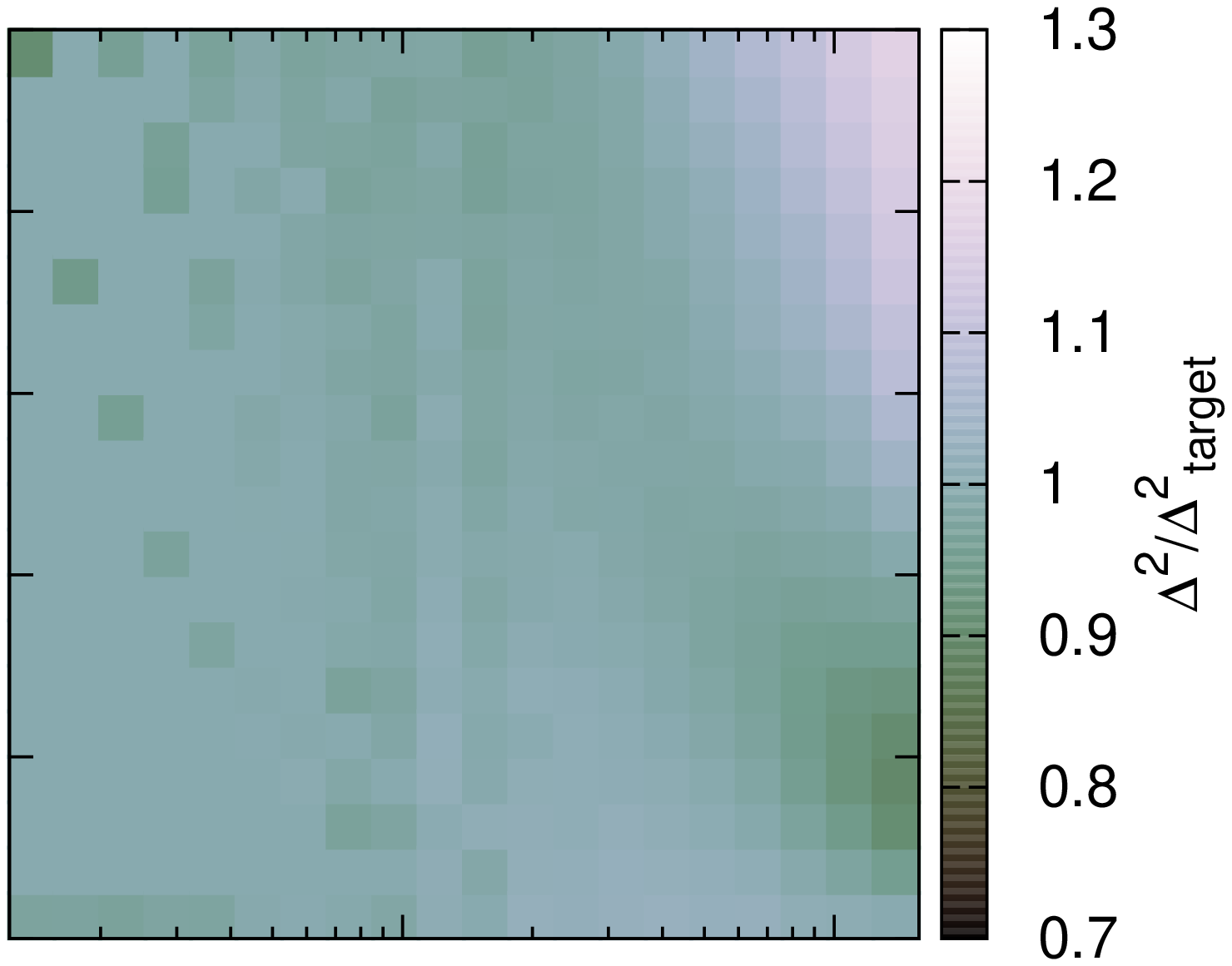}}}
\makebox[\textwidth][c]{
\subfloat{\includegraphics[width=52mm,trim=3cm 8.5cm 2cm 6.8cm,angle=270]{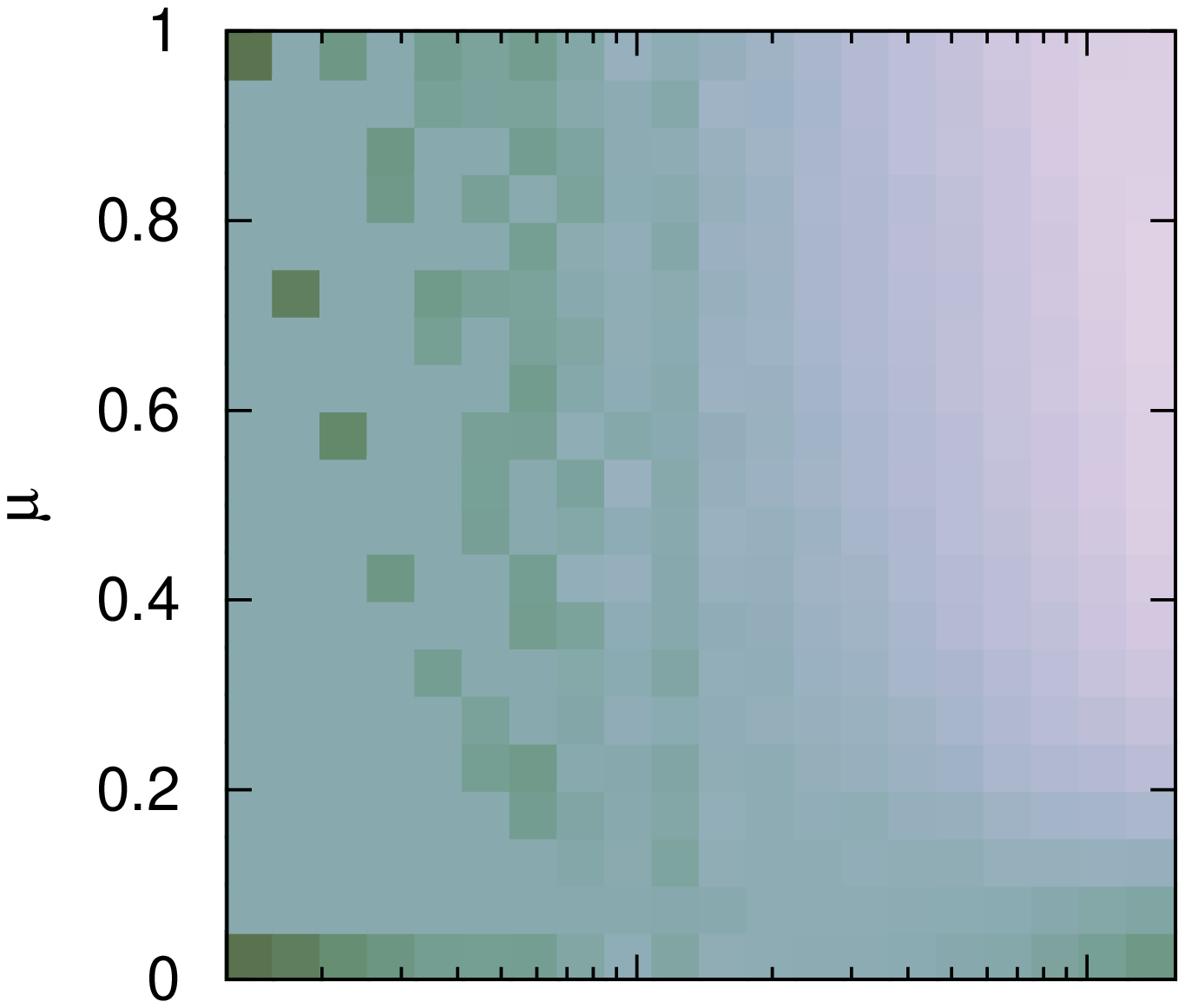}}\hspace{1cm}
\subfloat{\includegraphics[width=52mm,trim=3cm 8.5cm 2cm 6.8cm,angle=270]{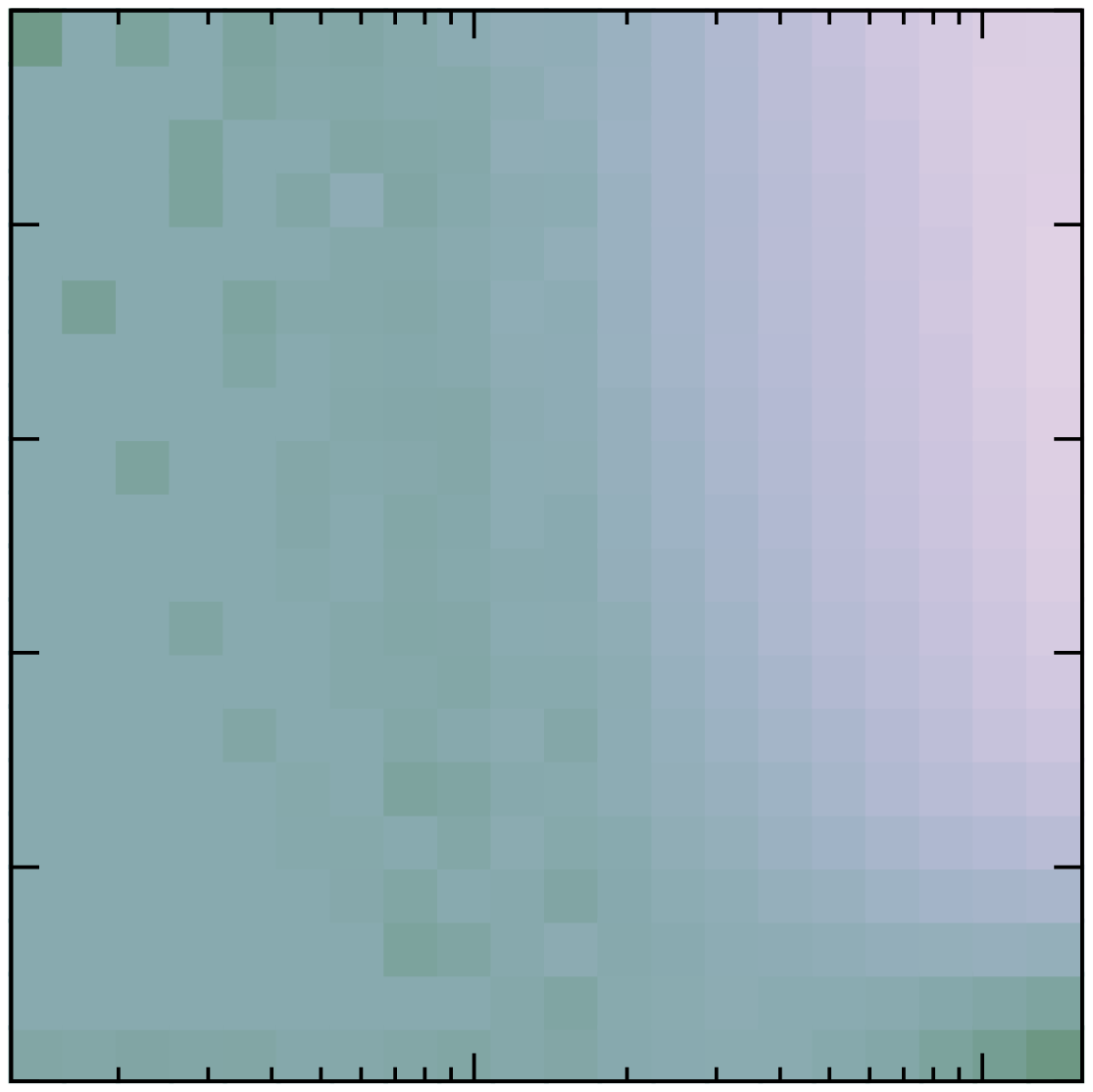}}\hspace{1cm}
\subfloat{\includegraphics[width=52mm,trim=3cm 8.5cm 2cm 6.8cm,angle=270]{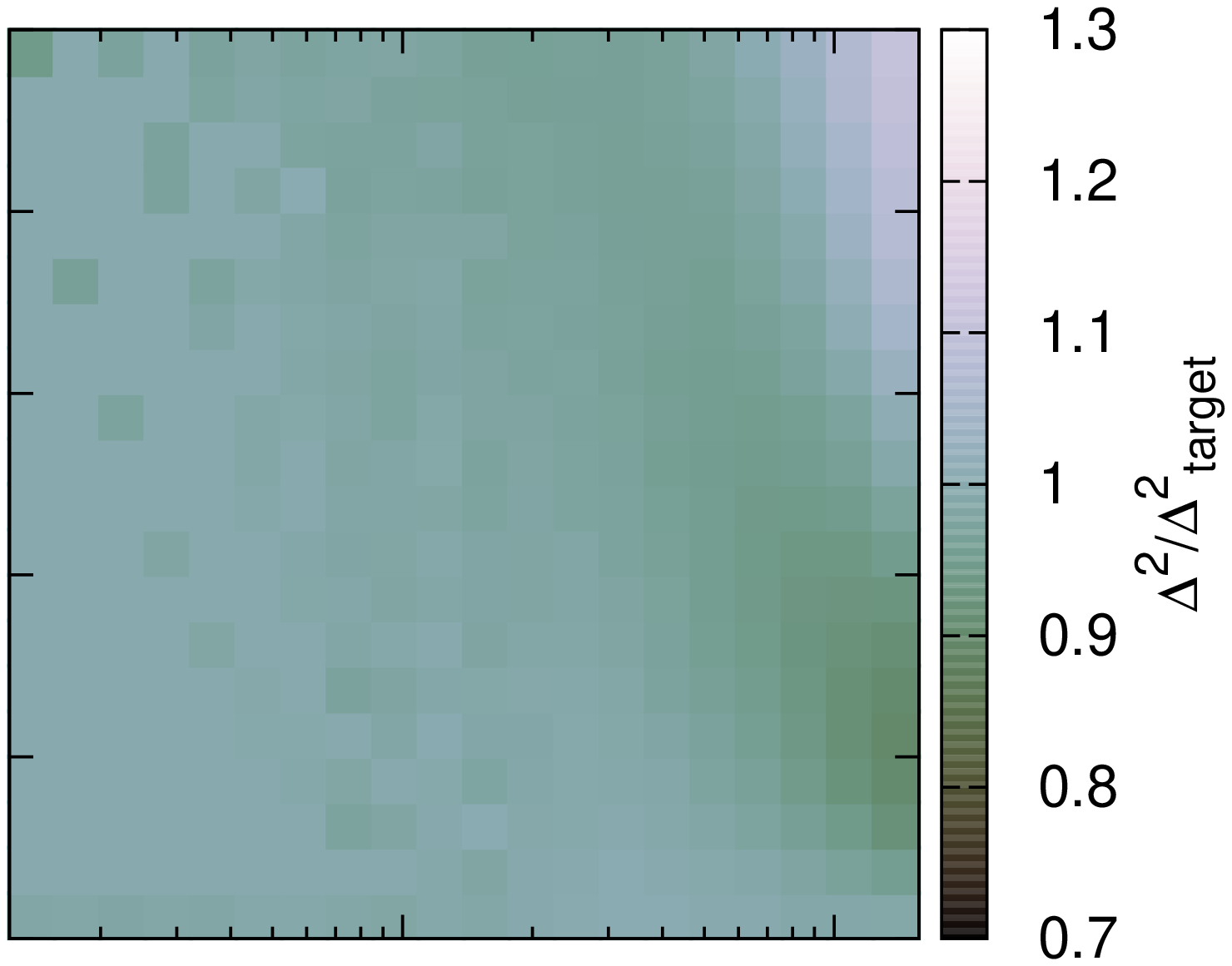}}}
\makebox[\textwidth][c]{
\subfloat{\includegraphics[width=52mm,trim=3cm 8.5cm 2cm 6.8cm,angle=270]{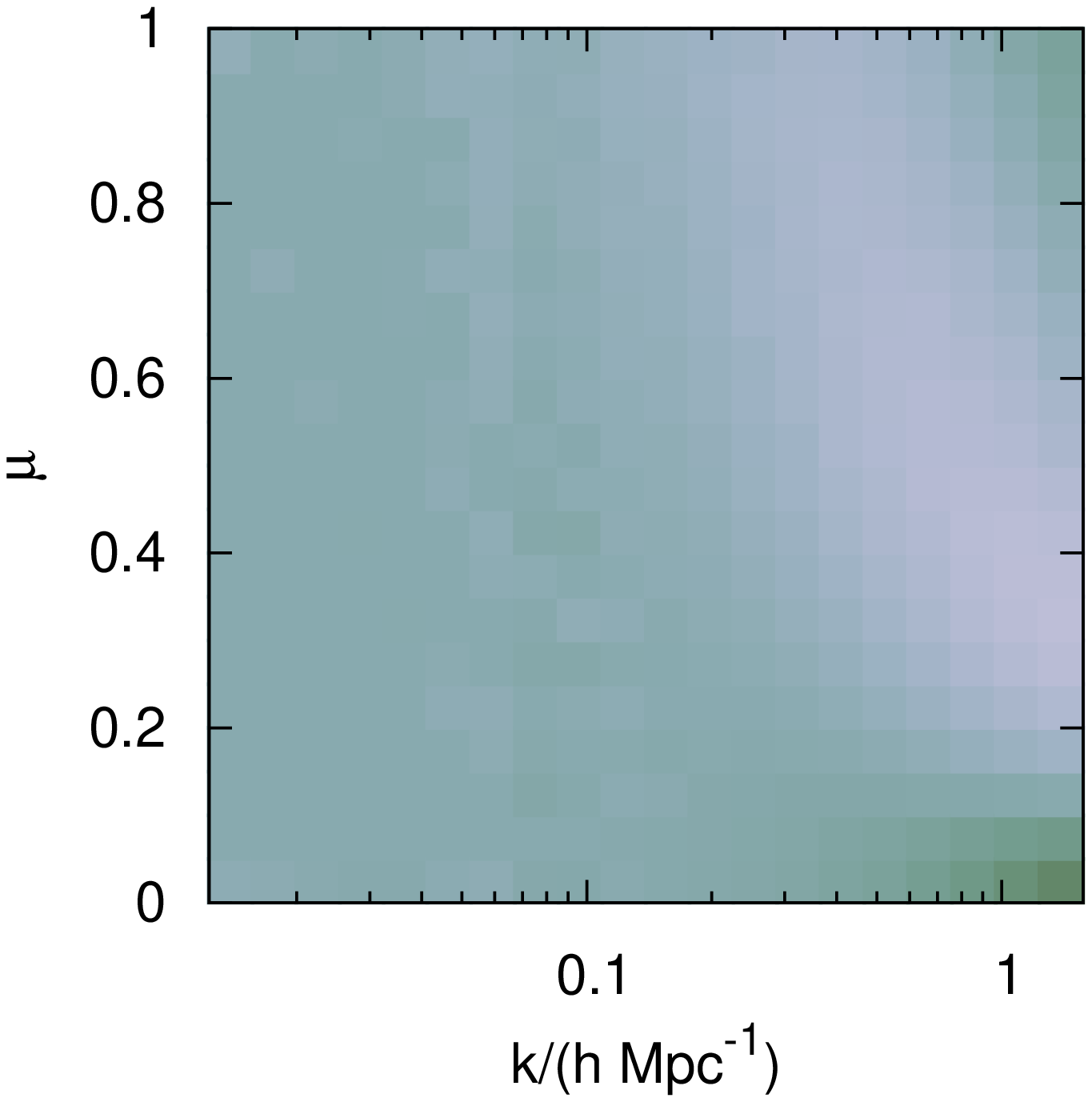}}\hspace{1cm}
\subfloat{\includegraphics[width=52mm,trim=3cm 8.5cm 2cm 6.8cm,angle=270]{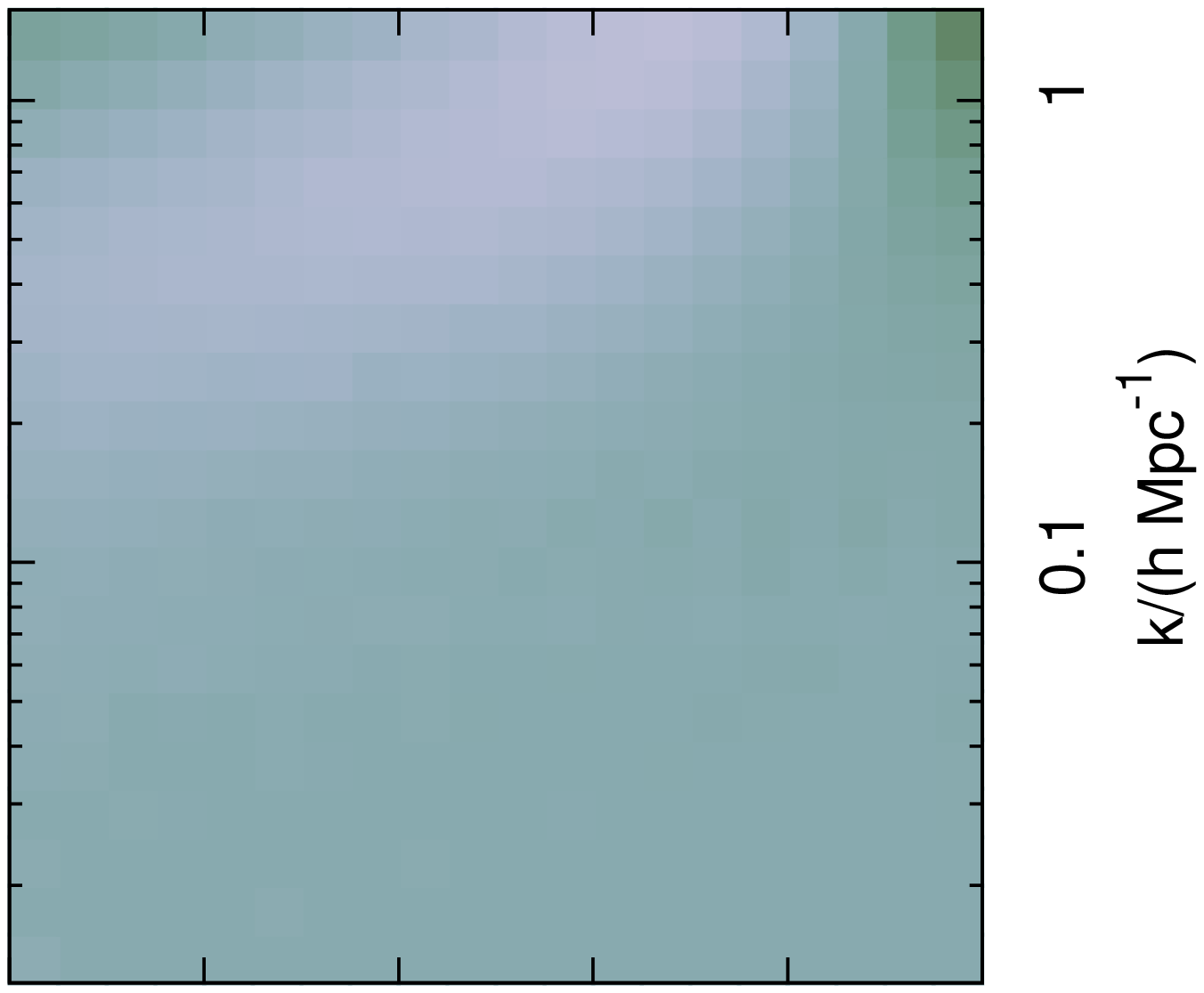}}\hspace{1cm}
\subfloat{\includegraphics[width=52mm,trim=3cm 8.5cm 2cm 6.8cm,angle=270]{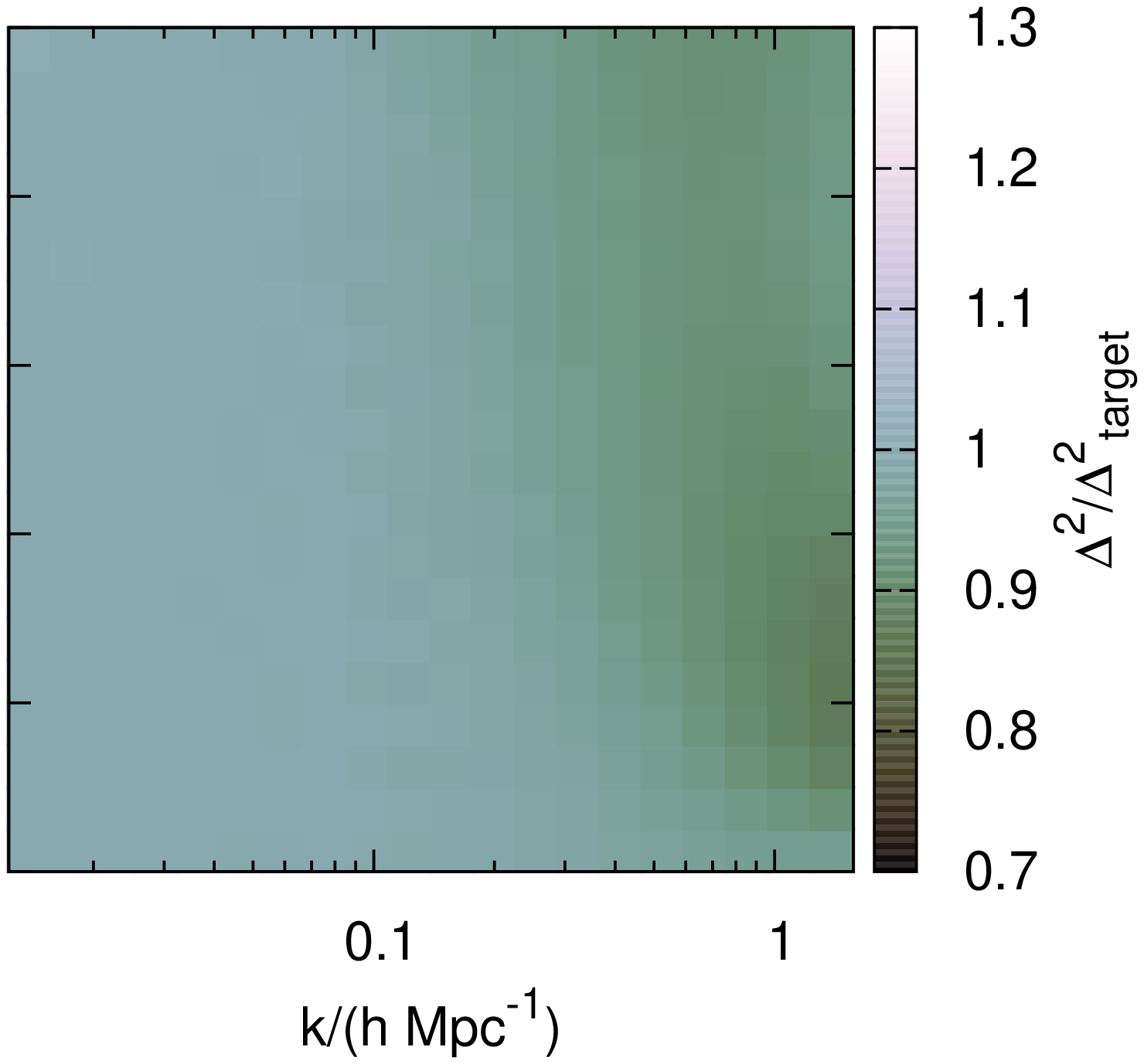}}}
\end{center}
\caption{The ratio of rescaled to target power in 2D redshift space for the F4 (top row), F5, F6 and GR (bottom row) models when the size and redshift parts of the rescaling method have been applied (left column), additionally modifying the displacement field (central column) and finally restructuring haloes (right column). The residual BAO seen noisily across all $\mu$ at large scales can be efficiently removed by the ZA correction (left to central column). Residual differences are then mainly concentrated high-$k$ values and these differences extend to larger scales for high-$\mu$ modes. Residuals are largely rectified by restructuring the halo particles in physical and velocity space; the correction is largest for high-$\mu$ regions, particularly for F4 and F5 models, that are dominated by non-linear FOG differences and is quite minor $\mu=0$ modes, which reflects the similarities in halo physical internal structure. That the residual changes sign as a function of $\mu$ for small scale ($k=1\iMpc$) modes in the F4, F5 and F6 cases is responsible for the poor quadrupole seen at these scales in Fig.~\ref{fig:scaling_power}, because the quadrupole essentially differences high and low $\mu$.}
\label{fig:scaling_rsdpower}
\end{figure*}

\begin{figure*}
\begin{center}
\makebox[\textwidth][c]{
\subfloat{\includegraphics[width=75mm,trim=2.5cm 8cm 2cm 6cm,angle=270]{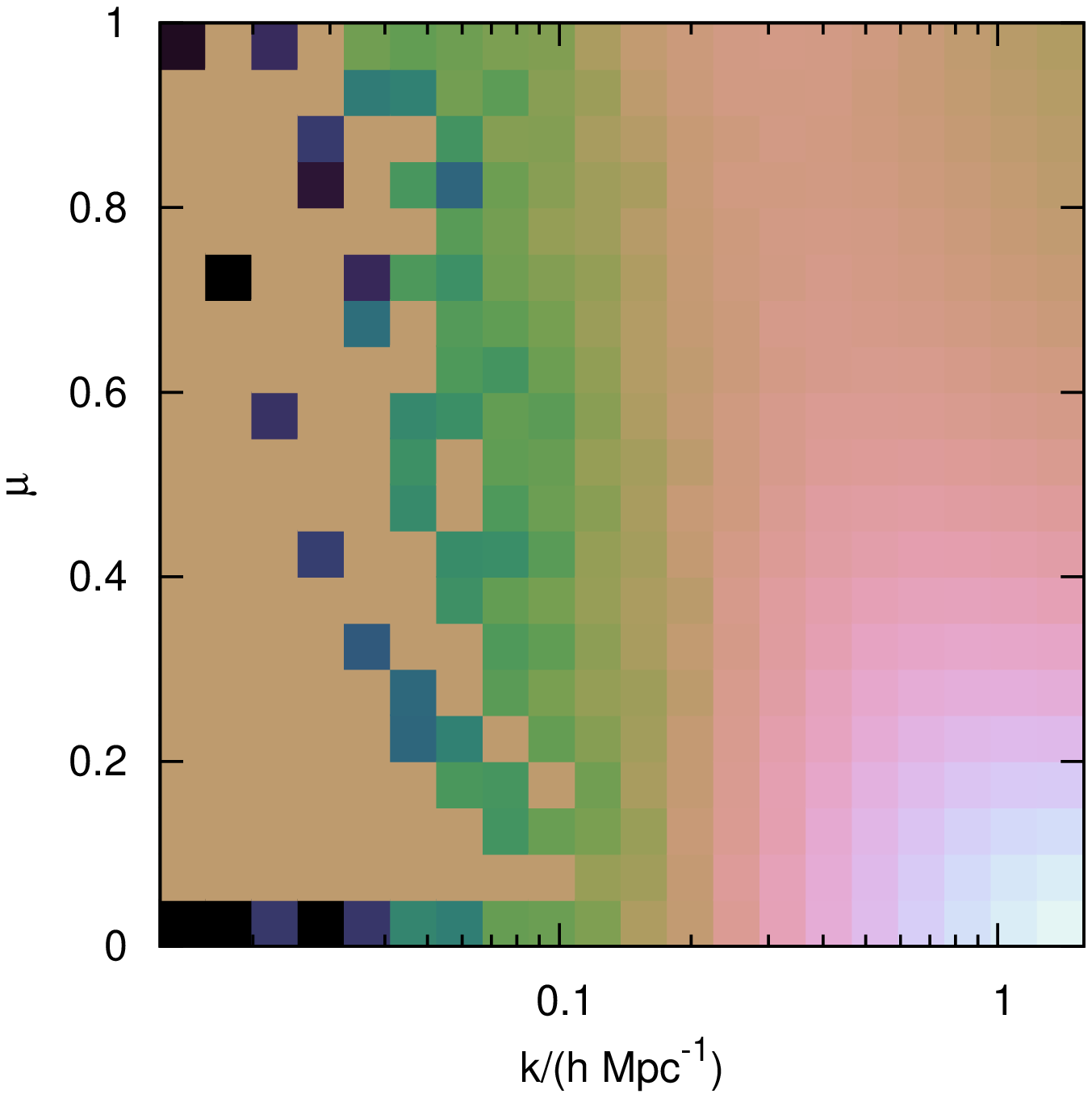}}\hspace{2cm}
\subfloat{\includegraphics[width=75mm,trim=2.5cm 9cm 2cm 6cm,angle=270]{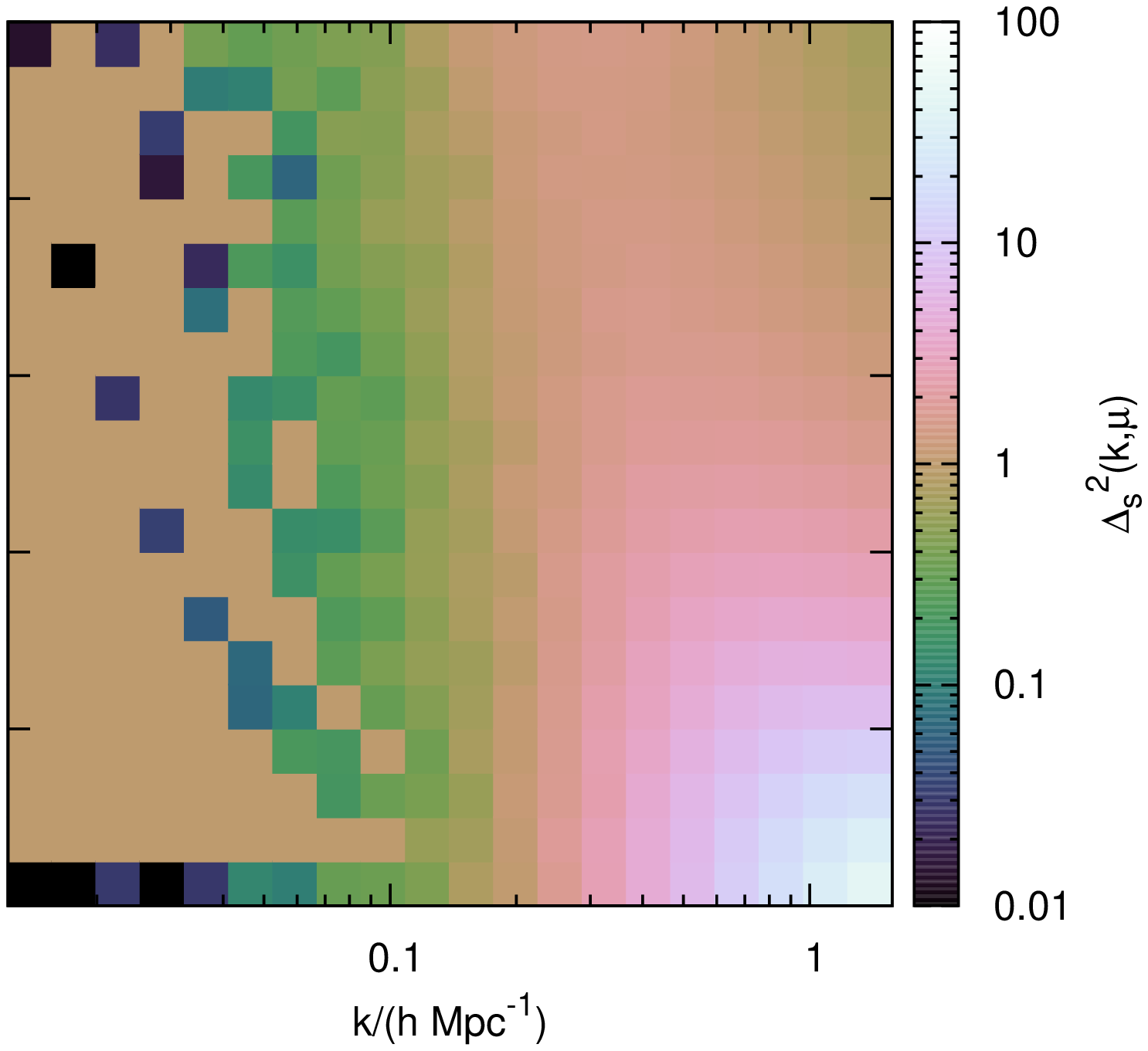}}}
\end{center}
\caption{The 2D redshift-space power spectrum as a function of $k$ and $\mu$, measured in a rescaled and restructured $\Lambda$CDM particle distribution (left) compared to the full simulation of the F5 model (right). The sparse sampling at low $k$ is due to the Cartesian geometry of the finite simulation cube. Differences are difficult to detect by eye and we therefore show the residuals in Fig. \ref{fig:scaling_rsdpower}.}
\label{fig:F5_rsdpower}
\end{figure*}

The scaling parameters $s$ and $z$ are chosen so as to match $\sigma(R)$ across a range of scales but this does not guarantee that the power spectrum will be exactly matched. Fig.~\ref{fig:linear_power_residuals} shows the residual theoretical \emph{linear} power, where it can be seen that the rescaled spectra match at the $\simeq 10$ per cent level around the Baryon Acoustic Oscillation (BAO) scale ($k\simeq 0.1\iMpc$) where residual BAO differences can be seen. However, the linear clustering is different by as much as $30$ per cent at scales around the box size at by as much $20$ per cent at small scales ($k=1\iMpc$).

Within the framework of the halo model (\eg \citealt{Peacock2000}; \citealt{Seljak2000}; \citealt{Cooray2002}) the full matter power spectrum can be considered to be a linear term (two-halo) plus a term due to haloes and their internal structure (one-halo). The first portion of the one-halo term should already be accounted for due to the mach in mass functions but the two-halo part has not yet been addressed. The linear differences seen in Fig.~\ref{fig:linear_power_residuals} can be corrected for by applying the \citeauthor{Zeldovich1970} Approximation (\citeyear{Zeldovich1970}; ZA) to perturb particle or halo positions using the displacement field: the phase of each mode of the field is preserved, but the amplitude is altered to match the target power spectrum.

The displacement field $\mathbf{f}$ is defined so as to move particles from their initial Lagrangian positions $\mathbf{q}$ to their comoving Eulerian positions $\mathbf{x}$: 
\begin{equation} 
\mathbf{x}=\mathbf{q}+\mathbf{f}\ .
\label{eq:q_to_x}
\end{equation}
Within the ZA the displacement field can be related to the over-density via
\begin{equation}
\delta=-\nabla\cdot\mathbf{f}\ .
\label{eq:za}
\end{equation}
If the displacement field in the original simulation is known (it may have been stored) then an additional displacement can be specified in Fourier Space to reflect the differences in the linear matter power spectra between the two cosmologies:
\begin{equation}
\delta\mathbf{f}'_{\mathbf{k'}}=\left[\sqrt{\frac{\Delta_{\rm{lin}}^{'2}(k',z')}{\Delta_{\rm{lin}}^2(sk',z)}}-1\right]\mathbf{f}'_{\mathbf{k'}}\ ,
\label{eq:move}
\end{equation}
where $\mathbf{f}'$ is the linear field in the original simulation \emph{after} it has been scaled. MG models have a scale-dependent growth factor which can be included in this step because the displacement field is scaled mode-by-mode. Particles can then be differentially displaced to account for the differing linear power spectra: $\mathbf{x}''=\mathbf{x}'+\delta\mathbf{f}'$.

If the displacement field is not known then it must be reconstructed from the evolved simulation output. This is discussed in detail in MP14a where it was shown that the particles in the evolved original simulation can be used to reconstruct the overdensity field via the linear relation between overdensity and displacement field in equation~(\ref{eq:za}). Since this is only valid for the linear components, the fields must be smoothed to remove the non-linear components. To do this we use a Gaussian filter, $\exp(-k^2 R_\mathrm{nl}^2 /2)$, and define a non-linear scale $R'_\mathrm{nl}$ such that 
\begin{equation}
\sigma'(R'_\mathrm{nl},z')=1\ ;
\label{eq:nl_scale}
\end{equation}
all fluctuations on scales larger than this are considered to be in the linear regime. A non-linear wavenumber can then be defined: $k_\mathrm{nl}=R_\mathrm{nl}^{-1}$ and this determines which Fourier components of the density field and displacement field are taken to be in the linear regime.

The ZA also allows residual differences in linear velocities to be corrected on a mode-by-mode basis. In the ZA the peculiar velocity field is related to the displacement field by $\mathbf{v}=aHf_\mathrm{g}\mathbf{f}$ and additional differential changes to the peculiar velocities of particles or haloes are then given by
\begin{equation}
\eqalign{
\delta\mathbf{v}'_{\mathbf{k}'}=a'H'f'_\mathrm{g}(k'_\mathrm{b}&,z')\times\cr
&\left[\frac{f'_\mathrm{g}(k',z')}{f_\mathrm{g}(sk',z)}\sqrt{\frac{\Delta_\mathrm{lin}^{'2}(k',z')}{\Delta_\mathrm{lin}^2(sk',z)}}-1\right]\mathbf{f}'_{\mathbf{k}'}\ .
}
\label{eq:move_v}
\end{equation}
Clearly the scale dependent growth rate in the MG models can be respected at this stage of the method. The final velocities after the displacement field step are then: $\mathbf{v}''=\mathbf{v}'+\delta\mathbf{v}'$.

After these manipulations the linear power and mass function ought to be very similar to those in the target cosmology. In AW10, MP14a and MP14b it was shown that the linear power ($k<0.1\iMpc$) can be matched at the $2$ per cent level in both real and redshift space. The results of rescaling the standard gravity particle distribution to HS07 models is shown in Fig.~\ref{fig:scaling_power} where the rescaled matter power spectra residuals are shown together with those of redshift-space monopole and quadrupole power. For the quadrupole we plot the residual $\delta Q/M$ where $\delta Q$ is the difference between rescaled and target quadrupole power and $M$ is the \emph{target} monopole power. This is more meaningful than the ratio of rescaled to target quadrupoles, which blows up around the non-linear scale where the quadrupole changes sign. Additionally Fig.~\ref{fig:scaling_rsdpower} shows the full redshift space residuals as a function of $k$ and $\mu=\cos\theta$ where $\theta$ is the angle of the mode to the line-of-sight.

In Fig.~\ref{fig:scaling_power} the matter power spectrum of particles can be seen to match the HS07 simulations at around the $3$ per cent level up to $k=0.1\iMpc$ for all models at all scales shown but there is an error that roughly scales in proportion to the severity of the necessary linear correction (green curve). The required $20$ per cent correction at the box scale in the F5 case (Fig.~\ref{fig:linear_power_residuals}) translates into a $3$ per cent under-prediction post-rescaling whereas in the GR case the match is better than $1$ per cent. Across the full range of scales, the F4 model seems to be best matched: this is unsurprising given that the chameleon effect is relatively unimportant in this model and it behaves simply as a $\Lambda$CDM model with an enhanced scale-dependent growth rate. However it is surprising that the F4 model is better matched than the GR model, which is the most discrepant of all the models ($10$ per cent at $k=1\iMpc$). Differences are more severe in redshift space where the F4 model disagrees at the level of $15$ per cent around $k=1\iMpc$ despite the near-perfect real space match. Linear scales in redshift space are also slightly less well matched than those in real space, with the maximum error being $4$ per cent in the F5 case at the box scale.

The good match to the HS07 models in real space probably arises because differences in halo physical structure are small when comparing HS07 models to an equivalent standard gravity model (\citealt{Schmidt2009}; \citealt{Lombriser2012a}; \citealt{Lombriser2012c}). However, the monopole and quadrupole display large differences at non-linear scales (around $15$ per cent at $k=1\iMpc$), and this plausibly reflects incorrect Fingers-Of-God (FOG) in the rescaled case, caused by the lack of an enhanced halo velocity dispersion in the rescaled $\Lambda$CDM simulations. This can be seen to be the case in the central column of Fig.~\ref{fig:scaling_rsdpower}, where the power in all non-transverse modes is strongly over-predicted by the scaling in the F4 and F5 cases.

This motivates restructuring halo particles to attempt to correct the small-scale properties. \new{This was achieved in \cite{Angulo2015} by the authors choosing $z$ and $s$ such that the growth history is matched together with $\sigma(R)$. This works because a haloes internal structure depends on its formation history. We take a more brute-force approach and take the NFW profile as a model for haloes in the HS07 cosmologies but manually change the concentration and halo internal velocity dispersions to account for the enhanced gravity (see Section \ref{sec:chameleon}).} The amount of mass enclosed by the NFW profile at a radius $r$ is given by
\begin{equation}
M_\mathrm{enc}(r)=M\frac{F(r/r_\mathrm{s})}{F(c)}\ ,
\label{eq:mass_enclosed}
\end{equation}
where $F$ is defined below equation~(\ref{eq:nfw_enclosed_mass}). Haloes can be reshaped by the ratio of mass enclosed at a radius $r$ from the halo centre in each case. A scaled particle originally at $r'$ should be moved to $r''$, given by
\begin{equation}
r''=F''^{-1}[F'(r')]\ ,
\label{eq:inverse_nfw}
\end{equation}
where $F^{-1}$ indicates the inverse function. $F''$ is the value calculated in the target cosmology whereas $F'$ is the value calculated for the original cosmology \emph{after it has been scaled}. Particle positions relative to the Centre of Mass (CM) $y$ can then be reassigned via
\begin{equation}
\mathbf{y''}=\frac{r''}{r'}\mathbf{y'}\ ,
\label{eq:cm_pos_scaling}
\end{equation}
so that they end up with the correct radial distribution for haloes in the new cosmology, while leaving asphericity unaltered. This also means that the haloes retain a dispersion in internal structure from the parent simulation.

In practice we implement this by calculating $r_\mathrm{v}$ from $M=4\pi r_\mathrm{v}^3\bar{\rho}_\mathrm{m}\Delta_\mathrm{v}/3$ with $\Delta_\mathrm{v}=178$ and using the $c(M)$ relation of \cite{Bullock2001}:
\begin{equation}
c(M,z)=\frac{9}{1+z}\left(\frac{M}{M_*(z)}\right)^{-0.13}\ ,
\label{eq:bullock}
\end{equation}
where $\sigma(M_*,z)=1.686$. This $c(M)$ relation is not the most accurate in the literature but was tuned to simulations with a wide variety of cosmological parameters. Because the changes we implement are differential we view the coverage of parameter space in the \cite{Bullock2001} relation as more important than accuracy. The concentration of haloes has been found to be only slightly enhanced in HS07 haloes compared to those in GR (\citealt{Lombriser2012a}) and an enhancement comes out of equation~(\ref{eq:bullock}) naturally because $M_*$ is lower in these models compared to the equivalent standard gravity model. Physically, the increased concentration can be attributed to haloes forming at slightly earlier times when gravity is enhanced. But differences should be also arise due to the different gravity law and halo velocity structure.


An enhanced halo velocity dispersion $\sigma_v$ for unscreened haloes can be seen in simulations (\eg \citealt{Schmidt2010}; \citealt{Arnold2014}) and this can be attributed to enhanced gravitational forces. Therefore halo particle peculiar velocities, $\mathbf{u}$, relative to the CM velocity, can be reassigned via
\begin{equation}
\mathbf{u}''=\frac{\sigma''_v}{\sigma'_v}\mathbf{u}'\ ,
\label{eq:cm_vel_scaling}
\end{equation}
where a theoretical $\sigma_v$ for an NFW profile can be calculated via the virial theorem (\eg MP14b):
\begin{equation}
\eqalign{
\sigma_v^2 &= {GM\over 3r_\mathrm{v}}\; {c[1-1/(1+c)^2-2\ln(1+c)/(1+c)]
\over 2[\ln(1+c) - c/(1+c)]^2} \cr
&\simeq \left[\frac{2}{3}+\frac{1}{3}\left(\frac{c}{4.62}\right)^{0.75}\right]\frac{GM}{3r_\mathrm{v}}\ .}
\label{eq:nfw_dispersion}
\end{equation}
$\sigma_v$ in haloes can then be boosted as a function of mass according to the simple screening model presented in Section~\ref{sec:chameleon} which gives the effective gravitational constant as a function of $M$ and HS07 model parameters.


An example of the full 2D redshift-space power spectrum for the rescaled and restructured $\Lambda$CDM particle distribution compared to the full F5 simulation is shown in Fig.~\ref{fig:F5_rsdpower} where differences are difficult to see by eye. We therefore show the power spectrum fractional residuals of the particle distributions after haloes have been restructured in Figs~\ref{fig:scaling_power} and~\ref{fig:scaling_rsdpower}. These show that restructuring has a small effect on the matter power spectrum, which is as expected given that the $c(M)$ relation changes only slightly. Unfortunately, the near perfect match in the F4 case is degraded slightly by restructuring, resulting in a $5$ per cent error at $k=1\iMpc$. However, the F5, F6 and GR cases are all improved by restructuring with F6 matched almost perfectly above the non-linear scale. There remains a $5$ per cent error post restructuring in the GR case. In contrast, the monopole is improved dramatically with the large non-linear residuals eradicated almost entirely in the F4 and F5 cases. F6 is also improved by the rescaling, but not quite to the same degree ($4$ per cent error at $k=1\iMpc$) whereas the match in the GR case is perversely worsened by restructuring, leaving an $8$ per cent error at the smallest scales shown. Restructuring improves the match to the quadrupole at quasi-linear scales but degrades it somewhat around $k=1\iMpc$. The improvement at quasi-linear scales stems from improving the $\sigma_v$ match, which effects quasi-linear scales for $\mu\sim1$. However at smaller scales Fig.~\ref{fig:scaling_rsdpower} shows that restructuring degrades the quadrupole match and one can see that this is because restructuring leaves errors of different sign at high and low $\mu$ around $k=1\iMpc$, which translate into quadrupole errors since this differences high and low $\mu$.

\subsection{Haloes}

\begin{figure*}
\includegraphics[width=16.2cm,angle=270]{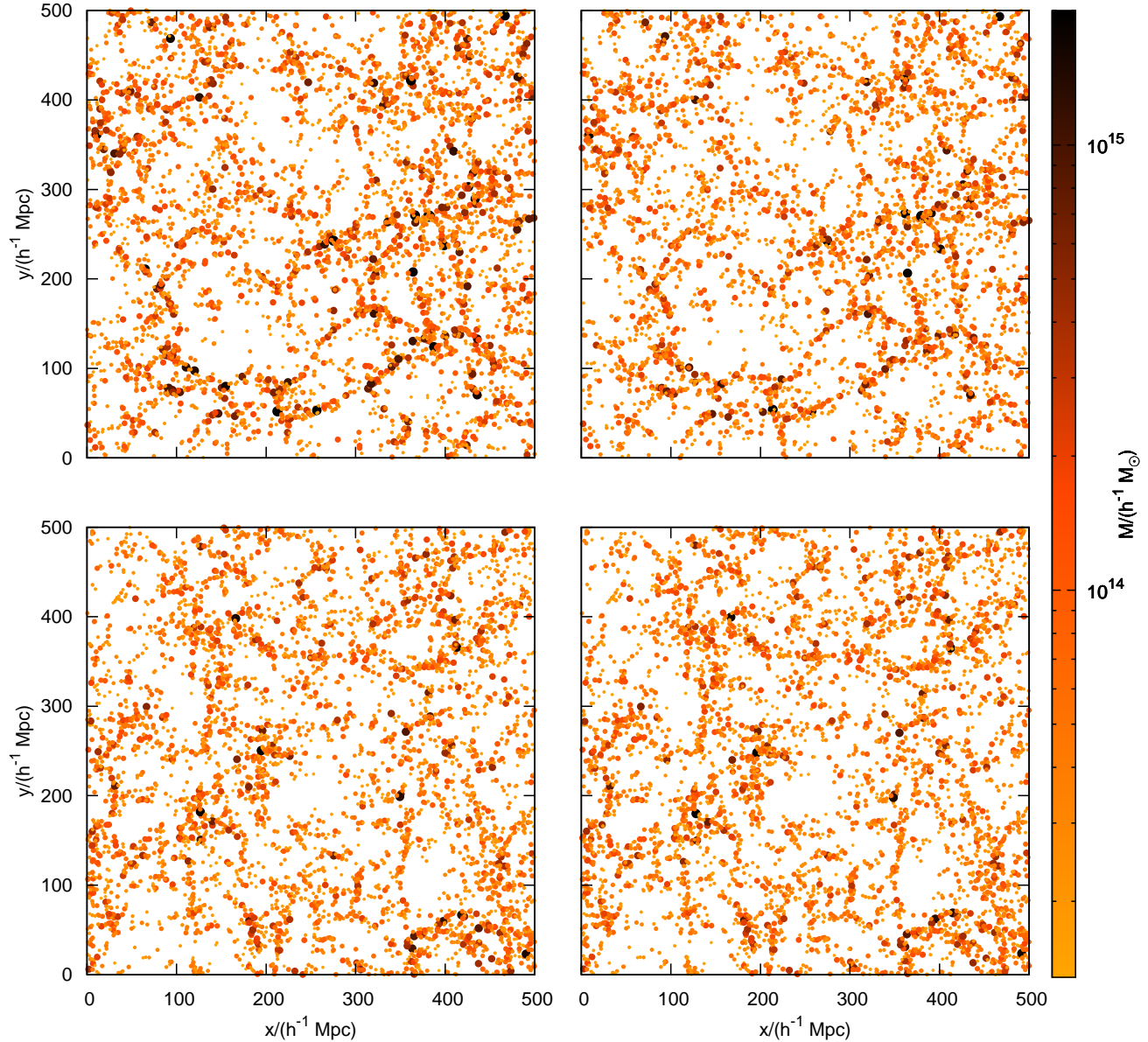}
\caption{A visual summary of the rescaling of a $\Lambda$CDM halo catalogue to an F5 halo catalogue. We show haloes with masses greater than $\sform{1.35}{13}\Msun$ in $50\Mpc$ slices through the simulation volume. Haloes have sizes and colours depending on their masses (small yellow $\sim10^{13}\Msun$; large black $\sim10^{15}\Msun$). The original $\Lambda$CDM simulation at $z=0$ is shown in the top left panel, the top right panel shows this at $z=0.38$, the bottom left panel then shows the result of scaling the box size and adding a displacement field correction. The bottom right panel shows the target halo catalogue from the F5 simulation which was run with the same random seed for the initial conditions. Differences between the lower two panels are difficult to identify visually.}
\label{fig:summary}
\end{figure*}

\begin{figure*}
\begin{center}
\makebox[\textwidth][c]{
\subfloat{\includegraphics[width=41.5mm,trim=1.cm 3.4cm 1.5cm 2.4cm,angle=270]{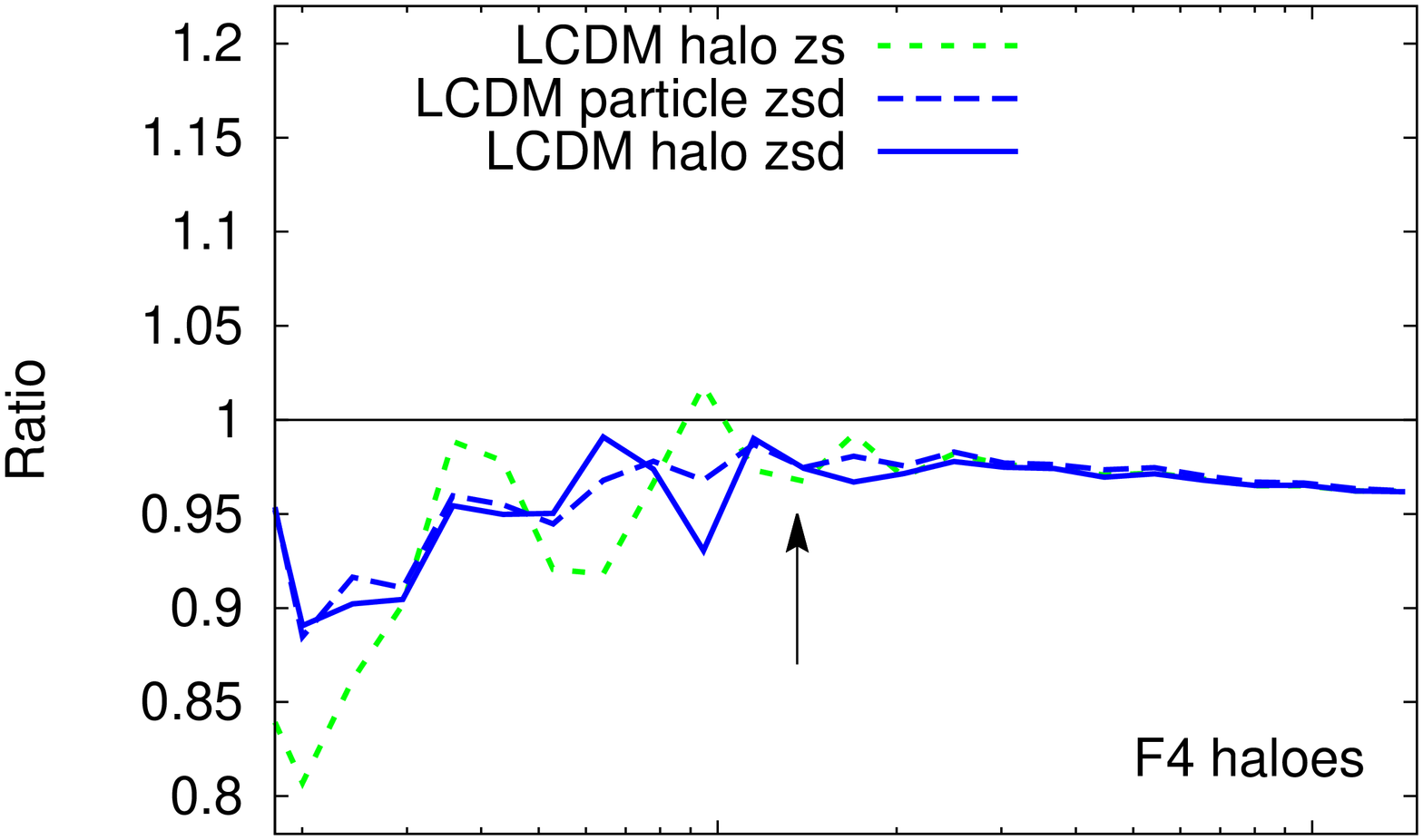}}\hspace{.1cm}
\subfloat{\includegraphics[width=41.5mm,trim=1.cm 3.4cm 1.5cm 2.4cm,angle=270]{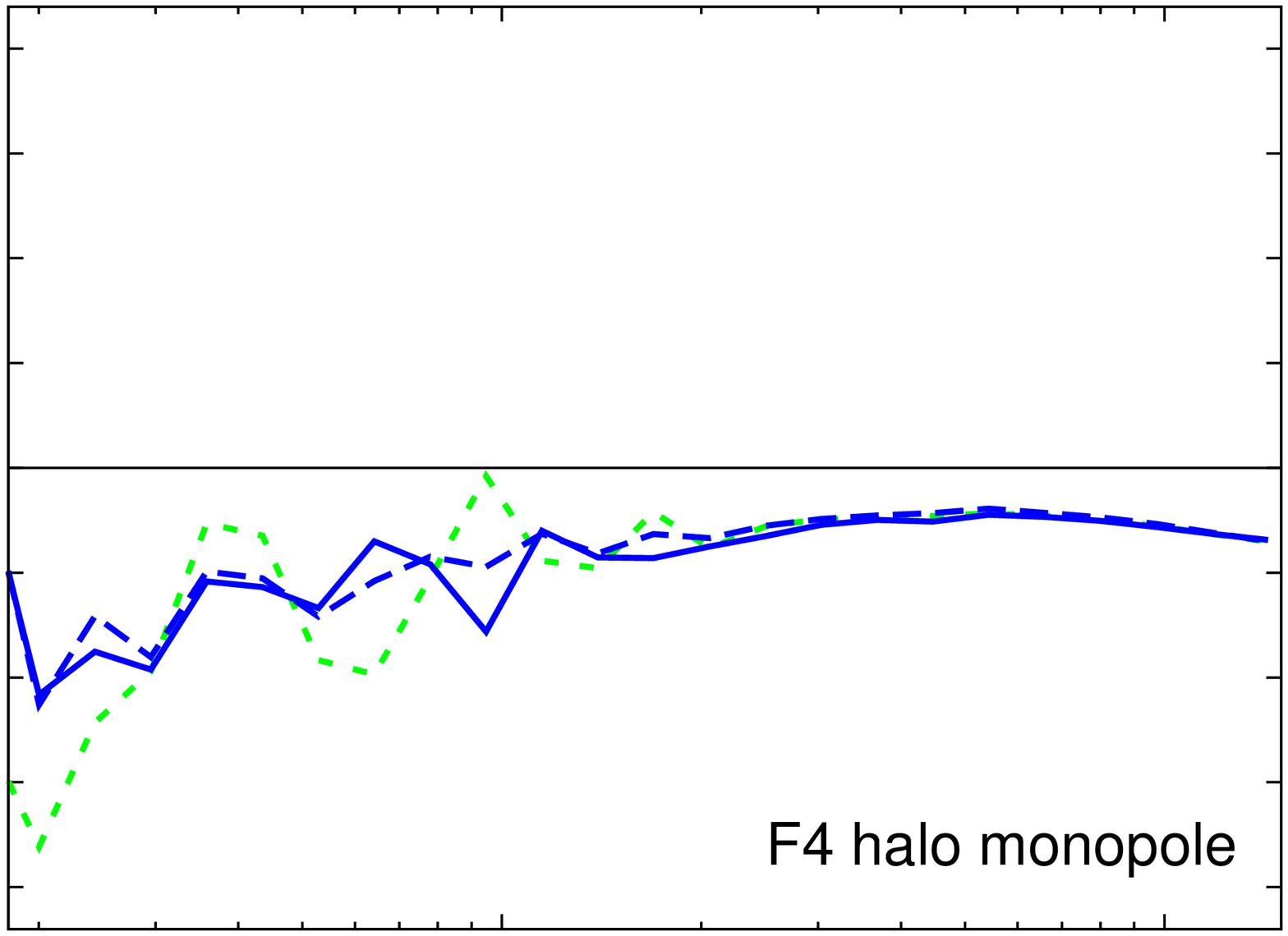}}\hspace{.1cm}
\subfloat{\includegraphics[width=41.5mm,trim=1.cm 3.4cm 1.5cm 2.4cm,angle=270]{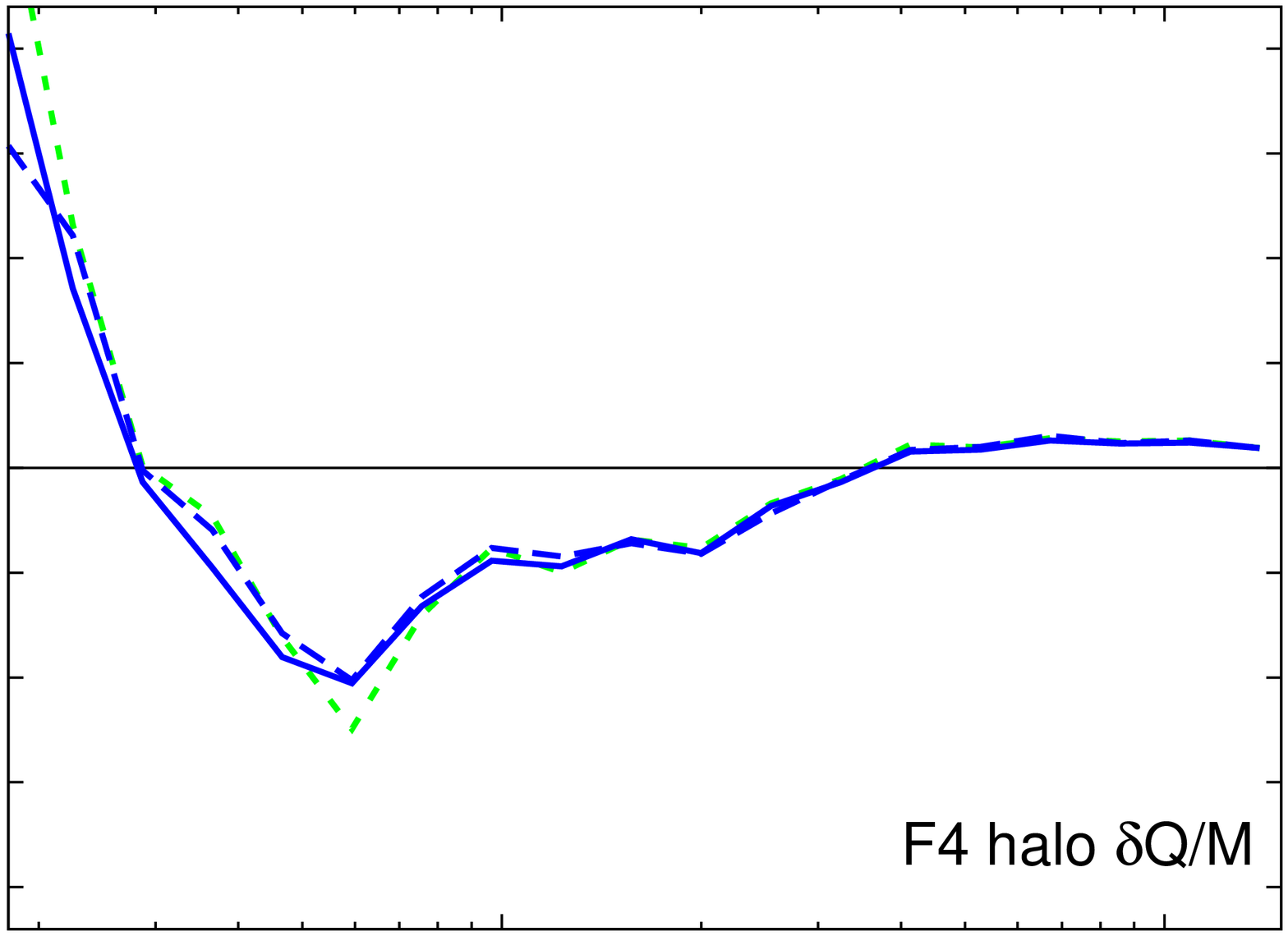}}}\\
\makebox[\textwidth][c]{
\subfloat{\includegraphics[width=41.5mm,trim=1.cm 3.4cm 1.5cm 2.4cm,angle=270]{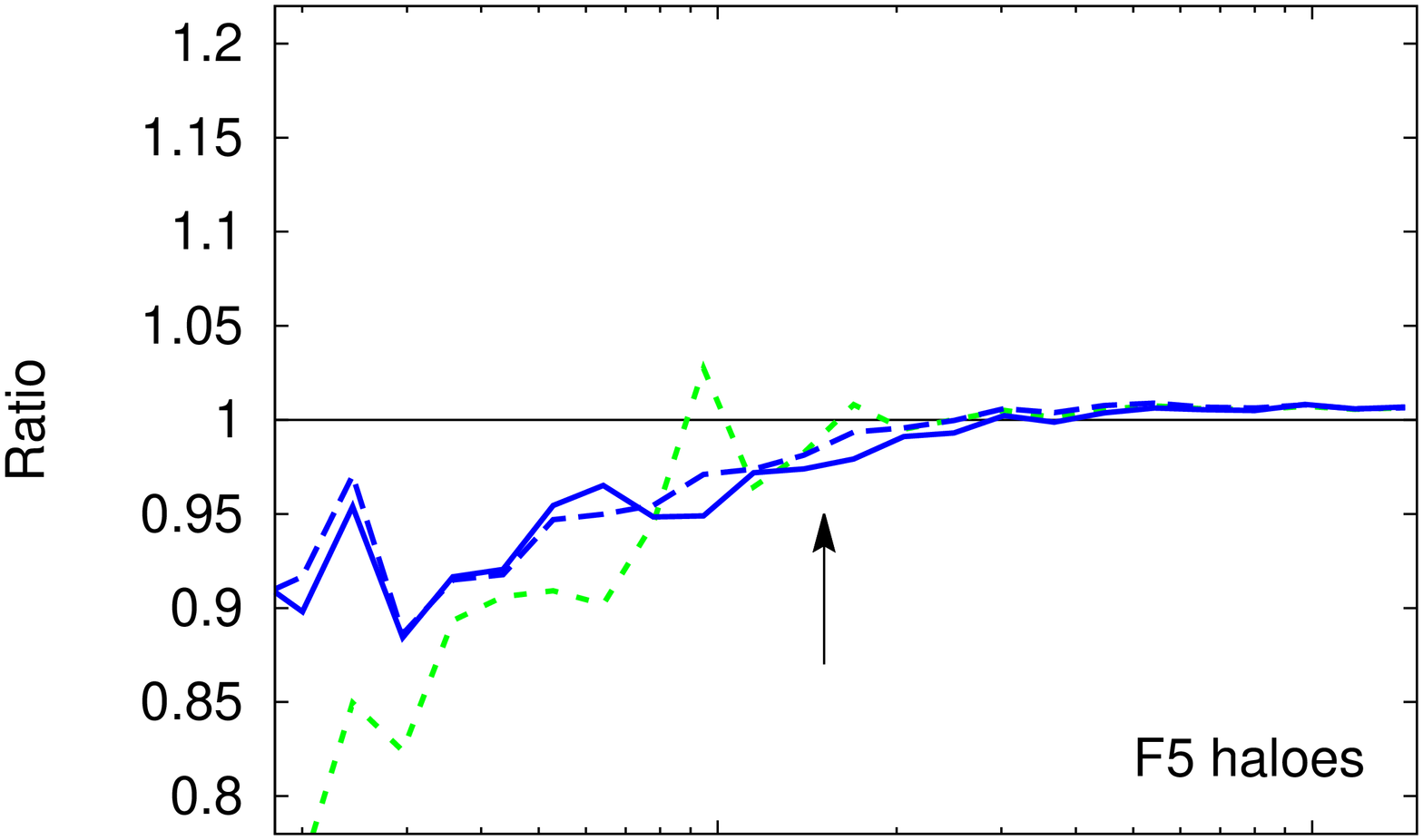}}\hspace{.1cm}
\subfloat{\includegraphics[width=41.5mm,trim=1.cm 3.4cm 1.5cm 2.4cm,angle=270]{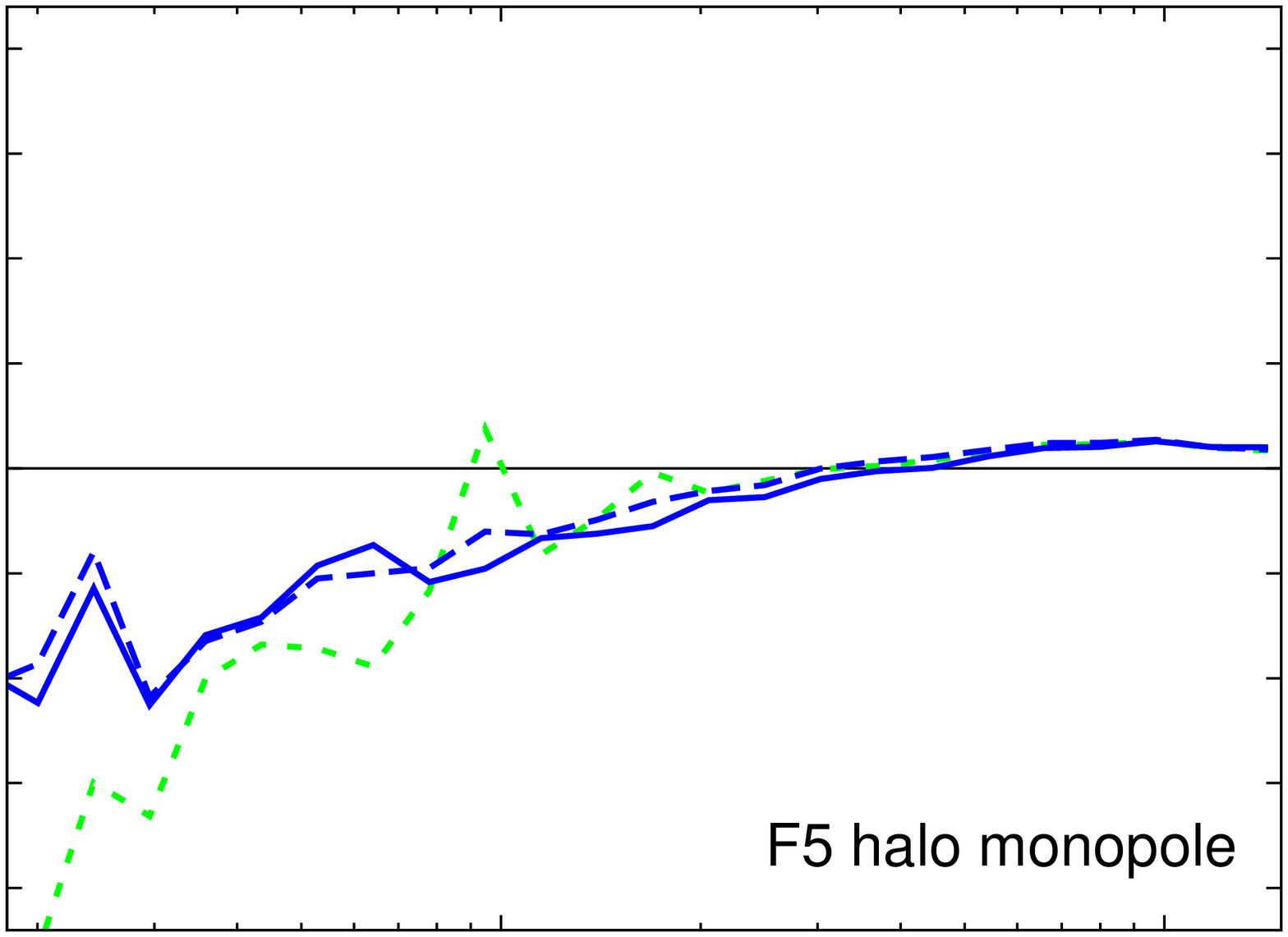}}\hspace{.1cm}
\subfloat{\includegraphics[width=41.5mm,trim=1.cm 3.4cm 1.5cm 2.4cm,angle=270]{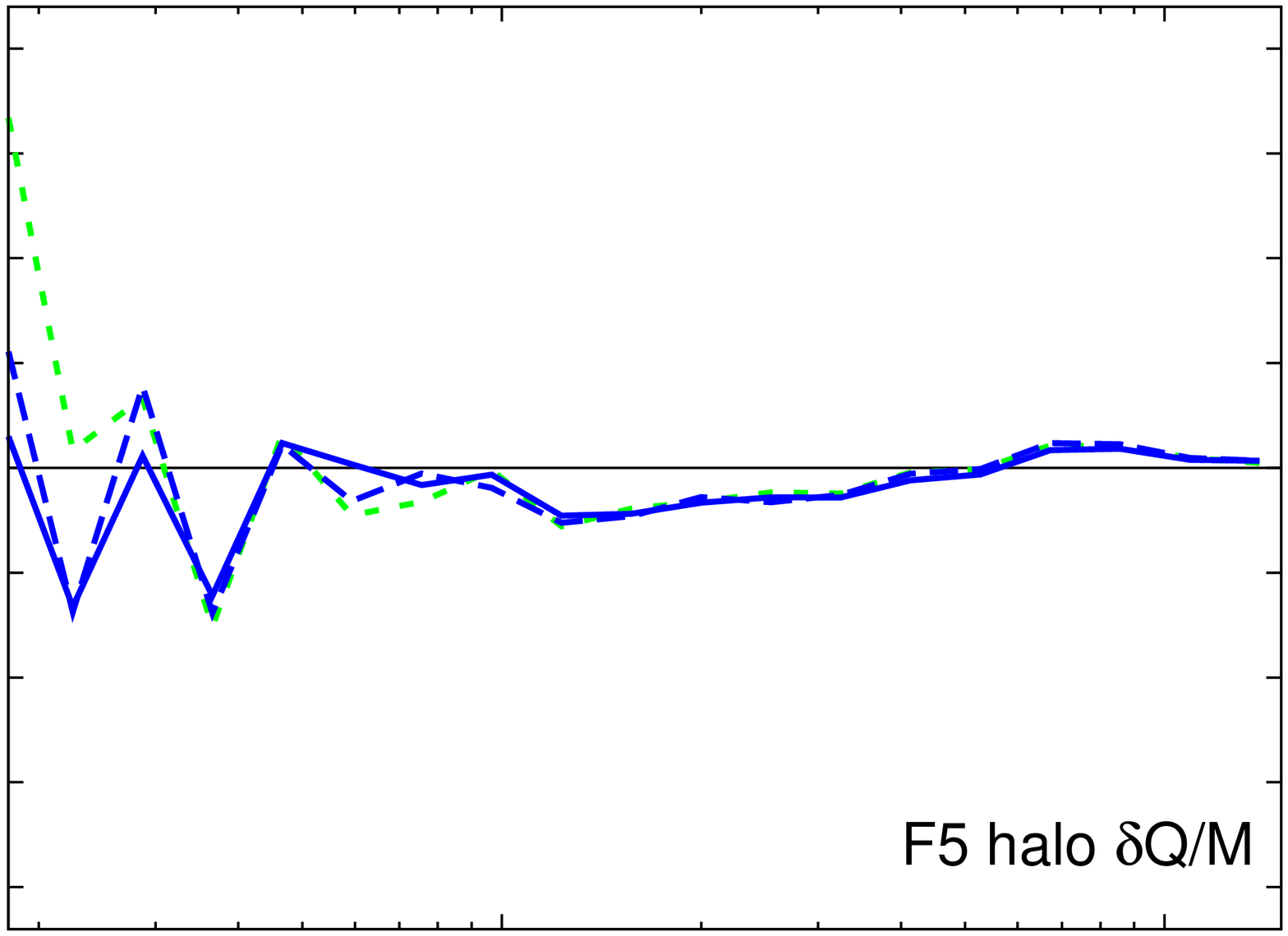}}}\\
\makebox[\textwidth][c]{
\subfloat{\includegraphics[width=41.5mm,trim=1.cm 3.4cm 1.5cm 2.4cm,angle=270]{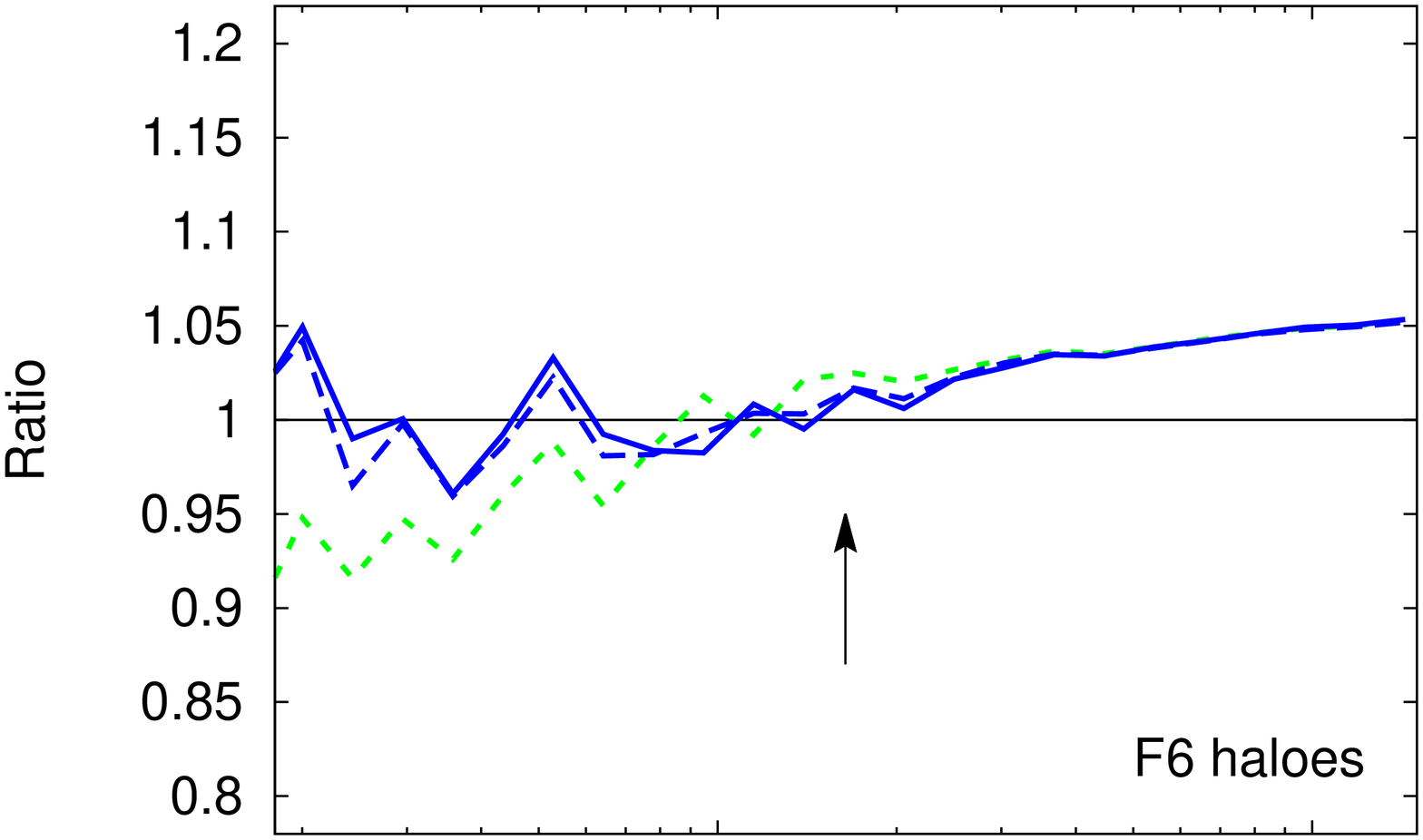}}\hspace{.1cm}
\subfloat{\includegraphics[width=41.5mm,trim=1.cm 3.4cm 1.5cm 2.4cm,angle=270]{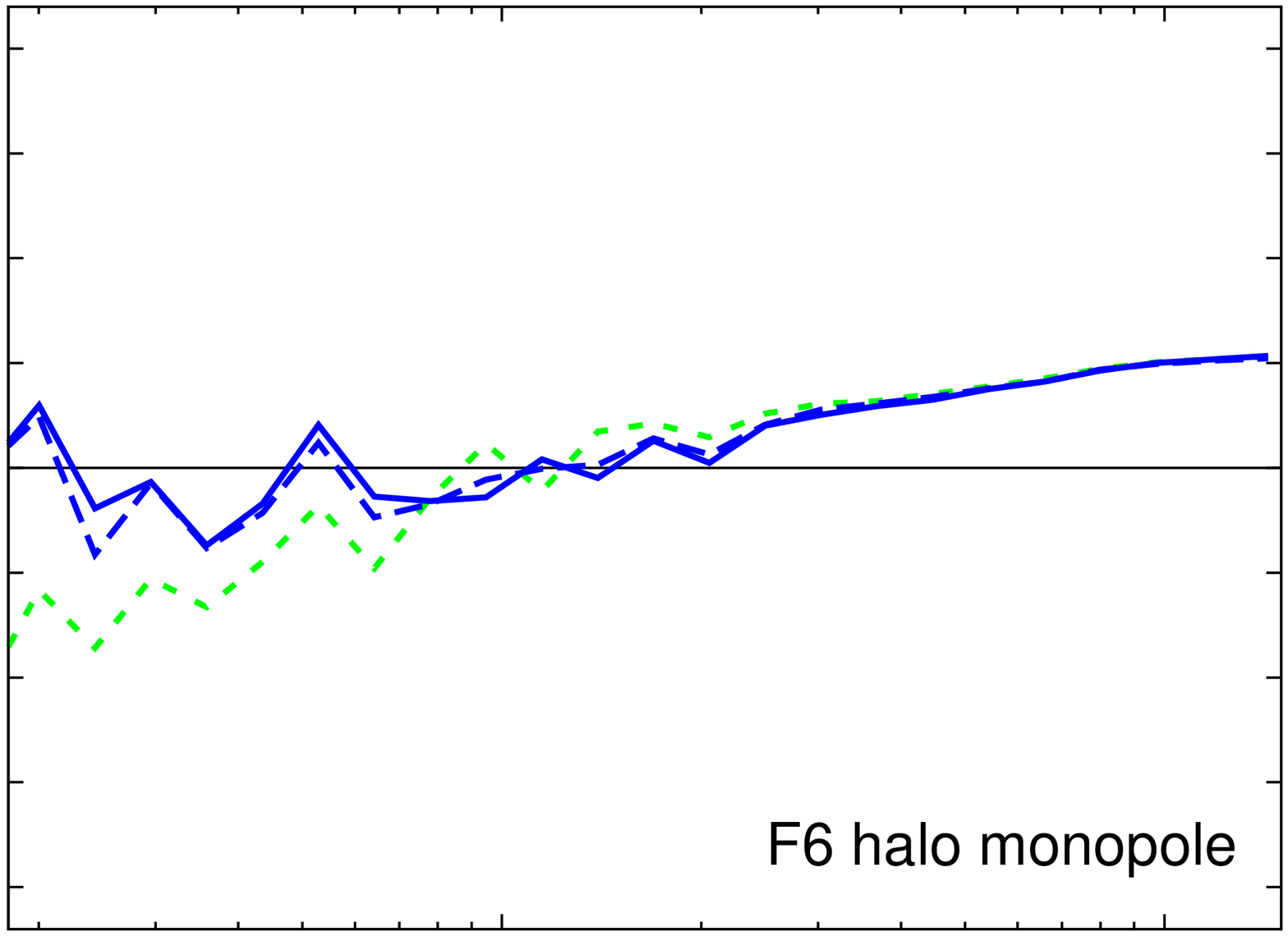}}\hspace{.1cm}
\subfloat{\includegraphics[width=41.5mm,trim=1.cm 3.4cm 1.5cm 2.4cm,angle=270]{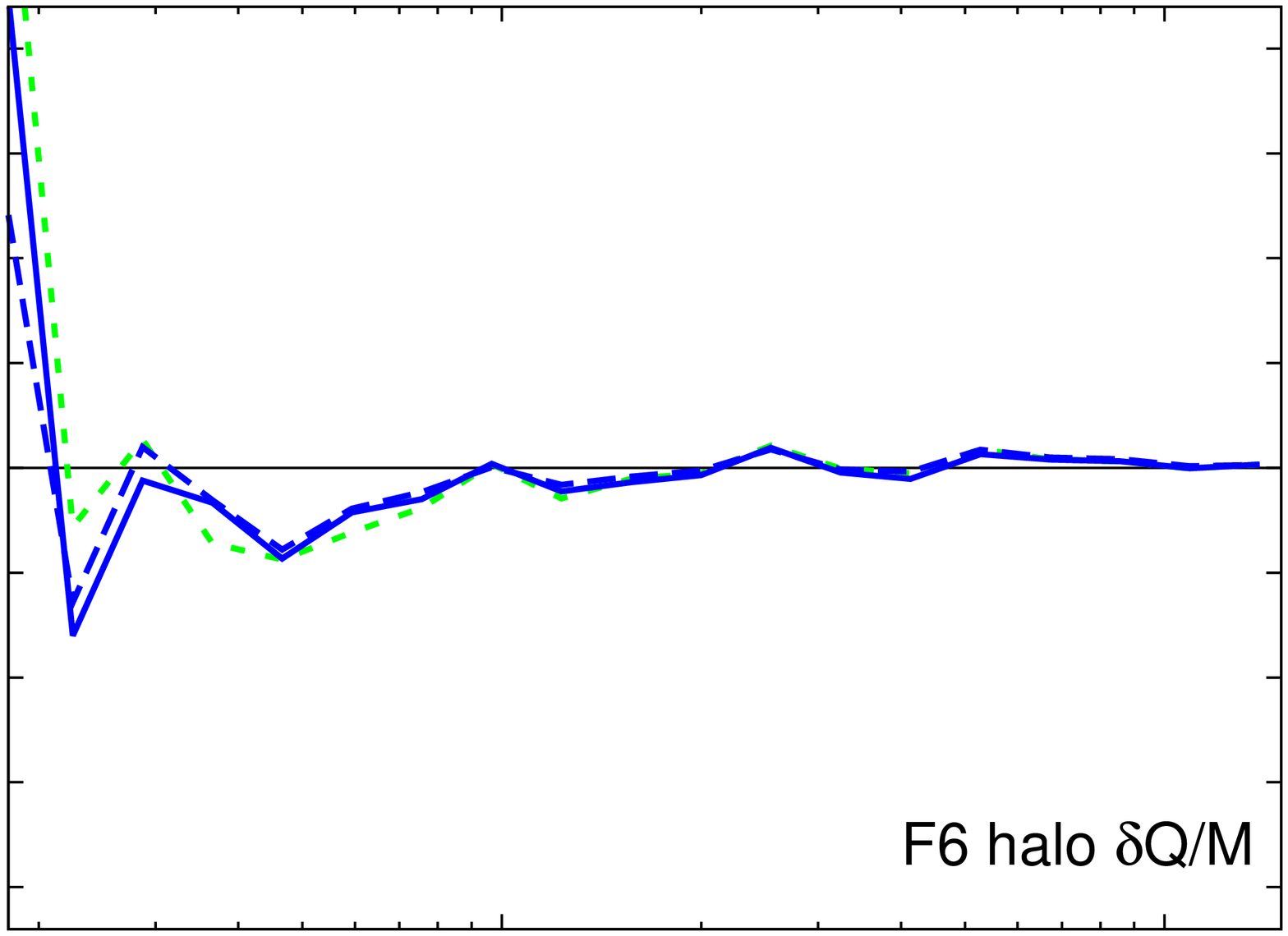}}}\\
\makebox[\textwidth][c]{
\subfloat{\includegraphics[width=41.5mm,trim=1.cm 3.4cm 1.5cm 2.4cm,angle=270]{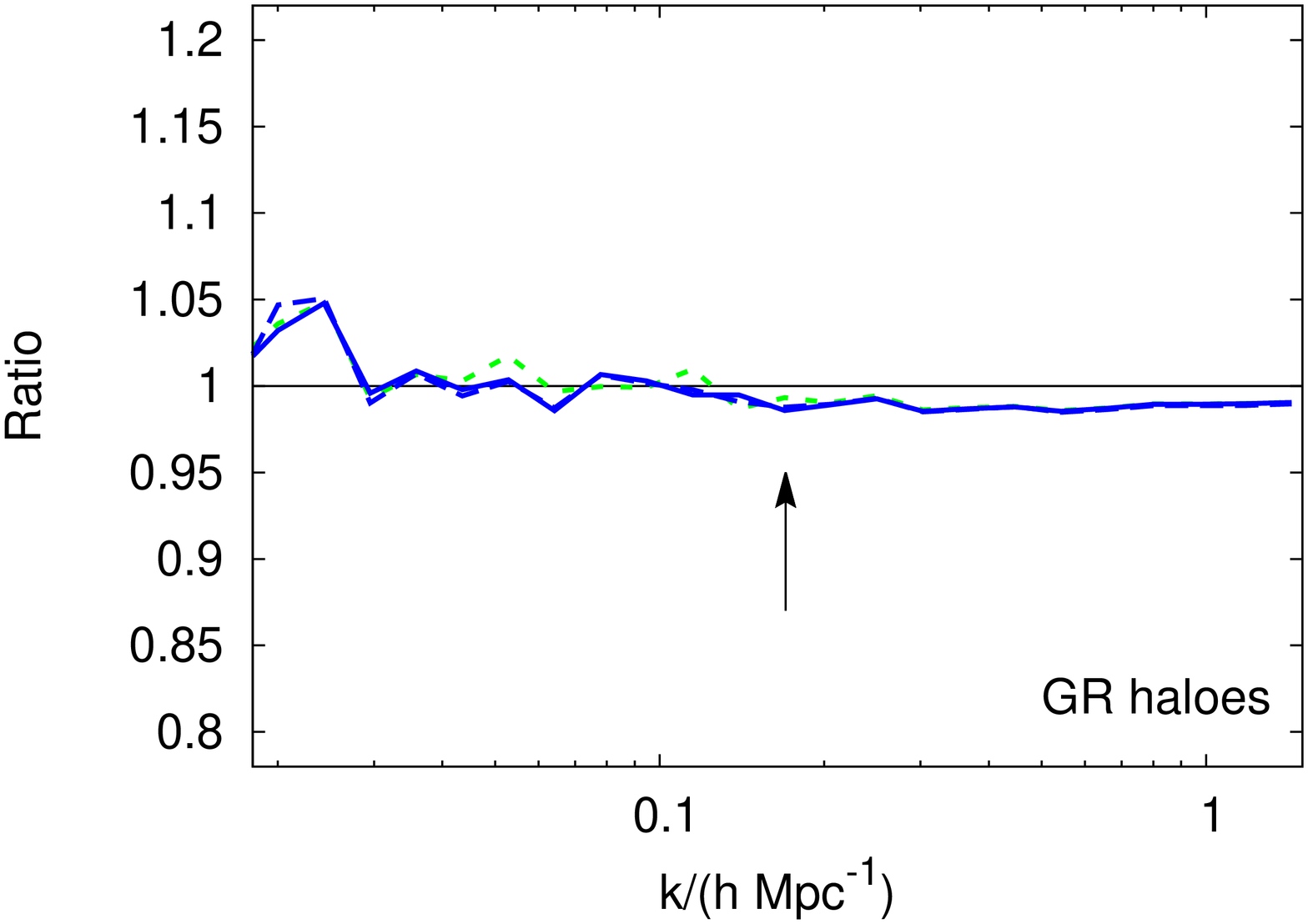}}\hspace{.1cm}
\subfloat{\includegraphics[width=41.5mm,trim=1.cm 3.4cm 1.5cm 2.4cm,angle=270]{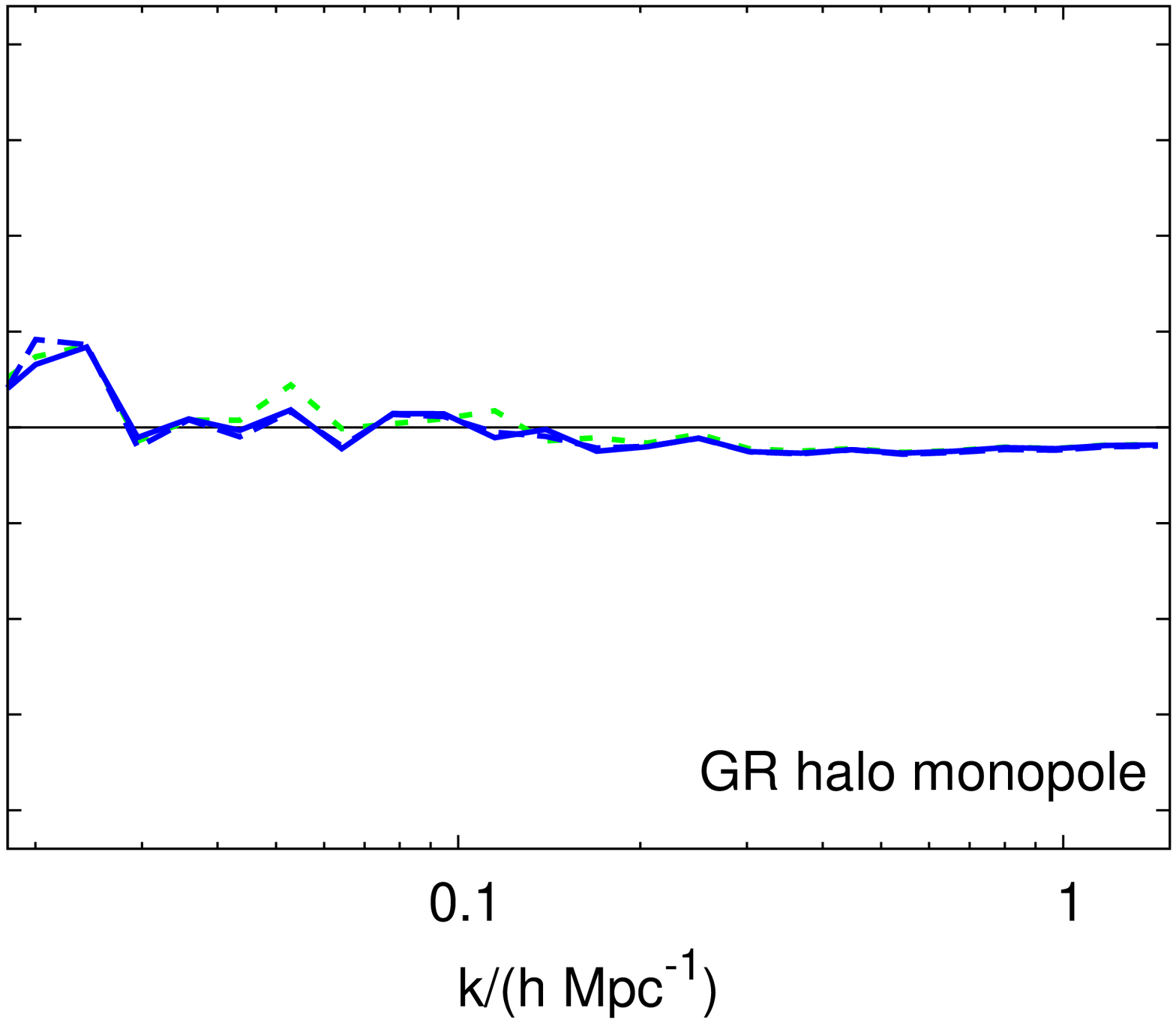}}\hspace{.1cm}
\subfloat{\includegraphics[width=41.5mm,trim=1.cm 3.4cm 1.5cm 2.4cm,angle=270]{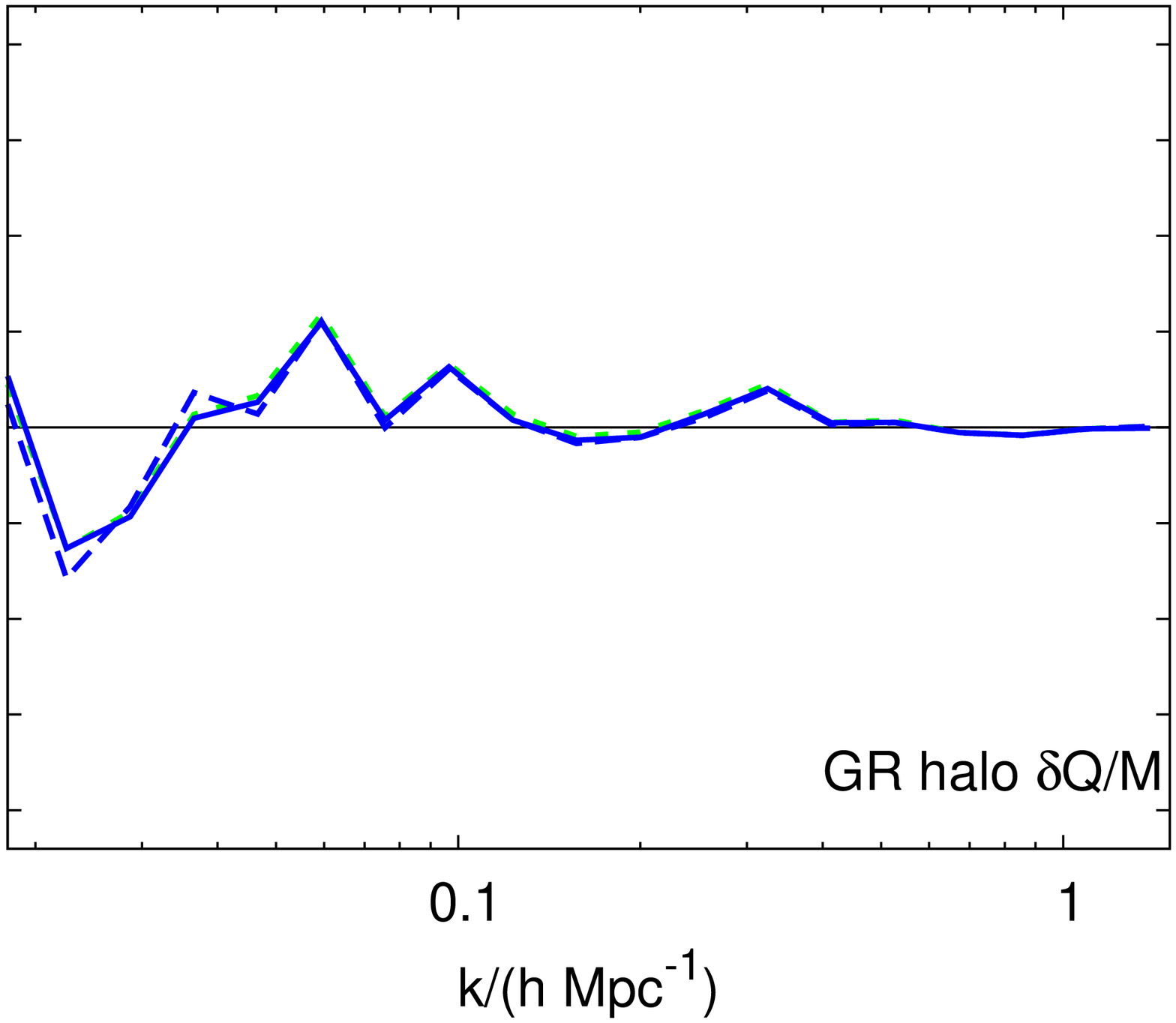}}}\\
\end{center}
\caption{The power spectra of haloes from scaling a $\Lambda$CDM halo catalogue to each of the F4 (top), F5, F6 and GR (bottom) models. The left column shows real space, the central column shows the redshift-space monopole and the right column showing the difference between rescaled and target quadrupole divided by the target monopole. The black arrow shows the non-linear scale that is slightly different for each model. In each case the short-dashed green curve shows the scaling in size and redshift (LCDM $zs$) while the solid blue curve shows the result of applying the additional extra (biased) displacements (LCDM $zsd$). The power measured in the distribution of haloes identified in the rescaled \emph{particle} distribution is also shown (blue long-dash -- particle $zsd$). After the full scaling the power is mainly matched at the $5$ per cent level for most of the scales shown (in both real space and the monopole) but the match is best at small scales and somewhat surprisingly the largest deviations are seen around $k_\mathrm{box}$, the reason for this is unknown. The match obtained using haloes identified in the rescaled particle distribution is very similar to that obtained using the haloes directly.}
\label{fig:halo_scaling_power}
\end{figure*}

The case of rescaling halo catalogues directly is detailed in MP14a and is subtly different from rescaling a full particle distribution: If one is reconstructing the matter displacement field from haloes the halo over density field $\delta_\mathrm{H}$ must be debiased, respecting the relation
\begin{equation} 
\delta_\mathrm{H}=b(M)\delta\ .
\label{eq:density_bias}
\end{equation}
In this work we use the bias relations of \cite{Sheth1999}, within the peak-background split formalism the bias is given in terms of the mass function by:
\begin{equation}
b(\nu)=1-\frac{1}{\delta_{\mathrm c}}\left[1+\nu\frac{\mathrm{d}}{\mathrm{d}\nu}\ln f(\nu)\right]\ ,
\label{eq:PBS_bias}
\end{equation}
where we use the ST $f(\nu)$ given in equation~(\ref{eq:STmf}). The bias is calculated using the appropriate $\sigma(R)$ for each model and this has been shown  by \cite{Schmidt2009} to provide a good match to halo bias seen in HS07 simulations. We use $\delta_\mathrm{c}=1.686$ here although we acknowledge that the spherical model predictions in Fig.~\ref{fig:mg_dc} may be preferable in general. We ignore this because (a) the \cite{Sheth1999} bias has been shown to work well for HS07 models and (b) in any case the changes are quite small. 

In order to reconstruct the matter density field from the halo density field we define a number-weighted effective bias for all haloes
\begin{equation}
b_\mathrm{eff}=\frac{\int_{\nu_\mathrm{min}}^{\nu_\mathrm{max}} \,\mathrm{d}\nu\,
b(\nu)f(\nu)/M(\nu)}{\int_{\nu_\mathrm{min}}^{\nu_\mathrm{max}} \,\mathrm{d}\nu\,
f(\nu)/M(\nu)}\ ,
\label{eq:eff_bias}
\end{equation}
in order to debias, where $M(\nu)$ denotes halo mass as a function of $\nu$. When moving haloes according to the differential displacement field, their displacements must also be biased, so that $\mathbf{f}_\mathrm{H}=b(M)\mathbf{f}$ for each halo. This can be done as a function of mass for each halo individually. In MP14a, good results for the rescaled halo power spectrum were \emph{not} obtained unless a biased displacement field was used and we also bias the differential matter displacement field required to reposition haloes in this work: $\mathbf{x''}=\mathbf{x'}+b(m)\delta\mathbf{f'}$. In contrast to the displacement field, the halo velocity field is unbiased due to the equivalence principle.

In Fig.~\ref{fig:summary} we show a visual summary of the match to the F5 model halo catalogue at all stages of the rescaling. Differences in the final rescaled halo distribution compared to the target are difficult to identify visually. Power spectra of the halo distribution at all stages of the rescaling are shown in Fig.~\ref{fig:halo_scaling_power}. One can see that the halo power is mainly matched at the $5$ per cent level across all scales shown for all models, with the GR match being almost perfect. The largest deviations are seen at the largest scales with the rescaled F4 and F5 models showing a $\sim 10$ per cent deficiency in power in real and redshift space. A similar deviation was seen in MP14a and MP14b although in this paper the largest deviations coincide with the largest necessary displacement field correction. The match improves at small scales and this must in part reflect the lack of a strongly non-linear FOG in the halo distribution. At these scales the power in the halo distribution is governed by shot noise due to finite halo number density and a good small scale match here implies a good match of the number density. Note that at no stage in the rescaling of haloes has the chameleon effect been accounted for and the fact that we obtain good results suggests that the chameleon effect is less important for the halo distribution. The quadrupole of haloes is matched nearly perfectly in the F5, F6 and GR cases but shows $10$ per cent deviations for F4. We also show the power spectra of haloes that are identified in the rescaled \emph{particle} distributions discussed in the previous section. These match the results we obtain using only the haloes but for some noise at large scales and confirm that our methodology for working with only a halo catalogue is sound.


\subsection{Halo particles}

\begin{figure*}
\begin{center}
\makebox[\textwidth][c]{
\subfloat{\includegraphics[width=41.5mm,trim=1.cm 3.4cm 1.5cm 2.4cm,angle=270]{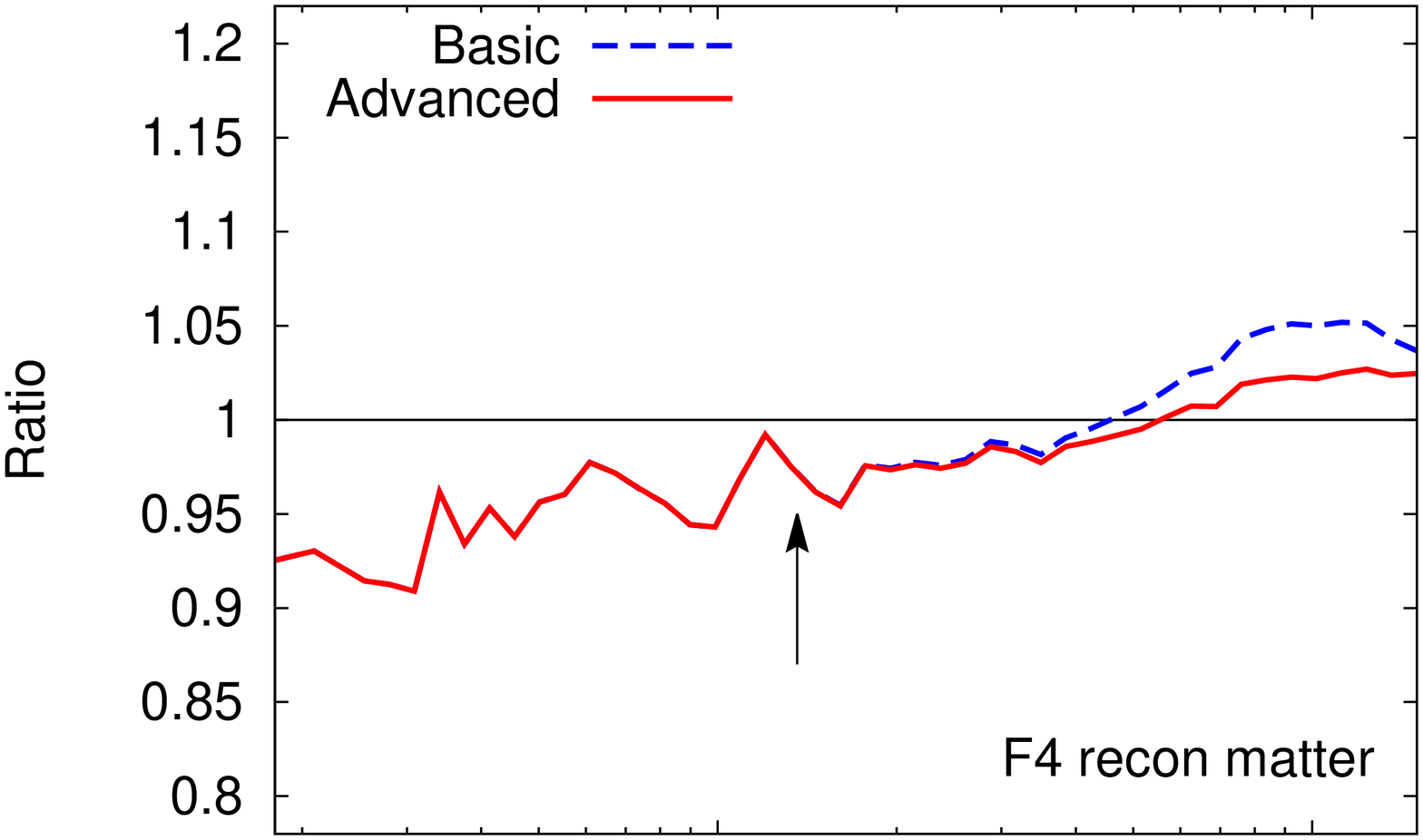}}\hspace{.1cm}
\subfloat{\includegraphics[width=41.5mm,trim=1.cm 3.4cm 1.5cm 2.4cm,angle=270]{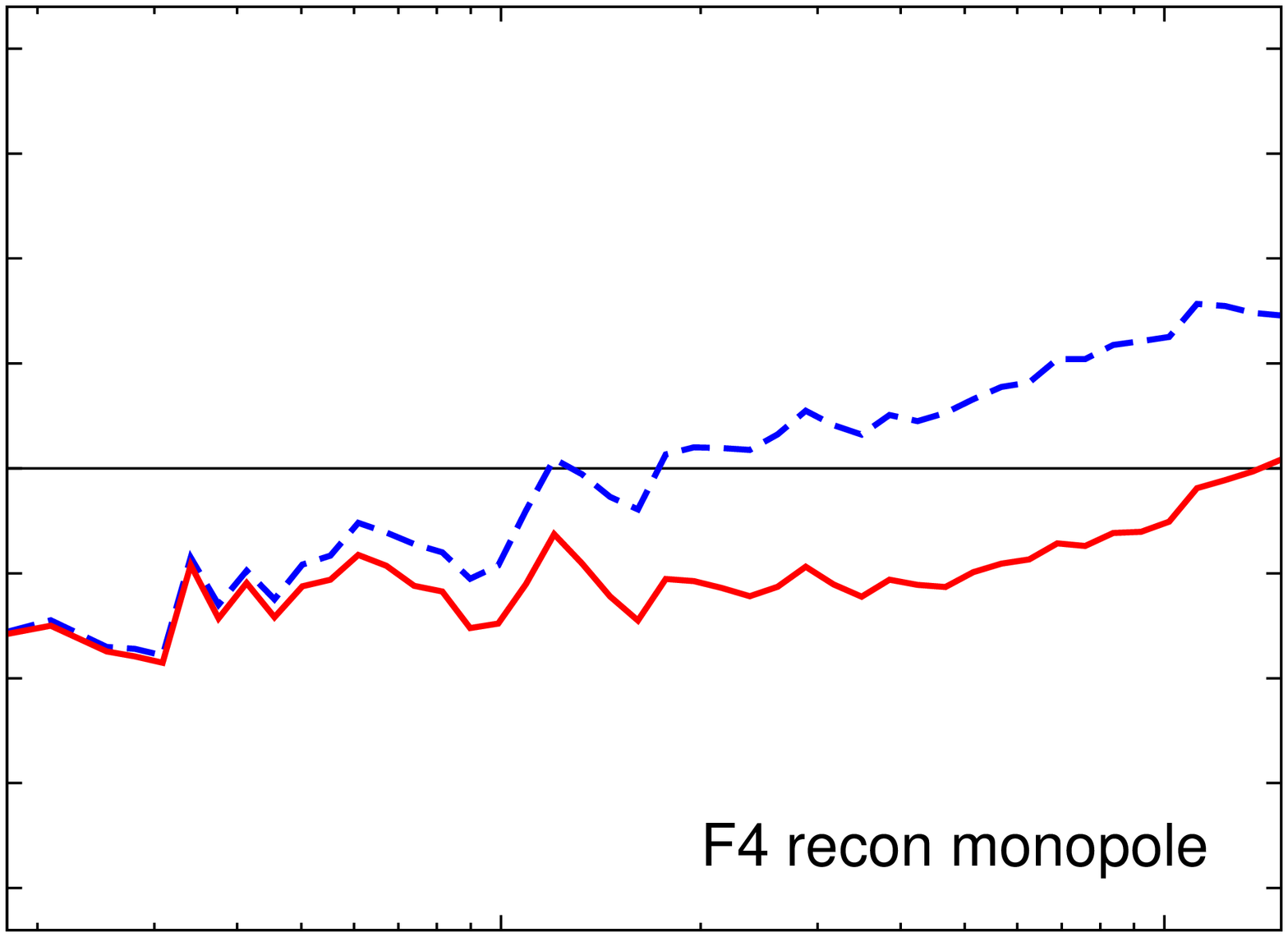}}\hspace{.1cm}
\subfloat{\includegraphics[width=41.5mm,trim=1.cm 3.4cm 1.5cm 2.4cm,angle=270]{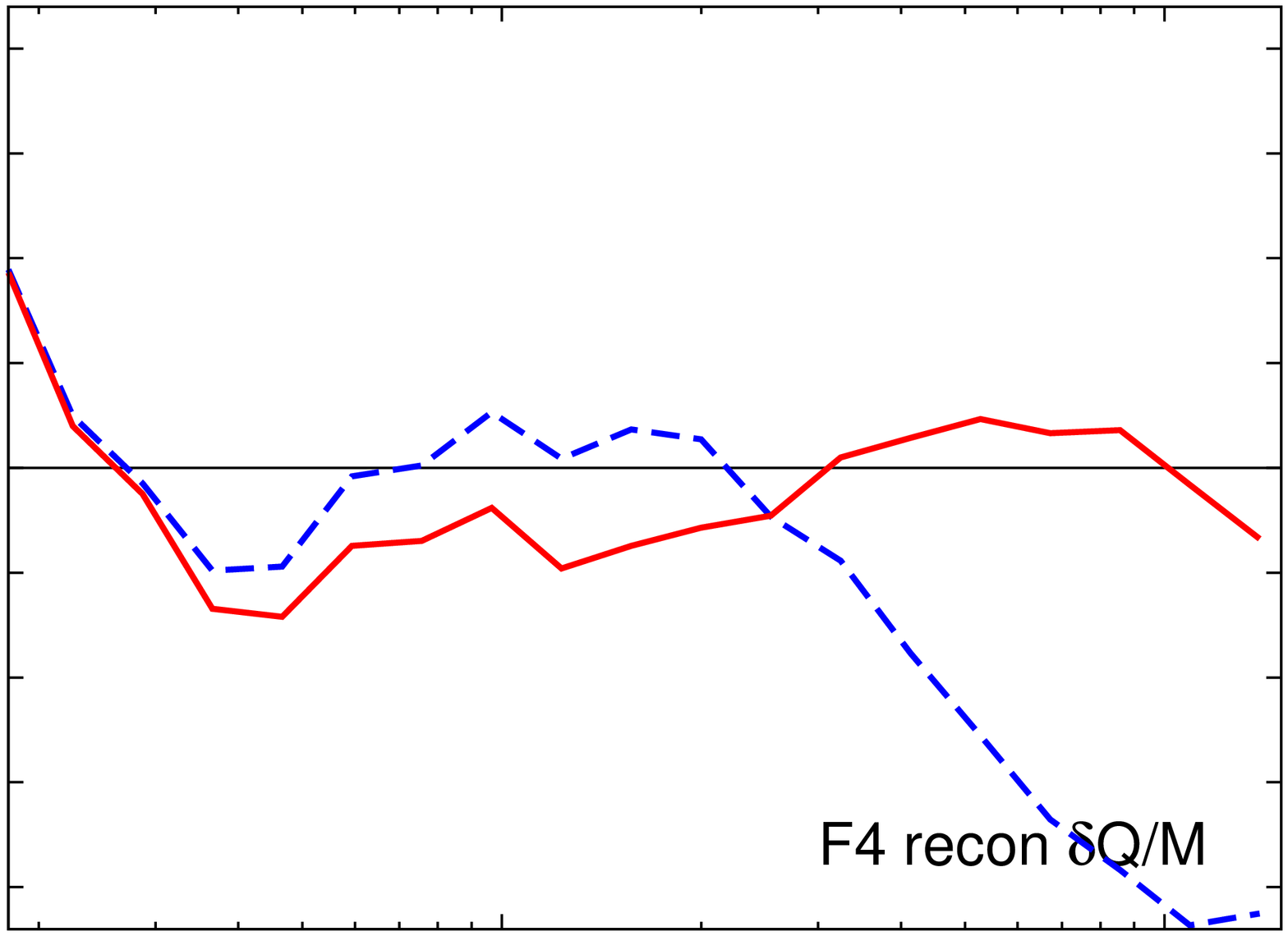}}}\\
\makebox[\textwidth][c]{
\subfloat{\includegraphics[width=41.5mm,trim=1.cm 3.4cm 1.5cm 2.4cm,angle=270]{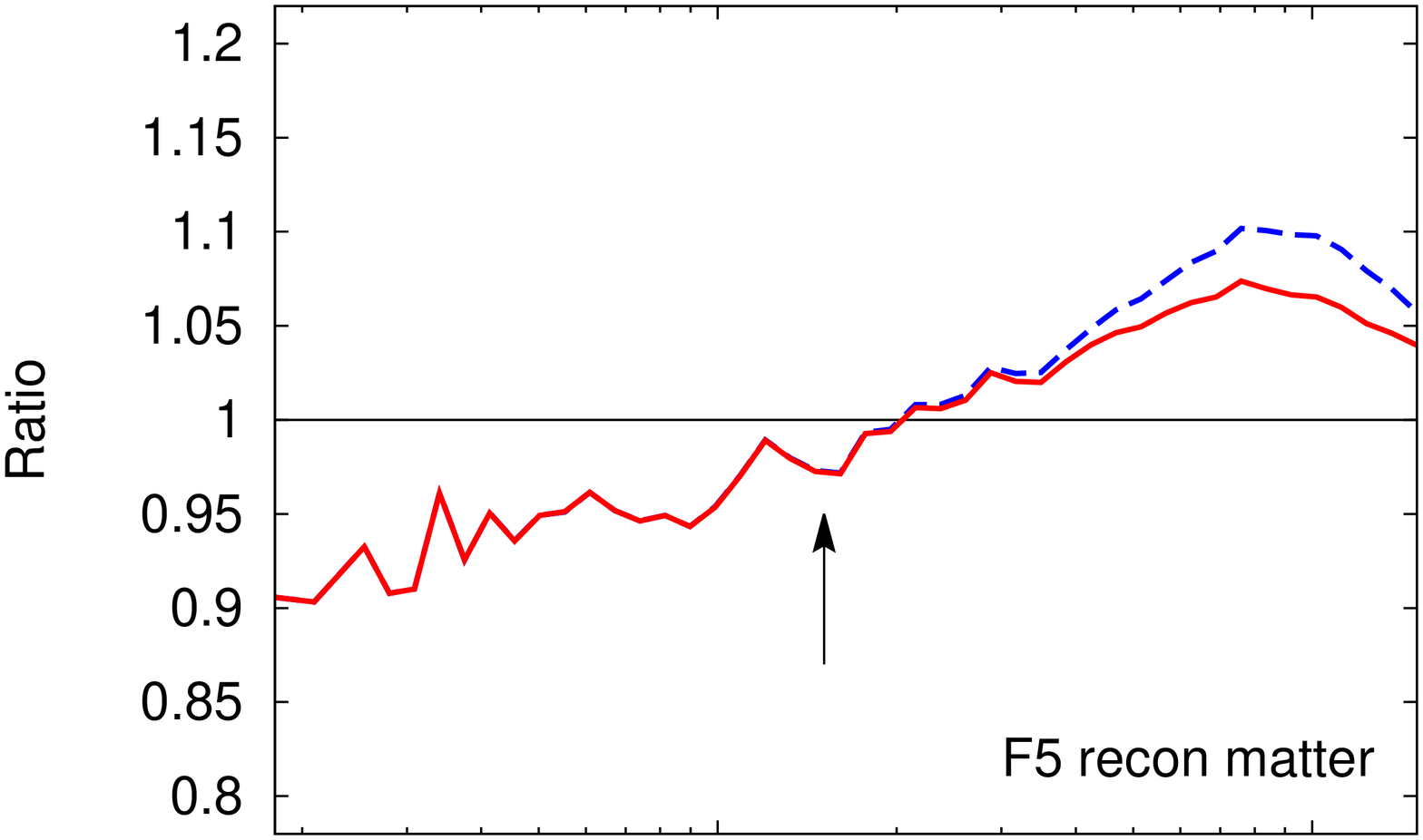}}\hspace{.1cm}
\subfloat{\includegraphics[width=41.5mm,trim=1.cm 3.4cm 1.5cm 2.4cm,angle=270]{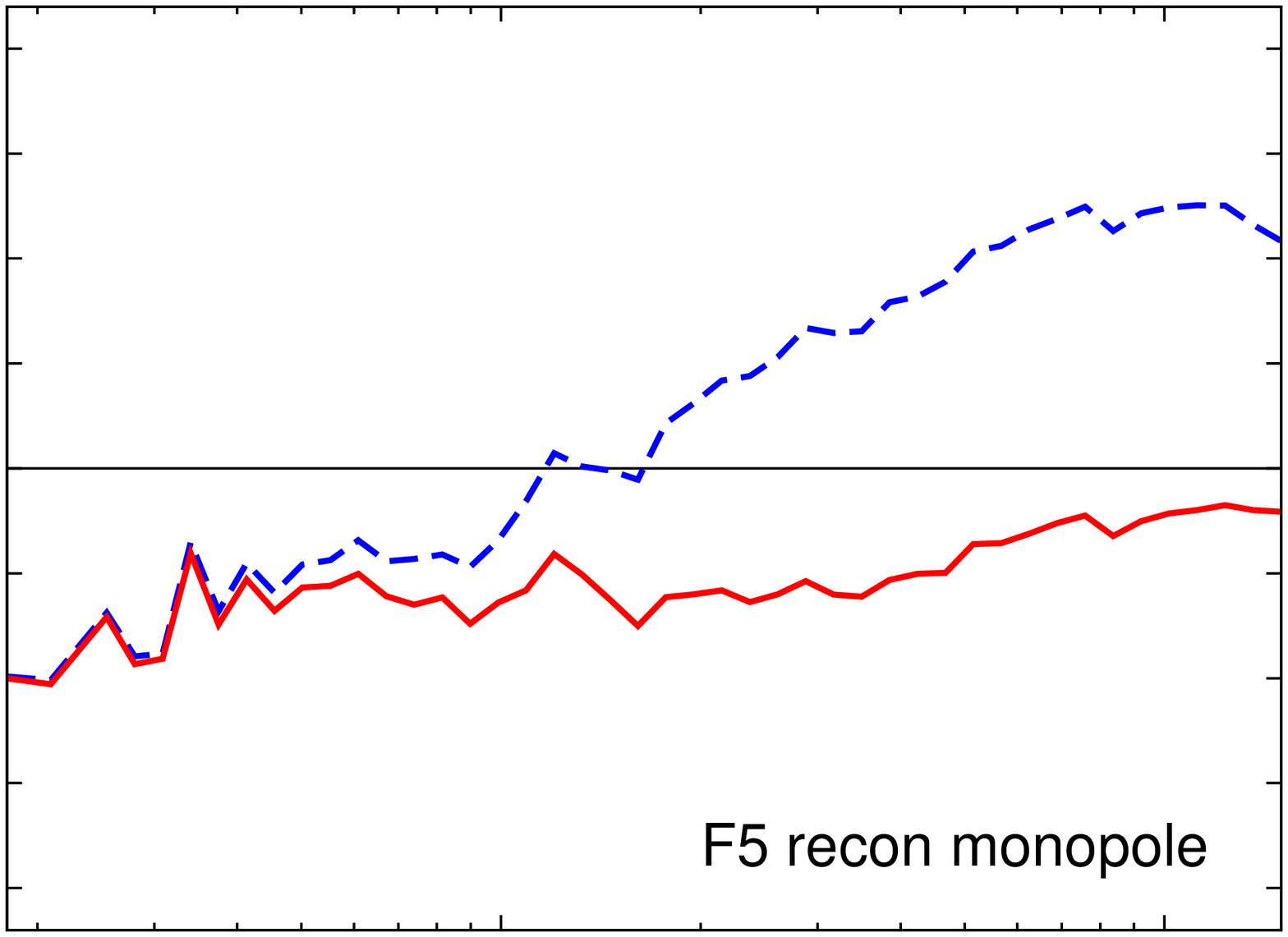}}\hspace{.1cm}
\subfloat{\includegraphics[width=41.5mm,trim=1.cm 3.4cm 1.5cm 2.4cm,angle=270]{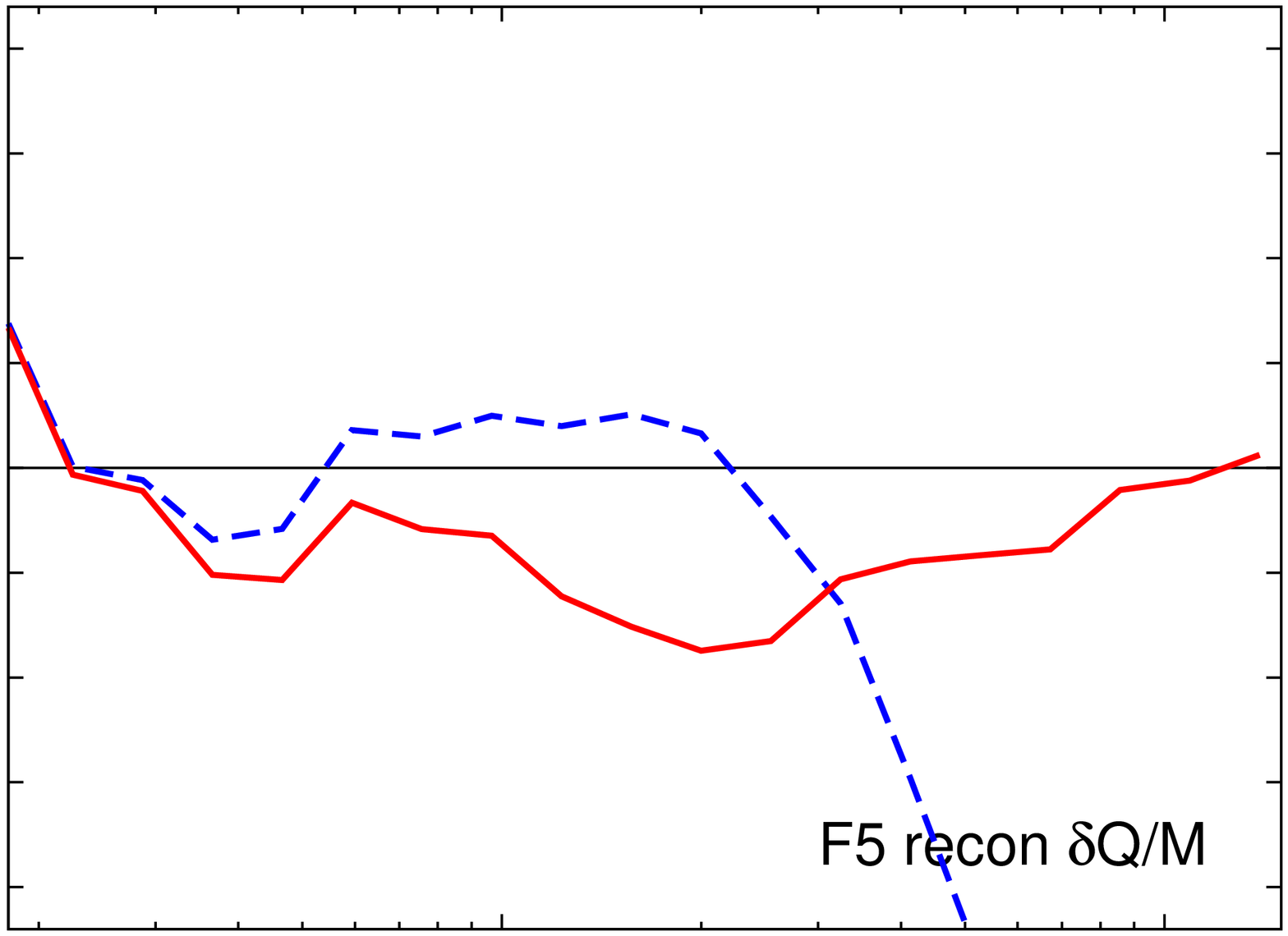}}}\\
\makebox[\textwidth][c]{
\subfloat{\includegraphics[width=41.5mm,trim=1.cm 3.4cm 1.5cm 2.4cm,angle=270]{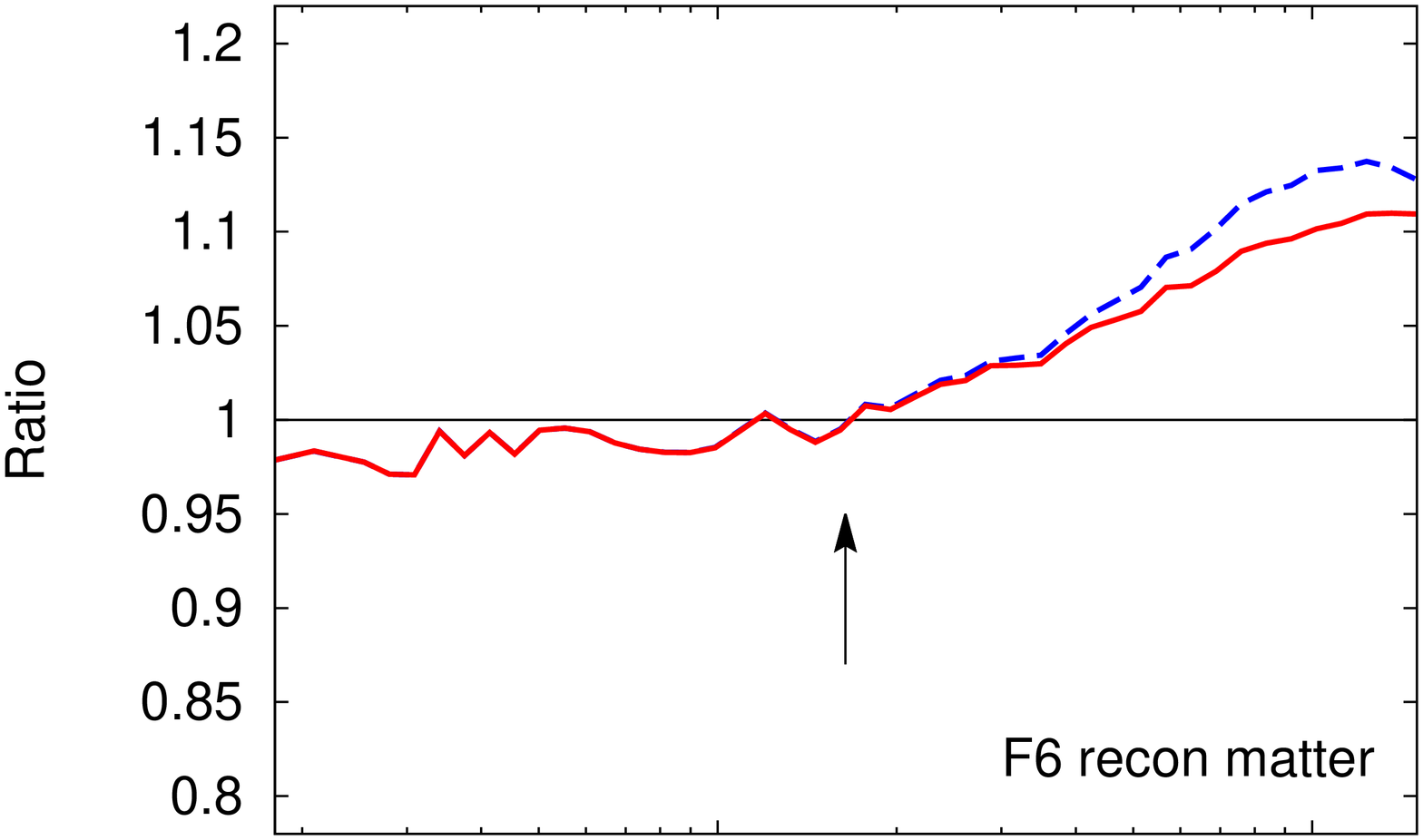}}\hspace{.1cm}
\subfloat{\includegraphics[width=41.5mm,trim=1.cm 3.4cm 1.5cm 2.4cm,angle=270]{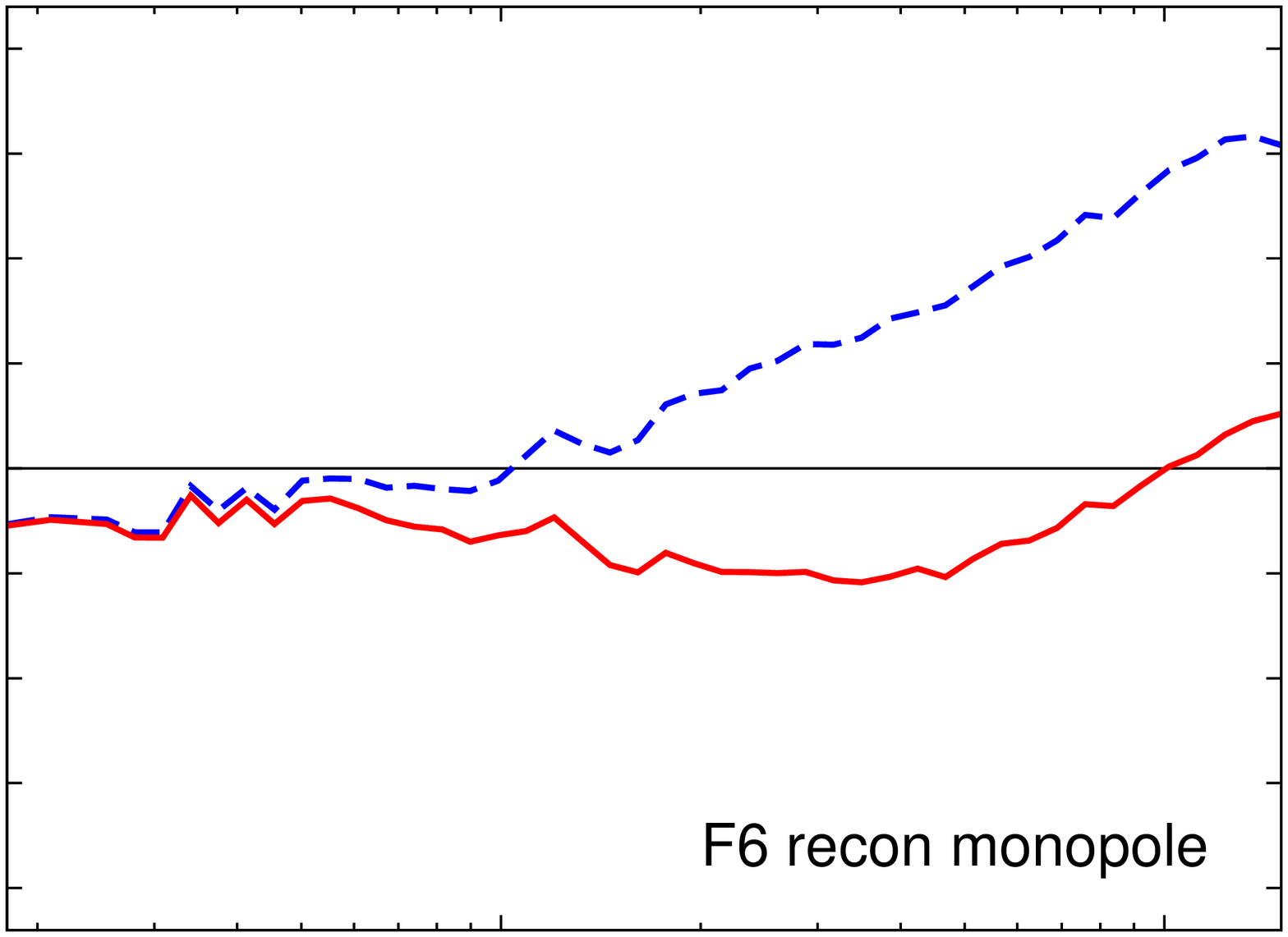}}\hspace{.1cm}
\subfloat{\includegraphics[width=41.5mm,trim=1.cm 3.4cm 1.5cm 2.4cm,angle=270]{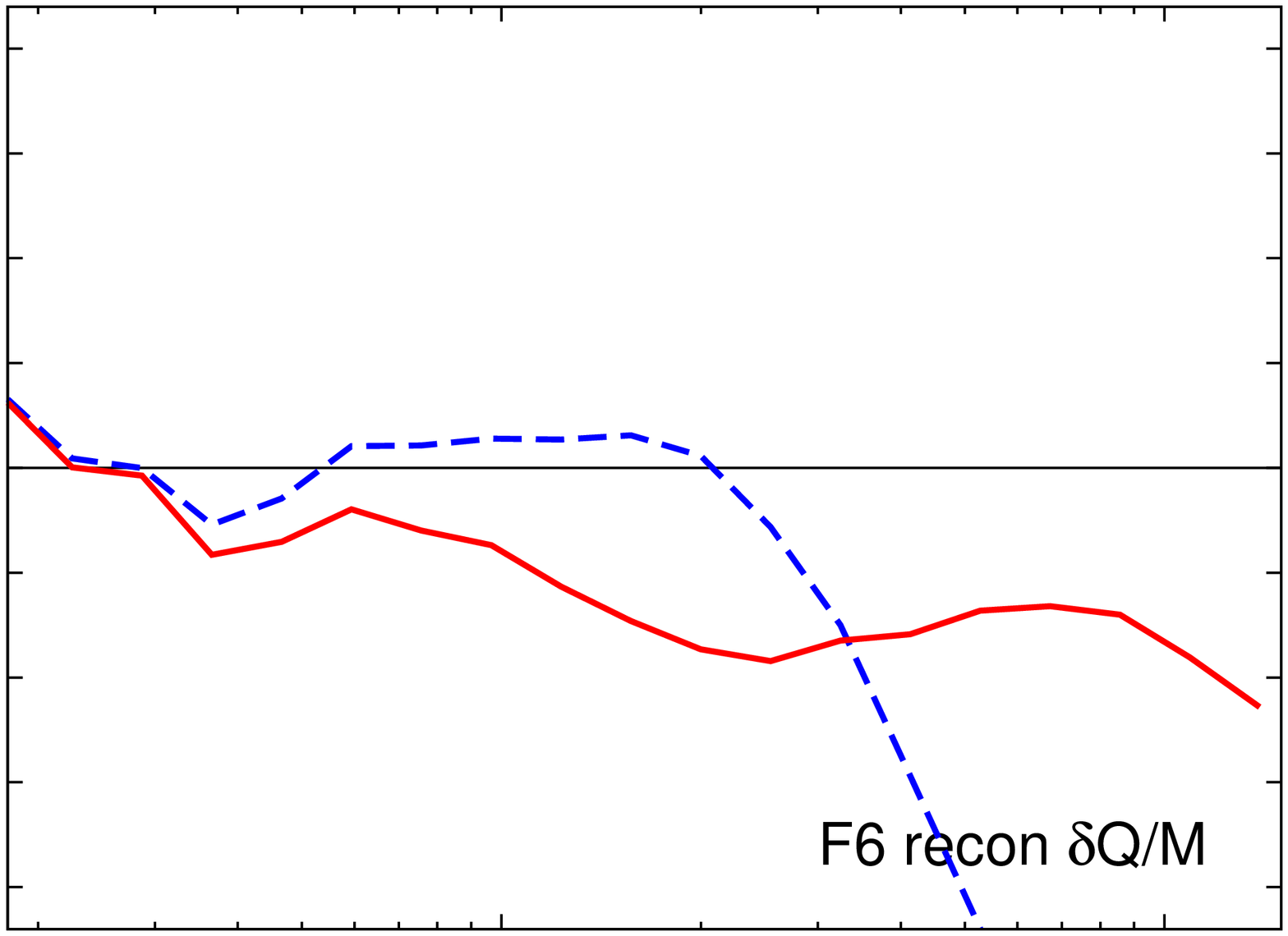}}}\\
\makebox[\textwidth][c]{
\subfloat{\includegraphics[width=41.5mm,trim=1.cm 3.4cm 1.5cm 2.4cm,angle=270]{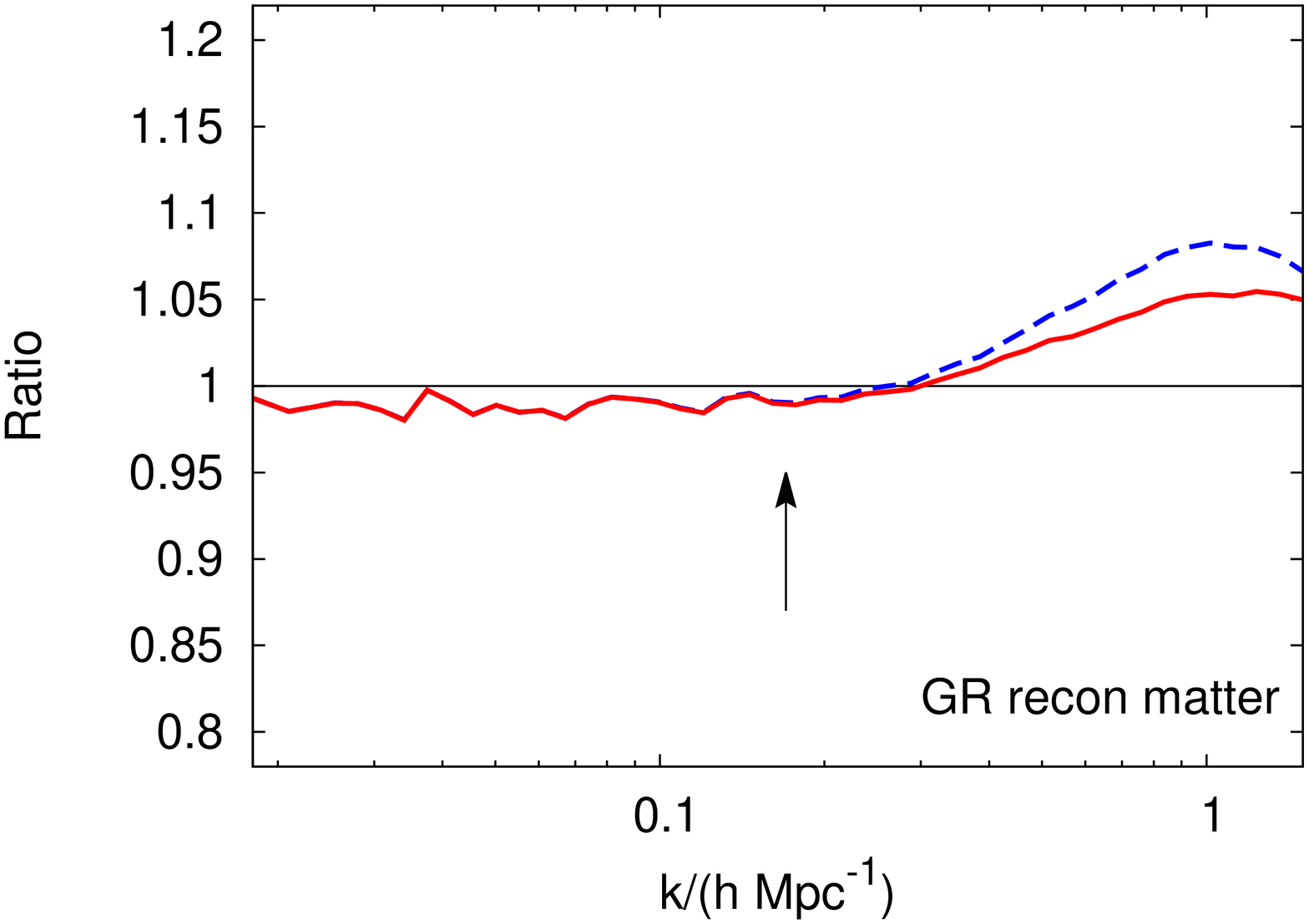}}\hspace{.1cm}
\subfloat{\includegraphics[width=41.5mm,trim=1.cm 3.4cm 1.5cm 2.4cm,angle=270]{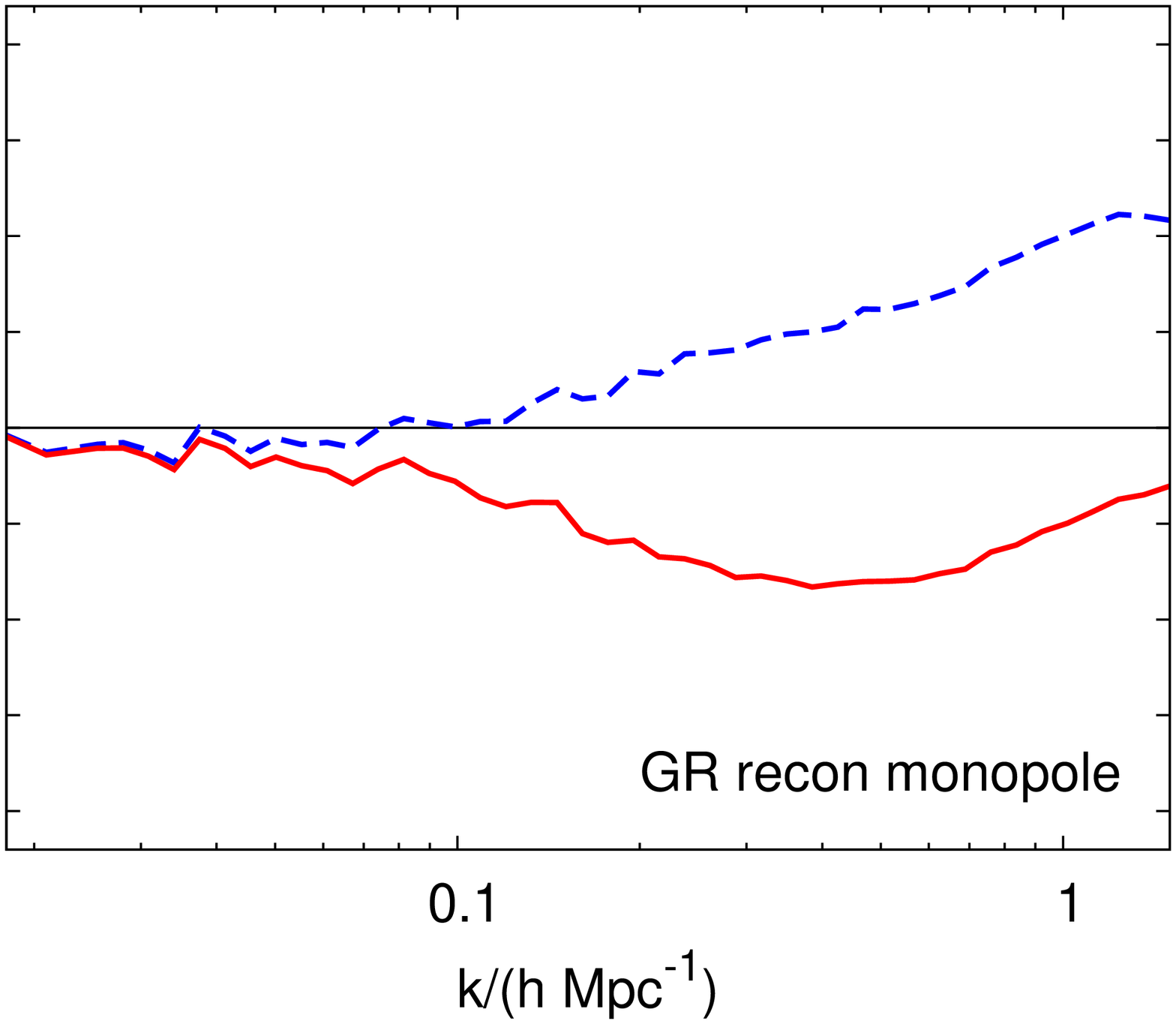}}\hspace{.1cm}
\subfloat{\includegraphics[width=41.5mm,trim=1.cm 3.4cm 1.5cm 2.4cm,angle=270]{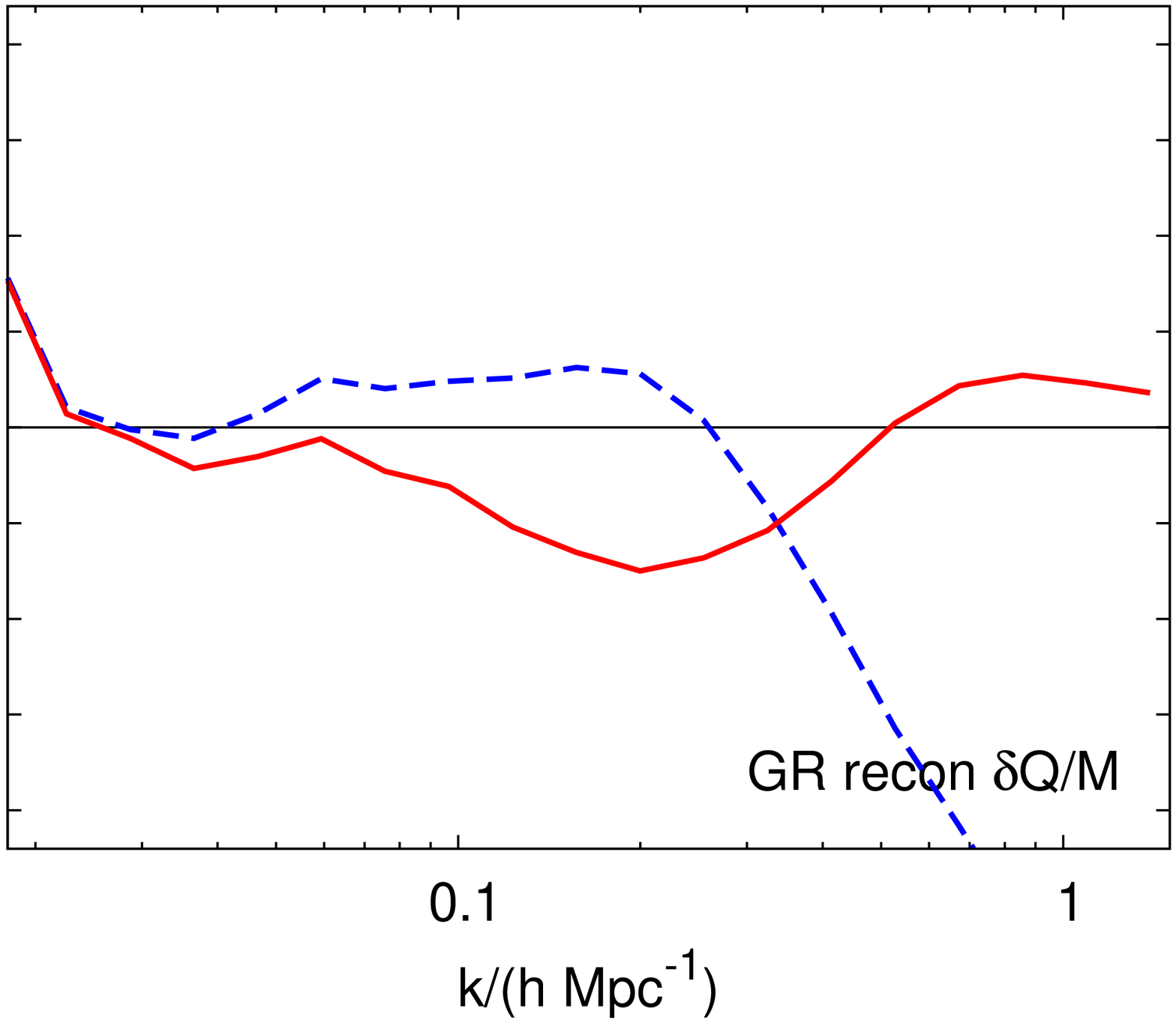}}}\\
\end{center}
\caption{The ratio of rescaled to target power of reconstituted particles in haloes rescaled from a $\Lambda$CDM halo catalogue compared to actual FOF tagged halo particles measured in the target F4 (top row), F5, F6 and GR (bottom row) simulations. We show the power spectra of halo particles (left column) and the redshift-space monopole (central column) and quadrupole to monopole ratio (right column). Two types of reconstituted haloes are shown; basic (dashed, blue) and advanced (solid, red) and these are described in the text. The matter and monopole spectra are low in the F4 and F5 cases at large scales by around $10$ per cent but the match improves somewhat at smaller scales. In contrast the GR and F6 cases are matched at around $3$ per cent at large scales but this degrades to $5$ and $10$ per cent at smaller scales. The quadrupole to monopole ratio is matched around $10$ per cent to $k=1\iMpc$ in each case. Overall the use of an advanced halo catalogue in reconstitution is preferable, but not by a large margin.}
\label{fig:recon_power}
\end{figure*}


The final comparison we make is that of halo particle distributions that are `reconstituted' from a rescaled catalogue compared to halo particles tagged by FOF in the target simulation. Simulated halo catalogues are commonly converted to galaxy catalogues by some halo-occupation-distribution prescription in which haloes are allocated a central galaxy and a number of satellites depending on their mass. The distribution of halo particles therefore stands as a proxy for a synthetic galaxy distribution.

We reconstitute haloes in two distinct ways. The first `basic' approach takes only the mass, position and velocity from the catalogues and then assumes that haloes are spherical objects with NFW profiles (equation~\ref{eq:nfw}) with concentrations from equation~(\ref{eq:bullock}) and velocity dispersions computed via the virial theorem (equation~\ref{eq:nfw_dispersion}) applied to the potential generated by a truncated NFW profile. For each halo in the catalogue, particles are then thrown down at random to fill up the density profile and each particle is given a velocity dispersion drawn from a Gaussian with width given by $\sigma_v$ that is independent of particle radius from the halo centre. See MP14a,b for more details of the reconstitution method.


The second `advanced' approach assumes that a catalogue contains more information, particularly a measured velocity dispersion and moment of inertia tensor. Diagonalizing this tensor provides the axis ratios of the halo (via the eigenvalues) and the orientation of the halo (via the eigenvectors). Haloes may be reconstituted with approximately correct aspherical orientations if the eigenvalues are scaled by a factor of $s$. The required scaling of the measured velocity dispersion can be computed from the ratio of target to original dispersions in equation~(\ref{eq:nfw_dispersion}) bearing in mind that the mass and dimensions of the halo have already been scaled by the earlier application of the method. If the FOF catalogue contained more information (for example concentrations and virial radii) then these can also be scaled using the ratio of some theoretical relationship (see MP14b).

In order to gauge how well our approach for reconstituting haloes works we compare the power spectra of reconstituted halo particles, generated from a rescaled halo catalogue, to those of the particles in FOF haloes measured in the target simulations. Results for the power spectra of reconstituted haloes compared to target haloes are shown in Fig.~\ref{fig:recon_power} for halo-matter and the redshift-space monopole and quadrupole. The error at large scales seen in the rescaled halo distribution persists, with up to $10$ per cent errors being present at large scales for the F4 and F5 cases, whereas the match is good to $3$ per cent for the F6 and GR cases. It is unsurprising that the error persists given that the power at large scales will simply be a re-weighted version of the large scale halo power shown in Fig.~\ref{fig:halo_scaling_power}. At smaller scales ($k>0.1\iMpc$) it is the F4 model that is best recovered, followed by the GR model. This makes sense if the F4 model is viewed simply as a GR model with a globally enhanced gravitational constant and no screening, the complexities of screening may make the F5 and F6 models more difficult to reconstitute. However, this conclusion is challenged in redshift space where the monopole and quadrupole are similarly recovered at around the $5$ per cent level for all models, irrespective of screening. The GR simulation is actually the least well reproduced at small scales in redshift space.

\section{Discussion}
\label{sec:discussion}


In this paper, we have demonstrated that reasonably accurate reproductions of simulations of \cite{Hu2007a} $f(R)$ MG models may be generated by rescaling standard gravity simulations using theoretical input to guide how changes between the cosmological models will manifest themselves.

If a simulation box is rescaled and displacement field altered for the full matter distribution, the power spectrum of matter in real space can be reproduced at the $\sim3$ per cent level out to $k=0.1\iMpc$. The match at non-linear scales can be slightly improved if one restructures halo density profiles; by doing so a $\sim3$ per cent match can be made in real space out to $k=1\iMpc$. The algorithm takes around 1 min to run on $512^3$ dark-matter particles and requires no information other than the initial and target cosmological parameters. This includes a step in which the displacement field is recreated from the evolved particle positions; run-time is reduced slightly if one is already in possession of a displacement field.

Similar results are produced at the level of the redshift-space monopole on linear scales, but for the stronger MG models the non-linear tail ($k>0.1\iMpc$) is grossly in error. We showed that this error must be due to incorrect FoG in the rescaled particles and that by artificially boosting the velocity dispersion in rescaled haloes to take account of the increased gravitational forces a $\sim5$ per cent match could be achieved to $k=1\iMpc$. Altering halo velocity structure also improves the match to the redshift-space quadrupole in the quasi-linear regime ($k\simeq 0.3\iMpc$). The quadrupole is an important quantity because the ratio between this and the monopole allows a measurement of the growth rate of density perturbations, which can be used to discriminate between gravity theories.

In restructuring haloes we chose to use a specific halo mass-concentration relation and velocity dispersion. It is possible that better results might have been obtained if we had chosen different relations or tuned parameters; but we avoided this as it was our goal to see how well rescaling works without additional parameter fitting. This is particularly because we aim to create a method that is applicable to models that have not yet been simulated and so tuning would not be possible in these cases.

In choosing the AW10 rescaling parameters $s$ and $z$ it was decided to use $\sigma(R)$ calculated from linear theory, which ignores the chameleon effect in the mass function. We showed that doing so would produce no discernible differences in the results -- $s$ and $z$ would have been very similar had we included collapse threshold ($\delta_\mathrm{c}$) differences induced by screening. However if we worked with simulations containing haloes of lower mass, particularly of the highly screened F6 model, we may have noticed differences. This may also have necessitated the use of a screened bias relation (which explicitly includes $\delta_\mathrm{c}$, rather than in combination $\nu=\delta_\mathrm{c}/\sigma$). On application of the MP14 method to haloes we reproduced the mass function in the MG models at the $\sim5$ per cent level. The clustering of haloes in real and redshift space was reproduced at the $\sim5$ per cent level but with the largest deviations at the largest scales, which are the scales at which the Zel'dovich correction was largest in our test example. No improvement at large scales was noticed if comparing to rescaled haloes measured from a rescaled \emph{particle} distribution (rather than just using the halo distribution) which leads us to believe the error must be due to the large ($\sim20$ per cent) displacement field correction required in the F4 and F5 case. The redshift-space quadrupole was reproduced particularly accurately in the halo distribution ($5$ per cent error at worst) for the F5, F6 and GR models but showed much larger errors in the F4 case.

We note that we generated our halo catalogues using the FOF method and did not attempt to unbind particles from FOF haloes that may not be gravitationally bound to the system. We do not believe that the unrelaxed haloes thus included will have a substantial impact upon our results, but in future work we intend to investigate the influence of the halo finder on the results of rescaling.

The final test we performed was to compare halo particle distributions reconstituted from a rescaled halo catalogue to halo particles measured in the target simulations. This is a similar test to comparing halo-occupation galaxies created from our rescaled halo catalogues to those that may be created from a full MG simulation. In this case the errors at large scales seen in the halo distributions were seen to persist in the F4 and F5 cases. This is expected because at large scales the power here is simply a re-weighted version of that in the case of the halo population. At non-linear scales ($k>0.1\iMpc$) the rescaled halo particles matched the target simulations at the $5-10$ per cent level in both real and redshift space. Results are slightly better if one uses more information from the halo catalogue (we used velocity dispersions and halo anisotropy information). Our halo catalogues did not include information such as virial radius and halo concentration and it remains to be seen if including these in a catalogue used in rescaling would further improve results.

So far we have ignored environmental dependence in applying rescaling; there is evidence of strong environment dependence in MG models for quantities such as the halo mass function because spherical model calculations depend on the local value of background $f_R$ field. If the rescaled halo catalogues and particle distributions respect this dependence has yet to be investigated, even in the case of standard gravity rescaling where environmental dependence is also expected. Given that AW10 rescaling only allows one to scale box quantities in a gross way, it is not obvious how one would include an environment dependence in this. But possibly some local rescaling of halo masses
as a function of environment could account for screening. We leave this to further work.

Finally, it should be pointed out that our reason for focusing on HS07 models in this work was purely because of the availability of simulations. There are no features of other MG models that obviously make them unsuitable for the type of methods used in this paper. In the appendix we include a brief discussion of the generalisation to chameleon theories that may be generated from a scalar-tensor action. Similarly, a straightforward next step would be to investigate how well rescaling works in models with different screening mechanisms, such as that of \cite{Vainshtein1972}. 


Overall, we find the results of this study encouraging, albeit with some reservations. We have shown that simulation rescaling via either the AW10 or MP14 approach is capable of capturing the low-order properties of models based on MG. As with previous applications to standard gravity, the approach is impressively rapid: To generate an $f(R)$ halo catalogue from a pre-existing standard-gravity halo catalogue takes a few seconds on a standard desktop computer for our test case of $\sim 100,000$ haloes and computational speed should scale linearly with halo number. The precision with which we can reproduce the power spectrum (particularly in the critical case of redshift space) is in the region of $5$ per cent, which is comparable to the accuracy achieved with standard gravity. This level of systematic error is barely tolerable with most current data sets, but for future studies it will not be sufficient. There can be various reactions to this. The first is to seek improvements of the method that will improve the precision; but it may be doubted whether one will ever achieve (say) $0.1$ per cent accuracy. Therefore, the utility of this approach is to be found in other ways. Even accepting the limit to precision delivered by rescaling, the method still permits a rapid exploration of parameter space, allowing a focus on a smaller sub-area for more detailed `exact' simulations. Furthermore, each rescaled simulation shares the virtue of detailed calculations in its ability to generate mock data that incorporate realistic non-linear effects in a cosmology of known background parameters. Thus the rescaling approach permits a large library of mock data with which the bias of practical parameter estimation schemes can be assessed. Both these aspects will be important tasks in the analysis of future large galaxy surveys, and we expect that rescaling will play a useful part in helping decide whether such data sets can robustly discriminate between alternative theories of gravity.

\section*{Acknowledgements}

AJM acknowledges the support of an STFC studentship and support from the European Research Council under the EC FP7 grant number 240185. LL has been supported by the STFC Consolidated Grant for Astronomy and Astrophysics at the University of Edinburgh. The simulations of this work used the COSMA Data Centric system at Durham University, operated by the Institute for Computational Cosmology on behalf of the STFC DiRAC HPC Facility (www.dirac.ac.uk). This equipment was funded by a BIS National E-infrastructure capital grant ST/K00042X/1, DiRAC Operations grant ST/K003267/1 and Durham University. DiRAC is part of the National E-Infrastructure. Please contact the author for access to research materials. 

\footnotesize{
\setlength{\bibhang}{2.0em}
\setlength\labelwidth{0.0em}
\bibliographystyle{mn2e_mead}
\bibliography{./../meadbib}
}

\appendix

\section{Scalar fields}
\label{sec:bd_gravity}

In this appendix we aim to demonstrate that it would be straightforward to apply the method outlined in this paper to more general MG theories. To this end we present the framework of a more general scalar-tensor model and discuss the \cite{Brans1961} model as an example. We use a metric convention of ($---$) defined by \cite*{b:MisnerThorneWheeler}.

Consider a seemingly distinct approach from that of $f(R)$ theories; coupling gravity to a scalar field in the action. A possible (Jordan frame) action is:
\begin{equation}
\eqalign{
S = \int\,\mathrm{d}^4x\frac{\sqrt{|g|}}{16\pi G}&\left[F(\phi)R+Z(\phi)\,\partial_a\phi\,\partial^a\phi-2V(\phi)\right]\cr&+\int\,\mathrm{d}^4x\sqrt{|g|}\lagr_\mathrm{m}(\psi_i,g_{ab})\ ,
}
\label{eq:scalar_tensor_action}
\end{equation}
where $\phi$ is the new field and $F$, $Z$, and $V$ are all arbitrary functions of $\phi$. Note that this model is but a sub-class of the \cite{Horndeski1974} model -- the most general scalar-tensor action that produces second order equations of motion in four dimensions while remaining Lorentz Covariant and local. If $F=1$ then the model becomes minimally coupled and so quintessence models are contained within the above action. However, in general the effective gravitational `constant' depends on the local value of $\phi$ via the function $F(\phi)$ and this function needs to be present for the theory to be an MG theory. Viable theories therefore require a screening mechanism whereby normal gravity may be recovered in regions such as the Solar system and Milky Way and the modification is confined to only being important on cosmological scales (\eg \citealt{Khoury2004}; \citealt{Hu2007a}).


The field equations of $g^{ab}$ and $\phi$ follow from variation of the action in equation~(\ref{eq:scalar_tensor_action}) (\eg \citealt{Esposito-Farese2001}). Variations with respect to $\phi$ lead to
\begin{equation}
2Z(\phi)\dalem\phi=F'(\phi)R-Z'(\phi)g^{ab}\partial_a\phi\,\partial_b\phi-2V'(\phi)\ ,
\label{eq:st_phi_equation}
\end{equation}
whereas variations with respect to $g^{ab}$ give
\begin{equation}
\eqalign{
&\left(R_{ab}-\frac{1}{2}g_{ab}R\right)F(\phi)+\left(g_{ab}\dalem-\nabla_a\,\nabla_b\right)F(\phi)\cr
&+Z(\phi)\partial_a\phi\,\partial_b\phi-g_{ab}\left[\frac{1}{2}Z(\phi)\partial_c\phi\,\partial^c\phi-V(\phi)\right] \cr
&=-8\pi GT_{ab}\ .
}
\label{eq:st_g_equation}
\end{equation}
The trace of this equation then gives energy conservation
\begin{equation}
(3\dalem-R)F(\phi)-Z(\phi)\partial_a\phi\,\partial^a\phi+4V(\phi)=-8\pi G T\ .
\label{eq:st_trace}
\end{equation}
As one can see, these equations couple the scalar to gravity in a non-trivial manner.


Any $f(R)$ theory can be mapped on to a scalar-tensor theory and thus $f(R)$ theories represent a sub class of scalar-tensor theories. If one defines $1+f'(R)=\phi$, where $f'(R)\equiv\mathrm{d}f/\mathrm{d}R$, and $-f'(R)R+f(R)=-2V(\phi)$ the action is left as that of a non-minimally coupled scalar-tensor theory:
\begin{equation}
S=\int\,\mathrm{d}^4x\,\sqrt{|g|}\left[\frac{\phi R-2V(\phi)}{16\pi G}+\lagr_\mathrm{m}(\psi_i,g_{ab})\right]\ ,
\label{eq:fr_action_scalar}
\end{equation}
with the restricted functional form $F=\phi$ and $Z=0$ in the action in equation~(\ref{eq:st_g_equation}). The function $V(\phi)$ required to directly map on to HS07 theories is
\begin{equation}
V(\phi)=\Lambda+\frac{1}{2}f_{R0}\bar{R}_0\left(1+\frac{1}{n}\right)\left(\frac{\phi-1}{f_{R0}}\right)^{n/(n+1)}\ ,
\end{equation}
which can be rewritten using $\alpha=n/(n+1)$ and $A=f_{R0}^{1-\alpha}\bar{R}_0/\alpha$:
\begin{equation}
V(\phi)=\Lambda+\frac{1}{2}A(\phi-1)^\alpha\ ,
\label{eq:fr_Vphi}
\end{equation}
note that $n\in[0,\infty]$ maps to $\alpha\in[0,1]$.

A straight-forward extension of the work presented in this paper that adds only a single additional degree of freedom is to consider \cite{Brans1961} type theories with constant $\omega$, which involves adding a non-canonical kinetic term to the action~(\ref{eq:fr_action_scalar}) of $Z(\phi)=\omega/\phi$ (see equation~\ref{eq:scalar_tensor_action}) but keeping $F(\phi)=\phi$ and $V(\phi)$ as in equation~(\ref{eq:fr_Vphi}). \cite{Lombriser2014a} provides a review of structure formation in such models. These models still exhibit chameleon screening but allow for modifications to the gravitational field strength other than just a factor $4/3$. These theories have linear enhancements of a factor $1+1/(3+2\omega)$ where $\omega>-3/2$ must hold. Note that $\omega=0$ corresponds to $f(R)$ gravity. The enhancement can be arbitrarily large but gravity is not allowed to be weakened.



\label{lastpage}
\end{document}